\documentclass[CRPHYS,Unicode,manuscript]{cedram}

\DeclareMathOperator{\Arcsin}{Arcsin}
\DeclareMathOperator{\Arccos}{Arccos}
\usepackage{comment}
\usepackage{multirow}
\usepackage{colortbl}
\usepackage{amssymb}
\usepackage{romanbar}

\usepackage{nicematrix}
\usepackage{tikz,colortbl}
  \usetikzlibrary{shapes.misc}
\usetikzlibrary{calc}
\usepackage{zref-savepos}

\newcounter{NoTableEntry}
\renewcommand*{\theNoTableEntry}{NTE-\the\value{NoTableEntry}}

\newcommand*{\notableentry}{%
  \multicolumn{1}{@{}c@{}|}{%
    \stepcounter{NoTableEntry}%
    \vadjust pre{\zsavepos{\theNoTableEntry t}}
    \vadjust{\zsavepos{\theNoTableEntry b}}
    \zsavepos{\theNoTableEntry l}
    \hspace{0pt plus 1filll}%
    \zsavepos{\theNoTableEntry r}
    \tikz[overlay]{%
      \draw[black]
        let
          \n{llx}={\zposx{\theNoTableEntry l}sp-\zposx{\theNoTableEntry r}sp},
          \n{urx}={0},
          \n{lly}={\zposy{\theNoTableEntry b}sp-\zposy{\theNoTableEntry r}sp},
          \n{ury}={\zposy{\theNoTableEntry t}sp-\zposy{\theNoTableEntry r}sp}
        in
        (\n{llx}, \n{lly}) -- (\n{urx}, \n{ury})
        (\n{llx}, \n{ury}) -- (\n{urx}, \n{lly})
      ;
    }%
  }%
}

\DeclareMathOperator{\Arccosh}{Arcosh}

\begin{document}
 
\title{On the art of designing effective space-times with free surface flows in Analogue Gravity}\ source
  file

\alttitle{De l'art de concevoir des espaces-temps effectifs avec des écoulements à surface libre en Gravitation Analogue}

\author{\firstname{Alexis}~\lastname{Bossard}}
\address{Institut Pprime, CNRS-Université de Poitiers-ISAE ENSMA, Poitiers, France}
\author{\firstname{Nicolas}~\lastname{James}}
\address{LMA, CNRS-Université de Poitiers, Poitiers, France}
\author{\firstname{Valentin}~\lastname{Jules}}
\addressSameAs{1}{Institut Pprime, CNRS-Université de Poitiers-ISAE ENSMA, Poitiers, France}
\author{\firstname{Johan}~\lastname{Fourdrinoy}}
\addressSameAs{1}{Institut Pprime, CNRS-Université de Poitiers-ISAE ENSMA, Poitiers, France}
\author{\firstname{Scott}~\lastname{Robertson}}
\addressSameAs{1}{Institut Pprime, CNRS-Université de Poitiers-ISAE ENSMA, Poitiers, France}
%
\author{\firstname{Germain}~\lastname{Rousseaux}\CDRorcid{0000-xxx-0000-yyyy}\IsCorresp}
\addressSameAs{1}{Institut Pprime, CNRS-Université de Poitiers-ISAE ENSMA, Poitiers, France}
\email[G. Rousseaux]{germain.rousseaux@cnrs.fr}


\ESM{Supplementary material for this article is supplied as a separate archive \cdrattach{Animation article.zip}}


\begin{abstract}

Accelerating/decelerating trans-critical flows (waterfalls/cataracts) are analogous to space-times of black holes/white fountains since the pioneering work of Schützhold \& Unruh in 2002. A single number is usually employed to classify trans-criticality namely the local depth Froude number which is the ratio between the local current speed and the local celerity of long gravity waves analogous to the light celerity. When the former reaches one, water waves are no more able to propagate upstream: the hydraulic black hole is a river of no return for them. At a higher level of understanding, two global dimensionless numbers, the upstream Froude number $Fr_{up}$ and the obstruction ratio $r_{up}$ (the height of a bottom obstacle, the underlying geometry inducing the effective space-time, divided by the upstream water depth) are essential to distinguish subcritical, trans-critical and supercritical zones in the - $Fr_{up}$ versus $r_{up}$ – hydraulic and non-dispersive diagram. The relationship between both global parameters for transcritical flows turns out to be a peculiar limit of the behaviour of boats navigating in confined media like canals or locks with a generalized obstruction factor based on the ratio between the boat section and the canal section. Here, we revisit the classification of flows over obstacles in open water channel taking into account both effects of dispersion and scale, two neglected topics so far. For the first time, we give a complete classification of flows in an open water channel based on sub-pixel detection method measurements of the free surface supported by numerical simulations. We generalized the obstruction factor by a filling factor taking into account the maximum height of the water channel, a crucial parameter that was overlooked so far. Our ultimate purpose is to understand how to reproduce in the laboratory analogues of curved space-times from the dynamical point of view.
\end{abstract}

\begin{altabstract}
Les écoulements transcritiques accélérés/décélérés (chutes d'eau/cataractes) sont analogues aux espaces-temps des trous noirs/fontaines blanches depuis les travaux pionniers de Schützhold \& Unruh en 2002. Un seul nombre est généralement utilisé pour classifier la transcriticité, à savoir le nombre de Froude de profondeur local qui est le rapport entre la vitesse locale du courant et la célérité locale des ondes longues de gravité analogue à la célérité de la lumière. Lorsque le premier atteint 1, les vagues ne peuvent plus se propager vers l’amont : le trou noir hydraulique est pour elles une rivière sans retour. À un niveau de compréhension supérieur, deux nombres globaux sans dimension, le nombre de Froude en amont $Fr_{up}$ et le taux d'obstruction $r_{up}$ (la hauteur d'un obstacle de fond, la géométrie sous-jacente induisant l'espace-temps effectif, divisé par la profondeur de l'eau en amont) sont essentiels pour distinguer les zones sous-critiques, trans-critiques et supercritiques dans le diagramme - $Fr_{up}$ versus $r_{up}$ - hydraulique et non dispersif. La relation entre les deux paramètres globaux pour les écoulements transcritiques s'avère être une limite particulière du comportement des bateaux naviguant dans des milieux confinés comme des canaux ou des écluses avec un facteur d'obstruction généralisé basé sur le rapport entre la section du bateau et la section du canal. Ici, nous revisitons la classification des écoulements au-dessus des obstacles dans les canaux à surface libre en prenant en compte à la fois les effets de dispersion et d'échelle, deux sujets négligés jusqu'à présent. Pour la première fois, nous donnons une classification complète des écoulements dans un canal à surface libre basée sur des mesures avec une méthode de détection sous-pixel de la surface libre appuyées par des simulations numériques. Nous avons généralisé le facteur d'obstruction par un facteur de remplissage prenant en compte la hauteur maximale du canal à surface libre, paramètre crucial et oublié jusqu'à présent. Notre objectif ultime est de comprendre comment reproduire en laboratoire des analogues d’espaces-temps courbes du point de vue dynamique.
\end{altabstract}



\maketitle
\section{Introduction}

Analogue Gravity aims to realise, in the laboratory, physical systems whose behaviour mimics that of fields propagating in a curved spacetime~\cite{LivingReview, Robertson-2012, faccio2013analogue, rousseaux2020classical, barcelo2019analogue, braunstein2023analogue, almeida2023analogue, field2024building}. 
If a system is divisible into a strong classical background plus weak fluctuations (either classical or quantum) on top of it, then the fluctuations will ``see'' the background as an effective spacetime.
In this way, what are often regarded as rather exotic features of curved spacetime can be realised in laboratory settings.  The most notable and extensively studied is the black hole spacetime: if the background is flowing, and this flow passes from subcritical to supercritical so that all fluctuations are dragged with the flow on the downstream side, then the point at which the passage occurs is analogous to a black hole horizon, for it is possible to fall past this point but impossible to climb back out.  Remarkably, such an analogue horizon is associated with an analogue of the Hawking effect, which causes the horizon to radiate.  This illustrates that Analogue Gravity is not merely a curious correspondence between radically different systems, but can reveal genuine physical phenomena.

There are many analogue systems currently under study, in several of which the analogue Hawking effect has been experimentally observed ({\it e.g.}, surface waves on flowing water \cite{euve2016observation}, phonons in an atomic condensate \cite{steinhauer2016observation}, and photons in a varying-index optical fiber \cite{Drori-et-al-2019}).
While the gravitational analogy is independent of the underlying dynamics responsible for the determination of the background~\cite{LivingReview, Robertson-2012, faccio2013analogue, rousseaux2020classical, barcelo2019analogue, braunstein2023analogue, almeida2023analogue}, any experimentalist in Analogue Gravity has to cope with the many dynamical regimes of the probed condensed matter systems before selecting the best ones that will mimic the original astrophysical system.

In this work, we study the particular case of open channel flows in the presence of surface waves, described by the Navier-Stokes equations of Classical Hydrodynamics \cite{Schuetzhold-Unruh-2002, Rousseaux-BASICS-2013, rousseaux2020classical}.
We present a variety of flow regimes above a fixed bottom obstacle in an open water channel amenable to a physical interpretation within Analogue Gravity and possibly to the discovery of generalized phenomena \cite{field2024building}.
We give, for the first time, a classification of the flow regimes that are amenable to a kinematical analogy with astrophysics, combining knowledge of hydraulic flow regimes~\cite{Faber,Baines,Long,houghton1968nonlinear,pratt1982dynamics,Pratt1984,lawrence1987steady,bouhadef1988contribution,Lowery-Liapis-1999,zerihun2004one,pompee2015modelling} with the effects of wave dispersion~\cite{rousseaux2020classical,Schuetzhold-Unruh-2002,Rousseaux-BASICS-2013,nardin2009wave,rousseaux2010horizon,weinfurtner2011measurement,chaline2013some,coutant2014undulations,Michel-Parentani-2014,Peloquin,robertson2016scattering,Euve-Rousseaux-2017,euve2017interactions,Michel-et-al-2018,Coutant-Weinfurtner-2018,Binder} to infer the most general flow diagrams.
We compare our results with numerical simulations and discuss their implications for Analogue Gravity from a theoretical point of view. 
As a final outcome, we classify the historical experiments in Analogue Gravity within the regimes described in this work.
Our hope is that this will guide practitioners to design Analogue Gravity experiments most suited to the realisation of non-dispersive and transcritical ``flows'' of gravitational spacetimes.

The paper is organised as follows.
In Section~\ref{sec:AG}, we present the basics of Analogue Gravity and describe how it applies to the particular case of interfacial hydrodynamics. In Section~\ref{sec:classification}, we theoretically describe the three hydrodynamical regimes 
characterised solely by the flow profile, which we link to the form of the free surface.  This description is purely hydraulic in nature, and does not take wave dispersion into account. 
In Section~\ref{sec:experiment} we describe our experimental setup, and point out that the experimentally observed surface classification is richer than that derived in Sec.~\ref{sec:classification}. 
In Sections~\ref{Th-disp} and~\ref{Exp-disp} we show, from a theoretical and experimental point of view (respectively), that with 
knowledge of the dispersive speeds we can provide a full classification of the underlying bulk speed regimes below the free surface to get a larger set of experimental flow regimes thanks to the measurements of the free surface heights for a given flow rate, downstream boundary condition and geometry of the obstacles probed.
We conclude in Section~\ref{sec:conclusion}. In addition, the theory developed here, supported by numerical simulations that validate the use of the acoustic metric, makes it possible to review the historical experiments in Analogue Gravity which had been implemented without any theoretical guidance and which are therefore being interpreted for the first time through the prism of this theory. We decided to append the discussion around these historical experiments as well the numerical justifications of our hypotheses (also a premiere in the literature so far) so as not to weight down the main body of the text. We would like to emphasise that the present authors were surprised by the very large number of hydrodynamic regimes identified both experimentally and theoretically, many of which are absent from the literature. This abundance of hydrodynamic regimes is similar to the well-known classification in crystallography with 230 space groups, and it must be acknowledged that this implies the unusual length of this paper.

\section{Analogue Gravity and its application to interfacial hydrodynamics}
\label{sec:AG}

\subsection{Analogue Gravity}

\subsubsection{The analogy}

The essential idea behind Analogue Gravity is that many physical systems permit a conceptual separation between a strong classical background and weak fluctuations on top of that background.  As long as the system is interacting in some way, and as long as the fluctuations are sufficiently weak, the fluctuations ``see'' and respond to the background without affecting it or each other.  The background can be thought of as engendering an effective curved spacetime, within which the fluctuations are propagating.

Unruh~\cite{Unruh-1981} developed this idea in the context of acoustic waves in a flowing fluid, showing explicitly that the wave equation describing the propagation of the acoustic waves takes exactly the same form as for a massless scalar field in a curved spacetime:
\begin{equation}
    \frac{1}{\sqrt{-\text{g}}}\partial_\mu\left(\sqrt{-\text{g}}\:g^{\mu\nu}\partial_\nu\delta\phi\right)=0\quad\text{with}\quad \text{g}=\det\left(g_{\mu\nu}\right) \,.
    \label{beltrami}
\end{equation}
Here, $\delta\phi$ is the fluctuation field in question: it describes the acoustic waves and plays the mathematical role of a massless scalar field.  The tensor $g_{\mu\nu}$ (with inverse $g^{\mu\nu}$) describes the influence of the background flow on the propagation of the acoustic waves; it plays the same mathematical role as the spacetime metric of General Relativity, and thus encodes the presence of an analogue ``gravitational'' field.

Unruh finds the following explicit form for the effective metric:
\begin{equation}
    {\rm d}s^{2} = g_{\mu\nu}{\rm d}x^{\mu}{\rm d}x^{\nu} 
    = \frac{\rho}{c} \left[ c^{2} {\rm d}t^{2} - \left( {\rm d}\vec{x} - \vec{v} \, {\rm d}t \right)^{2} \right] \,,
    \label{eq:PG}
\end{equation}
where $\rho$ is the density of the background fluid, $c$ is the local speed of acoustic waves with respect to the fluid, and $\vec{v}$ is the local flow velocity of the fluid.  Interestingly, up to the factor of $\rho/c$ out front, the metric has a rather straightforward interpretation: the acoustic wave speed $c$ plays the role of the speed of light (which is compatible with the fact that $\delta\phi$ satisfies the same equation~(\ref{beltrami}) as a massless scalar field), and the spatial interval undergoes a Galilean transformation due to the fluid velocity $\vec{v}$.  If we were to move {\it with} the fluid, then ${\rm d}\vec{x}-\vec{v} \, {\rm d}t$ would vanish and the interval would be purely time-like.  In other words, it is as if space were actually moving with the fluid, and spatial displacements are thus to be measured in a reference frame in which the fluid is locally at rest.  The most important consequence of this is that all speeds measured in this reference frame are shifted by the flow velocity $\vec{v}$ in the lab frame.  If the flow velocity is faster than the ``light speed'' $c$, then everything is necessarily dragged in the direction of the flow.

This simple and intuitive picture has far-reaching consequences.  For a start, we can model a black hole if we realise an accelerating flow that passes from a subcritical region (where $\left|\vec{v}\right| < c)$ to a supercritical region (where $\left|\vec{v}\right| > c)$.  Such a flow is said to be {\it transcritical}.  In such a scenario, when approaching the transition from the subcritical side, it becomes more and more difficult for waves to propagate back upstream, and once we cross into the supercritical region it is no longer possible for waves to propagate upstream, with everything being dragged downstream by the flow.  The transition point (where $\left|\vec{v}\right| = c$) is thus analogous to a black hole horizon.
Moreover, it is also possible in such analogue systems to realise the time-reversed version of a black hole, which occurs when the flow is decelerating and thus passes from a supercritical region to a subcritical one.  The transition point here is called a ``white hole'' horizon, or sometimes (as we will refer to it in the rest of this article) a ``white fountain'' horizon.

Intriguingly, in a coordinate system of a falling observer known as Painlev\'{e}-Gullstrand coordinates \cite{eisenstaedt1982histoire}, the spacetime metric of a Schwarzschild black hole (the simplest kind, with no rotation or electric charge) appears in exactly the form~(\ref{eq:PG}), with $\rho/c \to 1$ and $\vec{v} = - c\sqrt{r_{S}/r} \, \hat{r}=\sqrt{2GM/r}\, \hat{r}$, where $r_{S} = 2GM/c^{2}$ is the Schwarzschild radius and marks the point where $\left|\vec{v}\right| = c$ (the usual expression of the Schwarschild metric found in textbooks corresponds to the point of view of an asymptotic observer that is limited to physical effects occurring in the exterior region of the black hole horizon with a coordinate singularity removed by the Painlev\'{e}-Gullstrand choice of coordinates \cite{eisenstaedt1982histoire}). $\sqrt{2GM/r}$ is the well known Newton escape velocity that tells us if an object is able to escape from the gravitational attraction of the body with mass $M$ at a distance $r$ (where $G$ is the gravitational constant).
Moreover, while the more general Kerr (rotating) black hole does not map precisely onto a metric of the form~(\ref{eq:PG})~\cite{Hamilton-Lisle-2008}, the key additional features associated with rotation are reproduced by the metric~(\ref{eq:PG}) where the flow velocity $\vec{v}$ has a non-zero angular component~\cite{Visser-1998,Torres-2017}.

The above discussion implies that one can infer the existence of an effective horizon by tracking a single dimensionless quantity, the Mach number $Ma=v/c$ (also known as the Froude number in interfacial hydrodynamics; see below) that is equal to unity at the horizon.
The Mach/Froude number defines the effective spacetime seen by long-wavelength waves in the non-dispersive limit, and is thus a crucial parameter in the construction of the analogy with gravity.  In practice, however, experimentalists have to create transcritical flows (or their dispersive generalizations) by imposing constraints on their experimental setup, and the Mach/Froude number achieved depends nontrivially on those constraints.
For example, while in hydrodynamics a flow profile is generated by having a fixed obstacle in an open channel with a flowing fluid pushed by an externally-driven pump \cite{euve2016observation}, we might also consider moving the obstacle at constant speed in an initially static water column. The latter case is directly analogous to cold atom systems with a LASER step moving relative to the trapping well along the condensate \cite{steinhauer2016observation}: in the absence of any other constraints, the two situations would be equivalent according to the principle of relativity. In practice, however, these cases are not equivalent as the flows do not actually extend to infinity, and so boundary conditions at the ends of the flow may need to be taken into account: the water may have an initial static depth, and/or a downstream gate may affect the dynamical water depth for a given flow rate.
Furthermore, and as already noted, the Mach/Froude number is defined with respect to the speed of long-wavelength waves, and thus corresponds to a non-dispersive limit.  But it is often the case (especially in Analogue Gravity) that dispersion needs taking into account, and the dispersive properties of surface waves depend on characteristic length scales of the system.
Hence, in practice, the Mach/Froude number is not sufficient to fully describe the resulting dynamical flow regimes pertinent to the behaviour of surface waves. 

The vertical size of the obstacle (or the LASER step height in the analogous BEC experiments \cite{steinhauer2016observation}) was considered as a relevant additional parameter and was subsequently adimensionalized in previous theoretical studies so as to classify flow regimes.  In hydraulics, theoreticians have introduced the obstruction factor, namely the ratio between the maximum height of the obstacle to the upstream water depth \cite{Long, Baines}. Hence, the dispersionless and dimensionless flow diagram consists in plotting the upstream Froude number (of kinematics origin) as a function of the upstream obstruction factor (of geometrical origin).




\subsubsection{Negative-energy waves}

The motivation behind Unruh's formulation of the analogy between a flowing fluid and a gravitational field was to note that, since the wave kinematics should be the same in both systems (being sensitive only to the effective spacetime metric), then just as black holes are predicted to emit Hawking radiation, so {\it analogue} black holes in flowing fluids should be subject to an analogue Hawking effect.  To understand how this occurs and what it entails, we must introduce two new physical concepts.

First, the supercritical region (the Mach number is superior to one) where the waves are all dragged downstream by the flow is associated with the existence of {\it negative-energy waves}~\cite{musha1964,ostrovskii1986,stepanyants1989,fabrikant-chapter,bermudez2024}.  In particular, those waves whose velocity has been reversed by the strength of the flow are found to have negative energy in the laboratory reference frame.  This is allowed because the background is in a highly excited state, with a lot of kinetic energy.  The presence of counter-propagating waves, which are trying to propagate against the flow, actually reduces the total kinetic energy of the system so that the total energy is lower than what it would be in the absence of such waves.

To be more precise: Any plane wave is fully characterised by its frequency $\omega$ and its wave number $k$.   In a flowing fluid, we may also define a co-moving frequency $\omega_{\rm cm}$, which is the frequency measured in a reference frame in which the fluid is locally at rest.  These are related by a Doppler shift: $\omega_{\rm cm} = \omega - v k$, where $v$ is the local fluid velocity.  If $\omega$ and $\omega_{\rm cm}$ have the same sign (as is the case when $v$ is sufficiently small), then the wave has positive energy.  However, if the flow is sufficiently strong, then $\omega$ and $\omega_{\rm cm}$ can have opposite signs.  In this case, the wave has negative energy.

This distinction between the wave's frequency $\omega$ and its energy leads to interesting phenomena. On stationary flows, it is the frequency $\omega$ that is conserved by a linear scattering process.  This means that in certain situations, in particular where there is a horizon, positive- and negative-energy waves exist with the same frequency and can scatter into each other. Since total wave energy is conserved in such a process, it must be the case that a partial scattering of, say, a positive-energy wave into a negative-energy wave is accompanied by an amplification of the positive-energy waves. This is referred to as {\it anomalous scattering}, and can be thought of as the wave extracting energy from the background. This is exactly what happens in the Hawking effect: an incident wave (or a quantum fluctuation) stimulates the emission of positive- and negative-energy waves on either side of the horizon corresponding to a pair of particle and anti-particle also known as the Hawking mode and its negative partner. The total energy remains constant, but energy is emitted towards the black hole exterior and is accompanied by a {\it decrease} of energy within the black hole interior, eventually leading to the evaporation of the black hole.

The amount of stimulated emission that occurs generally depends on the form of the background flow ({\it i.e.}, on the variation of the effective spacetime) and also on the frequency of the incident wave. A final remarkable feature of the Hawking effect is that the anomalous scattering coefficient (or the amplification factor) shows a frequency-dependence that has the same form as a thermal (Planck) spectrum. This spectrum is characterised by a single quantity, the so-called Hawking temperature, which is itself proportional to the derivative of the wave speed at the horizon. So it is the behaviour of the flow at a single point, the effective horizon, that fully governs the Hawking effect. Moreover, in the quantum limit where the radiation is seeded only by vacuum fluctuations, this thermality of the scattering coefficients transfers exactly into a thermal radiation spectrum, so that the Hawking temperature can be regarded as the true physical temperature of the black hole.

\subsubsection{Dispersive waves}

A second important ingredient, which helps to bridge the gap between the gravitational scenario and the laboratory analogues, is the occurrence of dispersive waves, {\it i.e.}, waves whose speed depends on their frequency and/or wave number.  One can appreciate the importance of dispersion by noting that, in its absence, the frequency and wave number of a wave that is exactly counter-propagating against the flow are related by $\omega/k = v-c$, this being the total phase velocity of the wave.  Since $\omega$ is conserved in a stationary background, and since $v-c$ vanishes at a horizon, then the non-dispersive scenario implies that the wave number $k$ diverges as the horizon is approached; equivalently, the wavelength of a counter-propagating wave becomes arbitrarily small in the vicinity of the horizon.  This is a known issue in gravity, referred to as the {\it trans-Planckian problem} because it implies the existence of arbitrarily short wavelengths than are way below the Planck scale, and where we have good reason to doubt the validity of our current theories.  It is particularly relevant for the Hawking effect, since the radiation emitted from the horizon can be traced back to these arbitrarily short wavelengths.

In laboratory systems, this ``trans-Planckian'' issue is avoided by the ubiquity of dispersion~\cite{unruh1995,brout1995}.  The non-dispersive limit relevant to long wavelengths is an approximate behaviour, which is valid only when the wave is too long to resolve the short-scale physics at play in the analogue system.  But this implies that, when the wave is sufficiently short, it becomes sensitive to the short-scale physics and its behaviour changes.  This can typically be modelled as a non-trivial wave number-dependence of the frequency: $\omega = \omega(k)$, with the physically relevant wave speed being the group velocity $\partial \omega/\partial k$.  The change in the group velocity as $k$ becomes larger means that the wave is eventually ``tuned out'' of the horizon's grip, and the wave propagation carries on towards the asymptotic regions.  The implications of dispersion for the Hawking effect have been extensively studied in the theoretical Analogue Gravity literature, and remarkably, within an appropriate regime where the background varies relatively slowly, the effect is found to be robust against the introduction of dispersion.  The important novelty concerns the seeds and/or products of the process: at a black hole horizon, incident dispersive waves (of short wavelength) are scattered into outgoing hydrodynamical waves (of long wavelength), while at a white fountain horizon the inverse occurs.

Another interesting consequence of dispersion is that it can allow the existence of {\it trapped modes}.  These typically occur when two horizons are present, one a black hole and the other a white fountain, forming a cavity in between them; say, a supercritical region sandwiched between two subcritical regions, or indeed a subcritical region in between two supercritical regions.  Then there can be a continuous back-and-forth conversion between hydrodynamic and dispersive waves inside the cavity: a hydrodynamic wave is reflected as a dispersive wave at the white fountain horizon, and then this dispersive wave is reflected back as a hydrodynamic wave at the black hole horizon, and so on.  At each reflection, there can be anomalous scattering into other non-trapped modes, which means that the trapped modes are continuously amplified.  This is the so-called {\it black hole laser} (BHL) effect~\cite{corley1999}, which refers to a dynamical instability of the flow due to the simultaneous presence of trapped modes and anomalous scattering.

\subsection{Analogue Gravity in interfacial hydrodynamics}

As first demonstrated in~\cite{Schuetzhold-Unruh-2002}, the Analogue Gravity framework can be applied to surface waves on flowing water.  These surface waves live in an effective spacetime with two spatial dimensions (and of course one of time), though typically in channel flows we consider a non-trivial dependence in only the longitudinal direction. 

The relevant physical fields are the flow velocity $\vec{V}$ and the pressure $P$.  Let us assume that these do not depend on the transverse coordinate $y$, and moreover that $\vec{V}$ has no $y$-component.  We write the Cartesian coordinates of $\vec{V}$ explicitly as $\vec{V} = \left(V,0,V_{z}\right)$.  Assuming an incompressible, irrotational, inviscid flow, the following equations hold in the bulk of the fluid:
\begin{eqnarray}
    \left(\partial_{t} + V \partial_{x} + V_{z} \partial_{z}\right) V &=& - \partial_{x}P \,, \nonumber \\
    \left(\partial_{t} + V \partial_{x} + V_{z} \partial_{z}\right) V_{z} &=& - \partial_{z}P - g \,, \nonumber \\
    \partial_{x}V + \partial_{z}V_{z} &=& 0 \,, \nonumber \\
    \partial_{z}V - \partial_{x}V_{z} &=& 0 \,,
\end{eqnarray}
where $g$ is the acceleration due to gravity.  The first two equations are the Euler equations for the $x$ and $z$ components of the velocity field; the third equation encodes the incompressibility of the flow; and the fourth equation ensures that the flow is irrotational.  This last condition implies that we can write the velocity vector as the gradient of a scalar field, $\Phi$:
\begin{equation}
    \vec{V} = -\nabla \Phi \qquad \iff \qquad V = -\partial_{x}\Phi \,, \qquad V_{z} = -\partial_{z}\Phi \,.
\end{equation}
These equations are supplemented by kinematic boundary conditions on the bottom of the channel (which is assumed fixed with spatial profile $z = b(x)$) and on the free surface (at $z = H(t,x)$):
\begin{eqnarray}
    \left[ V_{z} - V \,\partial_{x}b \right]_{z = b(x)} &=& 0 \,, \nonumber \\
    \left[ V_{z} - V \, \partial_{x}b\right]_{z = H(t,x)} &=& \left[ \left(\partial_{t} + V \partial_{x}\right)H \right]_{z=H(t,x)} \,.
\end{eqnarray}

As mentioned above, the Analogue Gravity framework follows from an explicit separation of the system into a strong background plus weak perturbations on top of that background.  We thus decompose all the relevant fields accordingly: $V = v + \delta v$, $\Phi = \varphi + \delta\varphi$, {\it etc}.  We shall also assume that the background is stationary, so $v$, $\varphi$, {\it etc}. are all independent of time.  The equations above are arranged by perturbation order, and we keep only two orders: the zeroth-order equations, which are dynamical equations satisfied by the background, and the first-order equations, which are (linearized) equations of motion for the perturbation fields and depend on the background fields.  The equations can be further simplified by taking the long-wave limit, whereby all quantities are assumed to vary slowly in the $x$-direction.  Finally, we are led to the following equations: for the background,
\begin{eqnarray}
    \partial_{x}\left(\frac{1}{2} v^{2} + g \left(b+h\right) \right) &=& 0 \,, \nonumber \\
    \partial_{x}\left(v h\right) &=& 0 \,,
    \label{dynamic_backg}
\end{eqnarray}
and for the perturbations,
\begin{eqnarray}
    \left( \partial_{t} + v \partial_{x} \right) \delta \varphi &=&  g \, \delta h \,, \nonumber \\
    \left[ \left(\partial_{t}+ \partial_{x} v\right) \left(\partial_{t} + v \partial_{x}\right) - \partial_{x} g h \partial_{x} \right] \delta\varphi &=& 0 \,,
    \label{eq:wave_eqn}
\end{eqnarray}
where $h(x)$ is the local water depth, $\delta h(t,x)$ is its perturbation, and $g$ is the acceleration due to gravity.

The most important observation here, from the viewpoint of Analogue Gravity, is that the wave equation~(\ref{eq:wave_eqn}) is exactly the equation of motion for a massless scalar field living in the spacetime metric
\begin{equation}
    {\rm d}s^{2} = \frac{h(x)}{h_{0}} \left[ c^{2}(x) {\rm d}t^{2} - \left({\rm d}x - v(x) \, {\rm d}t\right)^{2} - {\rm d}y^{2} \right] \,,
    \label{metric hydraulic}
\end{equation}
where we have defined $c^{2}(x) = g h(x)$, and where $h_{0}$ is an arbitrary length introduced to make the metric dimensionally consistent (but has no bearing on the physics).  The metric~(\ref{metric hydraulic}) has the same form as that of Eq.~(\ref{eq:PG}), with only the overall factor in front taking a different form.  The background fields $c(x)$ and $v(x)$ can be written explicitly as
\begin{equation}
    v(x) = \frac{q}{h(x)} \,, \qquad c(x) = \sqrt{g \, h(x)} \,,
    \label{Profil_theo}
\end{equation}
where $q$ is the flow rate per unit width.
The nature of the flow can be parameterised by taking the ratio of these speeds, forming a dimensionless quantity known as the {\it Froude number}~\cite{rousseaux2020classical,Schuetzhold-Unruh-2002,Rousseaux-BASICS-2013}:
\begin{equation}
    Fr(x) = \frac{v(x)}{c(x)} = \frac{q}{\sqrt{g \, h^{3}(x)}} \,.
\end{equation}
Sub- and supercritical flows have $Fr < 1$ and $Fr > 1$, respectively, with the hydraulic horizon occurring at $Fr = 1$.
As mentioned above, an acceleration of the flow from subcritical to supercritical yields the analogue of a black hole horizon, while a deceleration of the flow from supercritical to subcritical yields the analogue of a white fountain horizon~\cite{Baines,Faber}.


The dispersion relation associated to the perturbations, in the long-wave (non-dispersive) limit, is
\begin{equation}
    \omega_{\rm cm}^{2} = c^{2} k^{2} = g h \, k^{2} \qquad {\rm with} \qquad \omega_{\rm cm} = \omega - v k \,.
    \label{dispersion_long-wave}
\end{equation}
This has two solutions, which are the left- and right-moving waves in the co-moving frame of the fluid.

Relaxing the long-wavelength limit for the perturbations gives a non-trivial dispersion relation $\omega(k)$ which can taking into account many effects like water depth, surface tension, transverse modes, and vorticity~\cite{rousseaux2020classical,Rousseaux-BASICS-2013,nardin2009wave,rousseaux2008observation,rousseaux2010horizon,weinfurtner2011measurement,chaline2013some,coutant2014undulations,Michel-Parentani-2014,Peloquin,robertson2016scattering,Euve-Rousseaux-2017,euve2017interactions,Michel-et-al-2018,Coutant-Weinfurtner-2018}.  In this paper we will concern ourselves only with the dispersive effects of finite water depth and surface tension, upon which the dispersion relation becomes
\begin{equation}
    \omega_\text{cm}^2=g k \left(1+\frac{\gamma}{\rho g}k^2\right)\tanh\left(h k\right)\qquad \text{with}\qquad \omega_\text{cm}=\omega-v k \,,
    \label{dispersion}
\end{equation}
where $\gamma$ is the surface tension and $\rho$ the density of the fluid.  This reduces to~(\ref{dispersion_long-wave}) when $ kh \ll 1$ and $k l_c\ll 1$ ($l_c=\sqrt{\gamma /\rho g}$ is the capillary length, see \cite{Rousseaux-BASICS-2013}
for a discussion of the many dispersive regimes in Analogue Gravity), but for larger $k$ ({\it i.e.}, smaller wavelength) it yields a different wave speed.  As discussed above, this short-wave dispersion provides exactly the mechanism that avoids the occurrence of arbitrarily short wavelengths in the vicinity of the horizon.

Moreover, as mentioned above, the energy associated with the perturbations is positive when $\omega$ and $\omega_{\rm cm}$ have the same sign, and negative when they have opposite signs.  This is generally true, even in the presence of dispersion.  So the counter-propagating wave always has negative energy in a supercritical region.  But dispersion allows the occurrence of negative-energy waves even if the flow is subcritical.  All that is required is that the phase velocity of the wave be reversed by the flow, {\it i.e.}, the flow speed $v$ must be faster than the phase velocity of the wave in the co-moving frame, $\omega_{\rm cm}/k$.

\section{Theoretical classification of dispersionless hydraulic regimes}
\label{sec:classification}

In this section, we neglect dispersive effects and consider three hydraulic regimes: subcritical, supercritical, and transcritical.  We shall give explicit formulae for the flows in these regimes, and show how they relate to the behaviour of astrophysical objects like black holes.

Let us consider further the possible profiles $h(x)$ that can be realised, since this determines the form of the effective spacetime in equation \ref{metric hydraulic}.
To do this, by combining equation \ref{dynamic_backg} (assumed to have no dependence on the transverse direction), we obtain a condition on the derivative of the water depth:
\begin{equation}
    \left(1-\frac{q^2}{gh^3}\right)\frac{dh}{dx}=\left(1-Fr(x)^2\right)\frac{dh}{dx}=-\frac{db}{dx}
    \label{deriv}
\end{equation}

With the help of equation \ref{deriv}, the critical condition at the horizon ($Fr(x)=1$) becomes \cite{Long, houghton1968nonlinear, pratt1982dynamics, Pratt1984, Faber, Baines}:
\begin{equation}
    h_c=\sqrt[3]{\frac{q^2}{g}}\quad\text{and}\quad\frac{db}{dx}=0\qquad\text{if} \qquad \frac{d^2b}{dx^2} \neq 0 
     \label{h_c}
\end{equation}
We can have then three different hydraulic regimes: the subcritical regime such that $Fr(x)<1$ everywhere, the supercritical regime such that $Fr(x)>1$ everywhere, and the transcritical regime where $Fr(x)$ crosses 1 at some position $x=x_{\rm hor}$ (where $db/dx=0$ condition \ref{h_c}), which we refer to as the horizon.
Equations~\ref{metric hydraulic} and~\ref{dynamic_backg} can be simplified and given the following 1D equation whose unknown is the water depth~\cite{Long, houghton1968nonlinear, pratt1982dynamics, Pratt1984, Faber, Baines}:
\begin{equation}
    \frac{q^2}{2h(x)^2}+g\left(h(x)+b(x)\right)=\frac{q^2}{2h^2_{\text{up}}}+gh_\text{up}\quad\text{with}\quad b_\text{up}=0 \,,
    \label{Bernou}
\end{equation}
where $h_\text{up}$ is the water depth far upstream from the obstacle, where the flow and the bottom is supposed to be flat ($b_\text{up}=0$).
Equation \ref{Bernou} can be written as a third order polynomial for the water depth scaled by the upstream dynamical water depth \cite{Larsen,Pratt1984}:
\begin{equation}
    \left(h^*\right)^3-\left(\frac{Fr^2}{2}+1-\frac{b(x)}{h_\text{up}}\right)\left(h^*\right)^2+\frac{Fr^2}{2}=0\quad\text{with}\quad h^*=\frac{h(x)}{h_\text{up}}\quad\text{and}\quad Fr=\frac{q}{\sqrt{g}h_\text{up}^{\frac{3}{2}}}
    \label{poly}
\end{equation}

\begin{figure}[!h]
    \centering
    \includegraphics[scale=0.3]{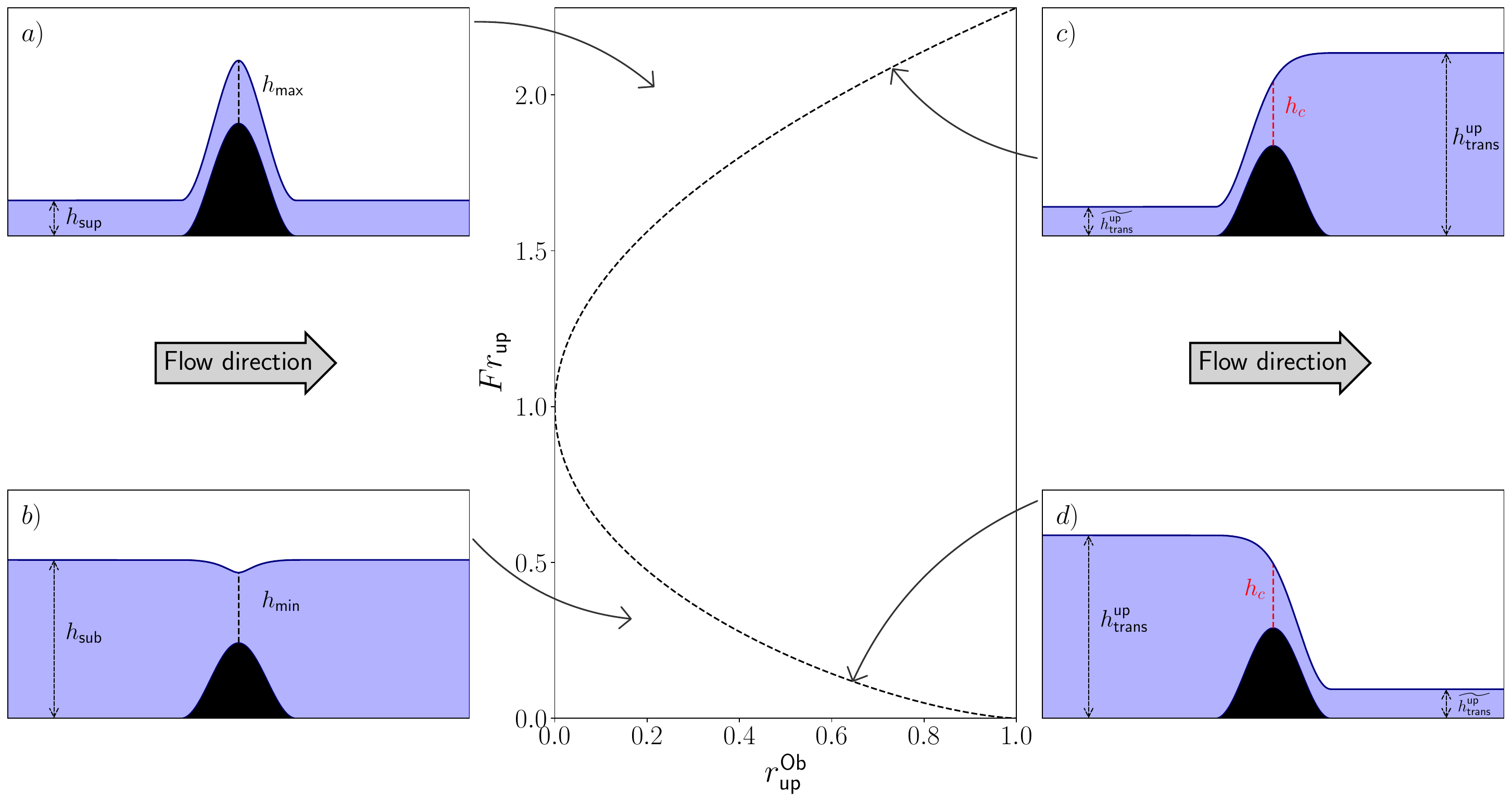}
    \caption{Graph summarising the different hydraulic regimes and their positions in relation to Long's law in the $\left(r^\text{Ob}_\text{up},Fr_\text{up}\right)$ plane. The dotted line corresponds to Long's law. The regime $a)$, above Long's law, is the supercritical regime; the regime $b)$, below Long's law, is the subcritical regime; the regime $c)$, which lies on Long's law for $Fr_\text{up}\geqslant1$, is the decelerating transcritical regime and the regime $d)$, which lies on Long's law for $Fr_\text{up}\leqslant1$, is the accelerating transcritical regime. For each figure, the blue coloured area represents the fluid and the obstacle corresponds to the black area, of equation $b(x)=b_\text{max}\cos\left(\pi/L x\right)^2\,\text{for}\, x\in[-L/2,L/2]$ and $b(x)=0$ otherwise.}
    \label{Long_regime}
\end{figure}

From now on, $Fr$ will denote the upstream Froude number computed with the upstream water depth. Following Pratt \cite{Pratt1984}, we use Cardan's method in order to look at the dimensionless discriminant of the polynomial (the formula given by Pratt has been corrected in this work):
\begin{equation}
    \Delta=-\frac{Fr^2}{2}\left(-4\left(\frac{Fr^2}{2}+1-\frac{b(x)}{h_\text{up}}\right)^3+27\frac{Fr^2}{2}\right)
    \label{discri}
\end{equation}
There is a regime change when $\Delta=0$, but according to the above criterion based on equation \ref{deriv}, a regime change occurs when the local Froude number is equal to 1: according to equation \ref{h_c}, this occurs for $b(x)=b_\text{max}$. By injecting the conditions $\Delta=0$ for $b(x)=b_\text{max}$ into the equation \ref{discri}, we obtain Long's law describing the transcritical condition as a relationship between the upstream Froude number and the so-called
obstruction ratio discussed above (see the dotted curve in the figure \ref{Long_regime}) but by a different path in comparison to Long since he was looking at the minimum of the specific energy 
\cite{Long, houghton1968nonlinear, pratt1982dynamics, Pratt1984, Faber, Baines}:
\begin{equation}
    r_\text{Long}=1+\frac{1}{2}Fr_\text{trans}^2-\frac{3}{2}Fr_\text{trans}^{\frac{2}{3}}\qquad\text{with}\qquad r_\text{Long}=r_\text{up} ^\text{Ob}=\frac{b_\text{max}}{h_\text{up}}\geqslant 0
    \label{Long}
\end{equation}
The obstruction ratio $r_\text{up} ^\text{Ob}$ compares the maximum heigh of the obstacle $b_\text{max}$ to the upstream water depth $h_\text{up}$. We can also obtain the reciprocal functions that are used in the theory of navigation in confined media to determine the maximum speed of a boat in restricted waterways (where the obstruction ratio is generalized as the ratio of the boat section to the channel section) \cite{pompee2015modelling}. To do this, we can rewrite the equation \ref{Long} as a 3rd degree polynomial:
\begin{equation}
    X^3-3X+2\left(1-r_\text{Long}\right)=0\quad\text{with}\quad X=Fr_\text{trans}^{\frac{2}{3}}
    \label{polyX}
\end{equation}
The discriminant of the polynomial \ref{polyX} gives  $\Delta_\text{inv}=-108r_\text{Long}\left(-2+r_\text{Long}\right)$. If $r_\text{Long}\in[0,2]$, then $\Delta_\text{inv}\geqslant 0$ and we see that the polynomial has 3 real solutions (using Cardan's formulae):
\begin{equation}
    X_n=2\cos\left(\frac{1}{3}\Arccos\left(r_\text{Long}-1\right)+\frac{2n\pi}{3}\right)\quad\text{with}\quad n\in\left\{0,1,2\right\}
\end{equation}
As $r_\text{Long}\in[0,2]$, we obtain the following inequality for the 3 solutions:
\begin{equation}
    X_0\geqslant 1\quad,\quad X_1< 0\quad\text{and}\quad -1\leqslant X_2\leqslant 1
\end{equation}
To describe the solutions for $0\leqslant Fr_\text{trans}\leqslant 1$, we need $r_\text{Long}\in[0,1]$ in order to have $0\leqslant X_2\leqslant 1$.
In addition, the solution for $n=1$ is therefore not interesting (because negative). Furthermore, the solution $X_0$ can be written as:
\begin{align}
    X_0&=2\cos\left(\frac{1}{3}\Arccos\left(r_\text{Long}-1\right)\right)\\
    &=2\cos\left(\frac{\pi}{6}-\frac{1}{3}\Arcsin\left(r_\text{Long}-1\right)\right)\quad\text{because}\quad \Arcsin(x)+\Arccos(x)=\frac{\pi}{2}\quad \forall x\in[-1,1]\\
    &=2\cos\left(\frac{\pi}{2}-\frac{\pi}{3}+\frac{1}{3}\Arcsin\left(1-r_\text{Long}\right)\right)\quad\text{because $\Arcsin$ is an odd function}\\
    &=-2\sin\left(-\frac{\pi}{3}+\frac{1}{3}\Arcsin\left(1-r_\text{Long}\right)\right)\\
    &=2\sin\left(\frac{\pi-\Arcsin\left(1-r_\text{Long}\right)}{3}\right)
\end{align}
By applying the same method to the solution $X_2$, we find the reciprocal functions for the equation \cite{Long, pompee2015modelling}:
\begin{equation}
    Fr_\text{trans}=\left(2\sin\left(\frac{\Arcsin\left(1-r_\text{Long}\right)}{3}\right)\right)^{\frac{3}{2}}\quad \text{for}\quad Fr_\text{trans}\leqslant 1\quad\text{and}\quad 0\leqslant r_\text{Long}\leqslant 1
    \label{Long_explicit_bas}
\end{equation}
In the case where $r_\text{Long}\geqslant 2$, the discriminant of \ref{polyX} is negative and we therefore obtain a real solution which extends the solution for $Fr_\text{trans}\geqslant1$ \cite{pompee2015modelling}:
\begin{equation}
    \begin{cases}
        Fr_\text{trans}=\left(2\sin\left(\frac{\pi-\Arcsin\left(1-r_\text{Long}\right)}{3}\right)\right)^{\frac{3}{2}}\quad \text{for}\quad Fr_\text{trans}\geqslant 1\quad\text{and}\quad 0\leqslant r_\text{Long}\leqslant 2\\
        Fr_\text{trans}=\left(\sqrt[3]{r_\text{Long}-1+\sqrt{r_\text{Long}\left(r_\text{Long}-2\right)}}+\sqrt[3]{r_\text{Long}-1-\sqrt{r_\text{Long}\left(r_\text{Long}-2\right)}}\right)^{\frac{3}{2}}\quad \text{for}\quad Fr_\text{trans}\geqslant 1\quad\text{and}\quad r_\text{Long}\geqslant 2
        \label{Long_explicit_haut}
    \end{cases}
\end{equation}

The equation \ref{Long_explicit_haut} when $Fr_\text{trans}\geqslant 1$ and $r_\text{Long}\geqslant 2$ can be written as a hyperbolic function using the following relationship:

\begin{equation}
    \cosh\left(\frac{\Arccosh(x)}{3}\right)=\frac{\sqrt[3]{x+\sqrt{x^2-1}}+\sqrt[3]{x-\sqrt{x^2-1}}}{2}\quad ,\quad \forall x\geqslant 1
\end{equation}

Long's law is a condition that must be satisfied by the upstream variables, i.e. the $Fr_\text{trans}$ and $r_\text{Long}$, in order to obtain a transcritical regime. 
The boat formulae reduced to the Long's law separating the subcritical and trancritical regimes when the geometry is rectangular and when the obstacle occupies the full width of the water channel. 
We may invert Long's law \ref{Long} to express the upstream water depth for a transcritical regime as a function of the experimentalist's control parameters $q$ and $b_\text{max}$ in a dimensionless complex version reported in \cite{lawrence1987steady} or with an unscaled explicit version derived recently in \cite{bossard2023create}:
\begin{equation}
    h_\text{trans}^\text{up}=\left(\frac{1}{2}\sqrt[3]{\frac{q^2}{g}}+\frac{b_\text{max}}{3}\right)\left [1+2\cos\left(\frac{1}{3}\Arccos\left(1-\frac{\frac{1}{4}\frac{q^2}{g}}{\left(\frac{1}{2}\sqrt[3]{\frac{q^2}{g}}+\frac{b_\text{max}}{3}\right)^3}\right)\right)\right]
    \label{hauteur Long}
\end{equation}
The equation \ref{hauteur Long} is the value of the upstream water depth (i.e. far from the obstacle) to obtain a transcritical regime at the obstacle. We can also obtain the value of the water depth far downstream of the obstacle, after the transcritical regime, using the equation \ref{hconjug}.
The equation \ref{hconjug} is a non-trivial solution of the equation \ref{poly} when $b=0$, i.e. a different solution of $h^*=1$ when $b=0$, which is present in an adimentioned form in \cite{lawrence1987steady} and dimensioned in \cite{bossard2023create}.
\begin{equation}
    \widetilde{h_\text{trans}^{\text{up}}}=h_\text{trans}^{\text{up}}\left(\frac{Fr_\infty^2}{4}+\frac{Fr_\infty}{4}\sqrt{Fr^2_\infty+8}\right)\quad\text{with}\quad Fr_\infty=\frac{q}{\sqrt{g}\left(h_\text{trans}^\text{up}\right)^{\frac{3}{2}}}
    \label{hconjug}
\end{equation}
Physically, it is the downstream water depth of an accelerating transcritical regime or the upstream water depth of a decelerating regime.
Finally, by using Cardan's formulae for the cubic polynomial (equation \ref{poly}) and the above initial conditions (equation \ref{hauteur Long} and the definition of each regime), we can obtain the analytical expression of the water depth as a function of the position for the main hydraulic regimes (the following formulae are used to construct the figure \ref{Long_regime}):

\begin{itemize}
    \item[a)] For a supercritical regime:
    \begin{equation}
    h(x)=\frac{\widetilde{h_\infty}}{3}\left(1+\frac{\widetilde{Fr_\infty}^2}{2}-\frac{b(x)}{\widetilde{h_\infty}}\right) \left[1+2 \cos\left(\frac{1}{3}\Arccos\left(1-\frac{\frac{27}{4}\widetilde{Fr_\infty}^2}{\left(1+\frac{\widetilde{Fr_\infty}^2}{2}-\frac{b(x)}{\widetilde{h_\infty}}\right)^3}\right)-\frac{2\pi}{3}\right) \right] \label{sup formule}
    \end{equation}
    with $\widetilde{h_\infty}=h_\text{sup}\left(\frac{Fr_\text{sup}^2}{4}+\frac{Fr_\text{sup}}{4}\sqrt{Fr^2_\text{sup}+8}\right)\quad\text{and}\quad h_\text{sup}\leqslant \widetilde{h_\text{trans}^{\text{up}}}$. Here, $\widetilde{h_\infty}$ is the conjugate water depth (using Lawrence's terms \cite{lawrence1987steady}) of the supercritical upstream water depth, $h_\text{sup}$ (see case a) in the figure \ref{Long_regime}). Once the obstacle is imposed, the supercritical regime has two degrees of freedom: the upstream water depth, $h_\text{sup}$, and the flow rate, $q$;
    
    \bigskip
    
    \item[b)] For a subcritical regime:
    \begin{equation}
        h(x)=\frac{h_\text{sub}}{3}\left(1+\frac{Fr_\text{sub}^2}{2}-\frac{b(x)}{h_\text{sub}}\right) \left[1+2 \cos\left(\frac{1}{3}\Arccos\left(1-\frac{\frac{27}{4}Fr_\text{sub}^2}{\left(1+\frac{Fr_\text{sub}^2}{2}-\frac{b(x)}{h_\text{sub}}\right)^3}\right)\right) \right]
        \label{sous formule}
    \end{equation}
    with $Fr_\text{sub}=q/\left(\sqrt{g}h_\text{sub}^{\frac{3}{2}}\right)$ and $h_\text{sub}\geqslant h_\text{trans}^\text{up}$ (see case b) in the figure \ref{Long_regime}). Once the obstacle is imposed, the subcritical regime has two degrees of freedom: the upstream water depth, $h_\text{sub}$, and the flow rate, $q$;
    
    \bigskip
    \item[c)] For a decelerating transcritical regime:
    \begin{equation}
        h(x)=\frac{h_\text{trans}^\text{up}}{3}\left(1+\frac{Fr_\infty^2}{2}-\frac{b(x)}{h_\text{trans}^\text{up}}\right) \left[1+2 \cos\left(\frac{1}{3}\Arccos\left(1-\frac{\frac{27}{4}Fr_\infty^2}{\left(1+\frac{Fr_\infty^2}{2}-\frac{b(x)}{h_\text{trans}^\text{up}}\right)^3}\right)-\frac{2\:\theta\left(x_{\rm hor}-x\right)\pi}{3}\right) \right] \label{trans decc formule}
    \end{equation}
    In the formula \ref{trans decc formule}, $\theta$ is the Heaviside function and $x_{\rm hor}$ is the position of the horizon ({\it i.e.}, following~(\ref{h_c}), we have $b\left(x_{\rm hor}\right) = b_{\rm max}$). This formula \ref{trans decc formule} characterises a flow analogue to a white fountain, because there is the presence of a horizon and the flow is decelerating (see case c) in the figure \ref{Long_regime}).
    \item[d)] For an accelerating transcritical regime:
     \begin{equation}
        h(x)=\frac{h_\text{trans}^\text{up}}{3}\left(1+\frac{Fr_\infty^2}{2}-\frac{b(x)}{h_\text{trans}^\text{up}}\right) \left[1+2 \cos\left(\frac{1}{3}\Arccos\left(1-\frac{\frac{27}{4}Fr_\infty^2}{\left(1+\frac{Fr_\infty^2}{2}-\frac{b(x)}{h_\text{trans}^\text{up}}\right)^3}\right)-\frac{2\:\theta\left(x-x_{\rm hor}\right)\pi}{3}\right) \right] \label{trans acc formule}
    \end{equation}
    Again in the formula \ref{trans acc formule}, $\theta$ is the Heaviside function and $x_{\rm hor}$ is the position of the horizon ({\it i.e.}, following~(\ref{h_c}), we have $b\left(x_{\rm hor}\right) = b_{\rm max}$). This formula \ref{trans acc formule} characterises a flow analogue to a black hole, because there is the presence of a horizon and the flow is accelerating (see case d) in the figure \ref{Long_regime}).
    
\end{itemize}
Once the obstacle is imposed, the accelerating or decelerating transcritical regime has only one degree of freedom: the flow rate $q$ or the upstream water depth $h_\text{trans}^\text{up}$, because of Long's law. In addition, if we take $x_{\rm hor}=0$ and if the obstacle is symmetrical, then the black hole type case is the spatial and temporal inverse of the white fountain case, as expected on general grounds.

\bigskip

In the rest of this work, we will not study supercritical regimes since they are not relevant for Analogue Gravity purposes. Although the previous classification is a theoretical one for the hydraulic regimes in the non-dispersive limit, we will confirm their existence experimentally with our experimental set-up.
 
\section{Experimental open channel flow and visualisations}
\label{sec:experiment}

\subsection{Description of the experimental setup}
\begin{figure}[!h]
    \centering
    \includegraphics[scale=0.7]{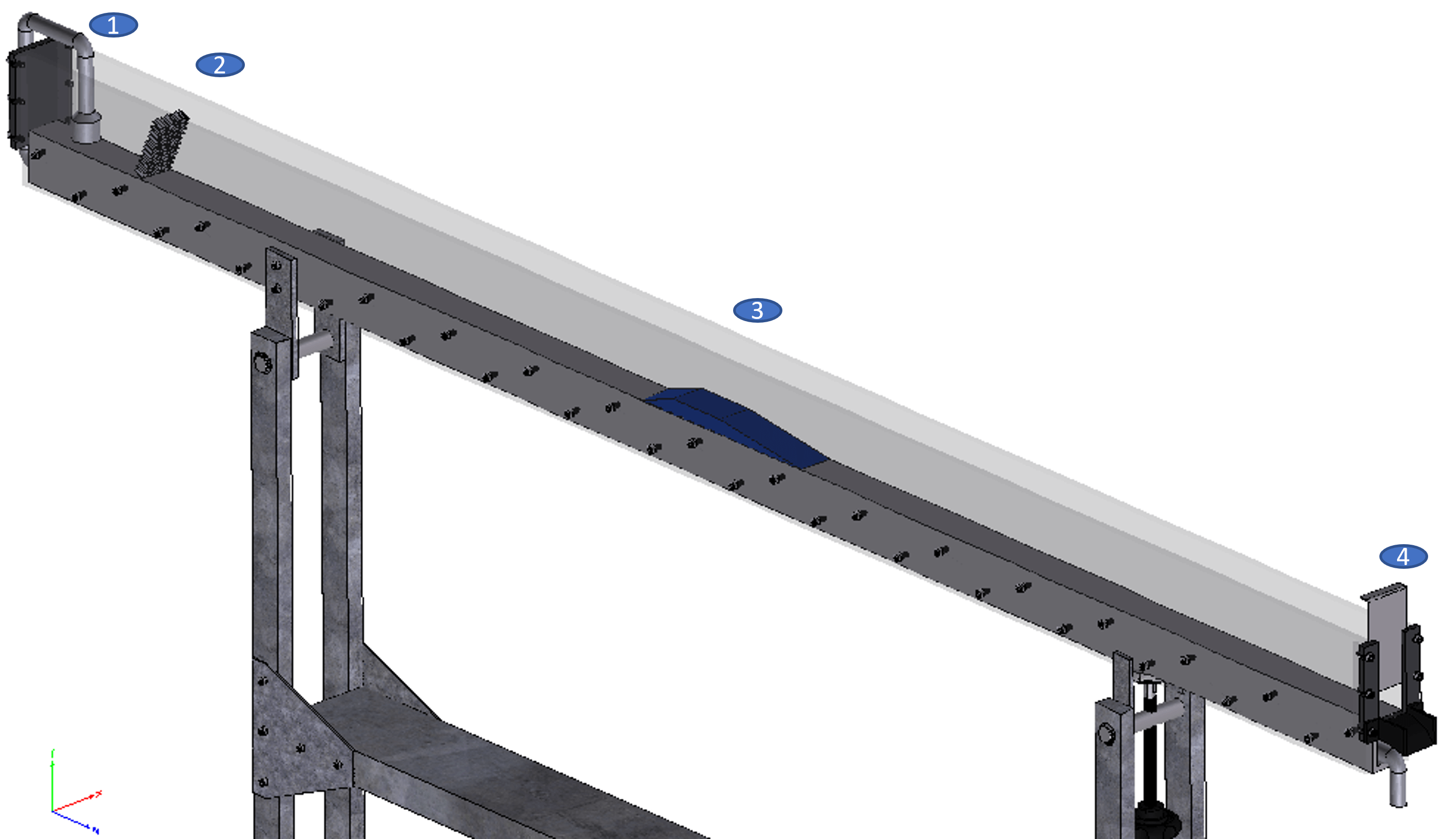}
    \caption{Computer-aided design (CAD) of the channel used. The numbers on the picture designate: 1) the flow inlet into the channel, 2) the honeycomb, 3) a bottom obstacle and 4) the downstream guillotine and flow outlet.}
    \label{CAO}
\end{figure}

\begin{figure}[!h]
    \centering
    \includegraphics[scale=0.6]{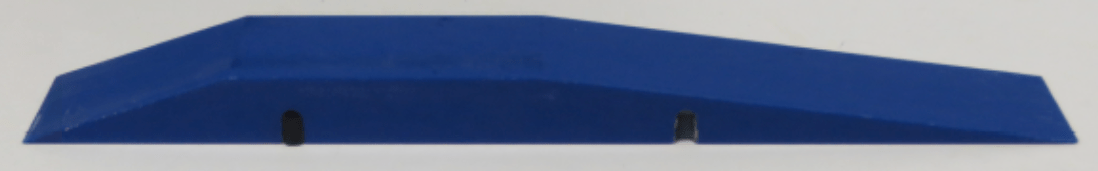}
    \caption{Image of an obstacle with the reference geometry ACRI 2010 \cite{chaline2013some}. The length of the obstacle is $L=32.2\:cm$ and the maximum height is $b_\text{max}=2.1\: cm$.}
    \label{ACRI2010}
\end{figure}

The experimental set-up consists of a TecQuipment FC50 open water channel. The channel is 2.5 m long, 0.12 m high and 0.053 m wide inside. The channel has two long transparent side walls to allow lateral visualisations (see figure \ref{CAO}). As the system is a closed circuit, the flow leaving the channel by a waterfall drops onto a disconnected reservoir, where there is a submersible pump which re-injects the water current at the entrance of the channel. An electronic flow-meter measures the outlet flow from the submersible pump. The submersible pump has a flow rate per unit width ($q=Q/W$ with W the width of the channel) range of $q\in[0.0006;0.0115]\: m^2.s^{-1}$. The signals from the flow-meter is converted into a digital display to show the flow rate. A hand-operated control valve adjusts the water flow rate from the pump. Finally, the channel was set to be horizontal with an electronic level system. LED lighting was placed at the top of the channel on the cameras side and along its length to illuminate the free surface though transmission in the side window. The opposite window is covered by white sheets of paper to enhance the meniscus contrast with the white background on the camera sensors.  A honeycomb is placed upstream at the entrance the channel (close to the vertical water injector oriented downwards) at an angle of $66^{\circ}$ as required by the designer, to suppress large vortices within the flow.  The obstacles placed in the middle of the channel length are manufactured using a 3D printer (Ultimaker 5S 3D printing machine) in Acrylonitrile butadienestyrene (ABS) of blue color. Black neoprene seals with a diameter of $6 mm$ from the CIEP company are glued to the sides of the obstacles within rectangular inner interstices to enable them to adhere to the channel by lateral compression. All the tested obstacles have the same geometrical aspect ratio ($\chi=L/b_\text{max}=15.3>>1$ with $L$ is the length of the obstacle) to avoid recirculations close to the obstacle and the same geometry but with different dimensions to allow comparison. The retained geometry is the ACRI 2010 geometry (see figure \ref{ACRI2010}) in reference to the shape of the obstacle introduced during the experiments in the ACRI coastal engineering company (\url{https://www.acri.fr/}) in 2010 \cite{chaline2013some}: it was an improvement of the upstream-downstream symmetric geometry used in 2008 \cite{rousseaux2008observation} plagued with a recirculation downstream of the obstacle but that was easier to build. A homothetic ratio will be applied to all the lengths of the geometry in order to observe a large set of hydrodynamic regimes. Compared to historical experiments in Analogue Gravity with open water channels discussed in the appendix, the present one is the smallest which will allow us to sweep all the flow regimes by a complete visualisation of the free surface on both side of the bottom obstacle.

\begin{figure}[!h]
    \centering
    \includegraphics[width=7.5cm, scale=0.15]{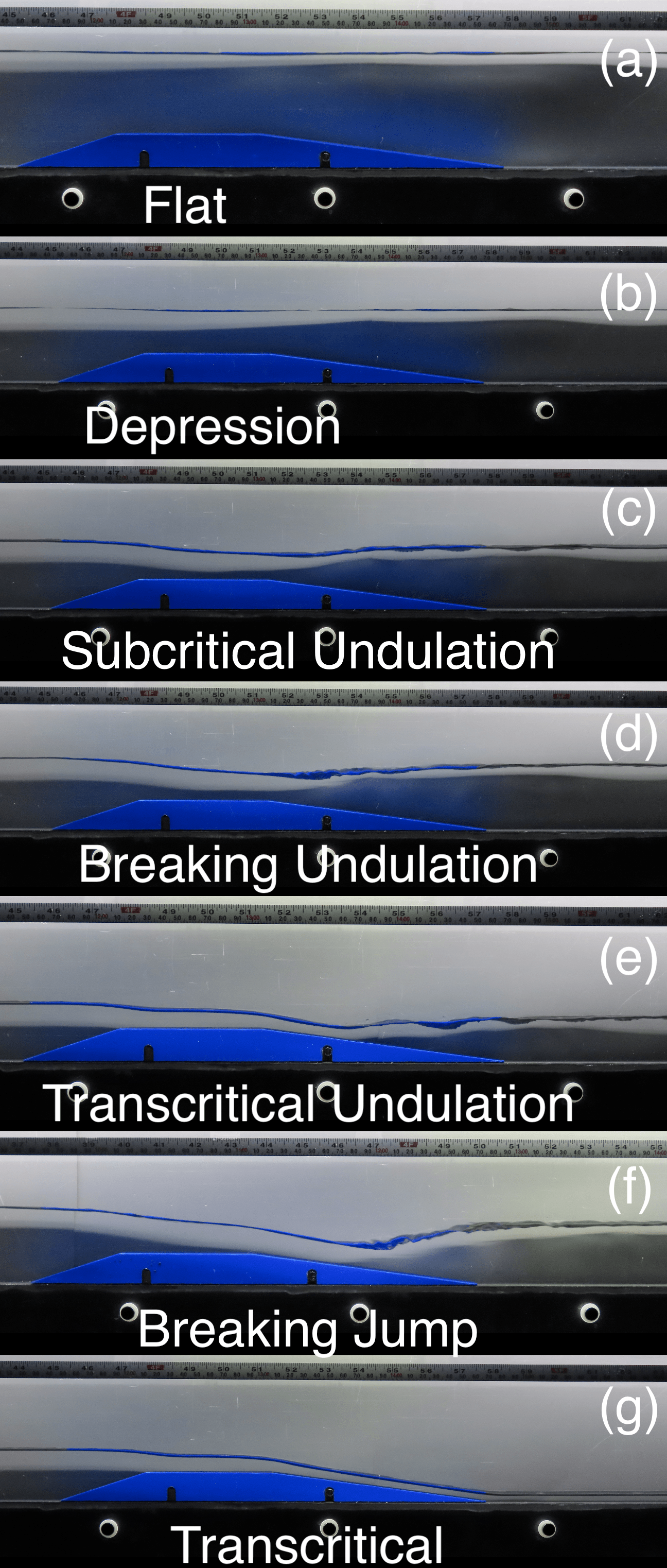}
    \caption{Photos illustrating the different hydrodynamic regimes that can be identified by visual observation of the free surface only in the water channel of Figure \ref{CAO}. The flow rate increased from case (a) to case (g) in presence of a gate except for the case (g ) that can be observed for whatever flow rate without a gate or for high flow rates with a gate (the channel was initially dry: no static water was present since it would be expulsed downstream).}
    \label{Planches_exp}
\end{figure}

The experimental protocol is as follows: we place the chosen obstacle at a fixed and arbitrary distance in the channel, with no initial static water depth in the channel (the canal is dry), and we switch on the pump to observe the resulting hydrodynamic regime for a given flow rate. We work with a null initial static water depth, to avoid threshold effects. In fact, when the static water depth is zero and a flow rate is imposed, the flow in the channel \ref{CAO} will fill the upstream part of the obstacle until it exceeds $b_\text{max}$: we obtain an accelerating transcritical flow (with no downstream condition). However, when the initial static head of water is sufficiently hight, for the same flow rate, we could obtain a subcritical regime; we would therefore have to increase the flow rate to obtain an accelerating transcritical regime. The non-zero static water depth could therefore limit the useful flow rate range for transcritical regimes. A quantitative study of this hydrodynamic regime is monitored by side cameras (2 grayscale Point Grey cameras with CMOS technology at an acquisition frequency of 25 images per second for a fixed duration corresponding to 8192 images) located 3.4 m perpendicularly to the channel on a displacement table from ISEL with actuators allowing an accurate positioning of the cameras with a set of staples, optical rails and ball joints. The combined field of view of both cameras is 2.04 m. The cameras allow us to measure the free surface through the Plexiglas side window. This measurement of the free surface allows us to trace the fluctuation and dispersion relation and thus to characterise the hydrodynamic regime as was done in the reference \cite{fourdrinoy2022correlations}. The channel has a downstream guillotine that controlled a back-reaction of the flow upstream of it and therefore allows all the regions of the hydrodynamic regimes to be enhanced, for example by observing transitions towards sub-critical regimes from transcritical waterfalls that are the natural regimes without the guillotine. In fact, to be more precise, for a so-called free flow regime (with no static water head or downstream condition), the regime obtained is a transcritical regime. However, if part of the guillotine is immersed, a mass accumulation occurs (upstream of the immersed part of the guillotine). This accumulation causes a non-linear wave to propagate upstream until an equilibrium position is reached. If the immersed part of the sash is too large, the non-linear wave can propagate upstream, against the current, and even exceed the horizon of the transcritical regime. In the latter case, as the horizon will have disappeared, the transcritical regime gives way to the subcritical regime. If we draw a parallel with the static water depth, we can say that lowering the guillotine has more or less the same effect as increasing the static water depth. The images taken by both cameras are processed by a Matlab program which adjusts the contrast of these images to exacerbate the meniscus. A  pixel-by-pixel detection of the meniscus is undertaken by looking for the maximum intensity on each column of the picture coded in grey levels. In all subsequent results, the pixel resolution is dx = 0.507 mm. In order to improve this resolution, we use the sub-pixel method by interpolating a Gaussian on each image column. We interpolate by a Gaussian function the intensity levels because according to the Figure \ref{Planches_exp}, the free surface, namely its side meniscus, appears black, surrounded by white bands. By inverting the contrast, we obtain an intensity distribution on one column that resembles a Gaussian function.

\subsection{Experimental flow description based on the free surface deformations only}

The walker along a river will classify the flow regime according to a visual inspection of the free surface she is looking at (see for instance \cite{zerihun2004one, rousseaux2020classical}): it is sufficient to think to some beautiful stationary waves surfed by kayakists for instance. However, as we will see shortly this naive classification will not sweep all possibilities because of our ignorance of the velocity distribution beneath the free surface. For now, let us classify the interfaces with simple arguments. In a laboratory open water channel, one can also use a downstream gate in addition to changing the flow rate within the channel by tuning the flow rate of a pump. We assume the river not to be dry initially with a given initial water depth. In the laboratory, both dry and wet initial conditions are possible: only the transient are different to reach the dynamical regimes we study.

First, when the water is at rest, the free surface is flat whatever the presence or not of a bottom obstacle and can be seen as a wave of infinite wavelength (or zero frequency) in accordance with the dispersion relation of water waves in the long wave-length limit ($\omega^2 \approx c^2k^2$ with $c=\sqrt{gh}$ when $kh<<1$). Obviously, when the water move slowly over the underlying obstruction, as a first approximation and to the visual accuracy of the observer, the free surface which is now moving with the bulk flow at a plug speed is still flat (see case (a) in figure \ref{Planches_exp}). Then, a depression appears for weak flows whose speeds are inferior to the long gravity speed of water waves $c$ (see case (b) in figure \ref{Planches_exp}). This depression is due to the fact that potential energy is converted into kinetic energy due to the conservation of mass (see equation \ref{Bernou}), resulting in a decrease in water depth \cite{Long, houghton1968nonlinear, pratt1982dynamics, Pratt1984, lawrence1987steady, bouhadef1988contribution, Faber, Lowery-Liapis-1999, zerihun2004one, pompee2015modelling, Baines}. Then, the effect of dispersion decorates the depression with a train of stationary waves downstream of the obstacle (see case (c) in figure \ref{Planches_exp}) \cite{Larsen, lawrence1987steady, bouhadef1988contribution, zerihun2004one, Rousseaux-BASICS-2013, coutant2014undulations, euve2016observation, Binder}. A Zero frequency solution ($\omega={\bf U.k}+gk\tanh (kh)= 0$) of the dispersion relation known as the undulation in the river banks frame of reference appears because of the Doppler effect (${\bf U.k}$) of the river current combined to the non-linear behavior ($gk\tanh (kh)$) of the dispersion relation for the sought wavenumber $k=k_z$ of the zero mode ($\omega= 0$) with a subluminal behaviour \cite{Rousseaux-BASICS-2013, coutant2014undulations, euve2016observation} ($d\omega/dk<0$). For harsher flow rates, the whelps (secondary crests) of the undulation are breaking (see case (d) in figure \ref{Planches_exp})\cite{Larsen, lawrence1987steady, zerihun2004one, rousseaux2020classical}. Then, for higher flow rates, a turbulent breaking jump appears on the obstacle above a supercritical plunging jet that follows the geometrical form of the obstacle (see case (f) in figure \ref{Planches_exp}) \cite{lawrence1987steady, zerihun2004one, rousseaux2020classical}. Increasing again the flow rate, the flow over the obstacle transforms into a waterfall that has pushed the hydraulic jump downstream of the obstruction (see case (g) in figure \ref{Planches_exp}). The waterfall relates both the upstream and downstream regions by a transcritical zone where the flow speed overcomes the speed of long gravity waves, an analogue black hole flow. Notwithstanding the effect of dispersion with the appearance of the undulation saturated by the non-linearity of the fluid dynamics equations, all the visually observed regimes was summarized so far in a hydraulic diagram that plot the upstream Froude number as a function of the dimensionless obstruction ratio \cite{Long, houghton1968nonlinear, bouhadef1988contribution,  Lowery-Liapis-1999, pompee2015modelling, Baines, rousseaux2020classical}. In the Royal society paper \cite{rousseaux2020classical}, an experimental hydraulic diagram -$Fr_{up}$ versus $r_{up}$- for the original Vancouver geometry and dimensions \cite{weinfurtner2013classical} was reported in presence of a a static water depth initially (that plays the role of a gate or of a secondary obstacle) with the inclusion of zones of appearance of the many free surface regimes reported in the Figure \ref{Planches_exp} (without a static depth) including the dispersive zones where the undulation was observed (subcritical dispersive regimes: (c) called U for a non-breaking undulation and (d) called UB for Undular-Breaking with a breaking whelp of the undulation in \cite{rousseaux2020classical}): the presence of the undulation is a sufficient but not a necessary condition of the possibility of seeing Hawking radiation in this regime \cite{rousseaux2010horizon, Rousseaux-BASICS-2013, coutant2014undulations, rousseaux2020classical}. Moreover, the boundary between subcritical and transcritical zones was plotted according to the Long's theory \cite{Long} that we were presented previously, and it was recently validated experimentally for different sizes and geometries of many bottom obstacles probed \cite{bossard2023create}. By selecting the couple (flow rate, position of the guillotine), one can stabilize and replace a subcritical zone connected to a breaking jump on the top of the obstacle (case (d) of Figure \ref{Planches_exp}) by a transcritical zone connected to a dispersive undular jump still located over the obstacle (case (e) of Figure \ref{Planches_exp}, the so-called transcritical gate case studied in \cite{fourdrinoy2022correlations}).

\section{Theoretical linear, geometrical and dispersive flow classification based on the hydrodynamic flow speed and the free surface deformations}
\label{Th-disp}

As we have seen, a purely hydraulic classification of regimes based only on the local Froude number encompasses sub-, trans-, and supercritical regimes as seen by long-wavelength waves, which live in a non-dispersive limit with wave speed $c=\sqrt{g h}$.  
The geometry is taken into account through the use of the obstruction factor $r$, which for transcritical flows is related to the upstream asymptotic value of $Fr(x)$ via Long's law~\ref{Long}. The latter tends to a global critical frontier $Fr_\text{up}=1$ when $r_\text{up}^\text{Ob}$ goes to zero in the -Fr versus r- diagram. Otherwise, the trans-critical line is given by the expressions \ref{Long_explicit_bas} and \ref{Long_explicit_haut} for a finite $r_\text{up}^\text{Ob}$ \cite{Long, houghton1968nonlinear, bouhadef1988contribution,  Lowery-Liapis-1999, pompee2015modelling, Baines, rousseaux2020classical}. For instance, in the classification of Baines \cite{Baines}, the cases (a), (c) and (e) of Figure \ref{Planches_exp} are not discussed. The case (b) corresponds to the so-called sub-critical flow regime. The case (d) is named a partially blocked flow with a lee jump whereas the case (g) is characterized as a partially blocked flow without a lee jump. Baines \cite{Baines} reported as well the supercritical flow that we did observed as well but that is not discussed in this work since it has no implication in Analogue Gravity.
The non-linear effect of saturation is not tackled by the following description neither harmonics generation, wave breaking nor the intensity of the Hawking conversion process.

\subsection{Dispersion and critical velocities}

The purely hydraulic classification is insufficient to describe the variety of possible wave behaviours when dispersion is taken into account.  Dispersive effects become important for waves of sufficiently short wavelength, so that they leave the long-wavelength limit characterised by the wave speed $c = \sqrt{g h}$.
 
Notice that dispersion relation~\ref{dispersion} has two limiting behaviours (using asymptotics developments of the hyperbolic tangent, $\tanh(kh)=kh-\frac{1}{3}(kh)^3+\underset{kh\rightarrow 0}{O}\left((kh)^5\right)$ and $\tanh(kh)\underset{}{\sim}1$ when $kh\rightarrow +\infty$):
\begin{equation}
    \omega_{\rm cm}^{2} \to \begin{cases}
    g h k^{2} \left(1 - \left[ \frac{h^{2}}{3} - \frac{\gamma}{\rho g}\right] k^{2} \right) +\underset{kh\rightarrow 0}{O}\left((kh)^6\right) \quad \text{(gravity/gravito-capillary waves)}\,,kh<<1\\
    \frac{\gamma}{\rho} k^{3} +\underset{kh\rightarrow +\infty}{O}\left(kh\right) \quad \text{(capillary waves)}\quad kh>>1\;\text{and}\; l_c^2k^2>>1
    \end{cases}
    \,.
\end{equation}
For large $k$, we always reach the capillary-wave regime, for which the (co-moving) phase and group velocities are always increasing with $k$.  On the other hand, since typically the water depth $h$ is much larger than the capillary length $l_{c} = \sqrt{\gamma/\rho g}$, the group and phase velocities are typically decreasing with $k$ in the gravity-wave regime.  Therefore, the phase and group velocities typically have minimum values at some finite $k$ on the order of $1/l_{c}$.  An exception occurs when $h < \sqrt{3}l_{c}$, whereupon the low-$k$ behaviour is switched and the dispersion is superluminal everywhere.

\begin{figure}[!h]
    \centering
    \includegraphics[scale=0.32]{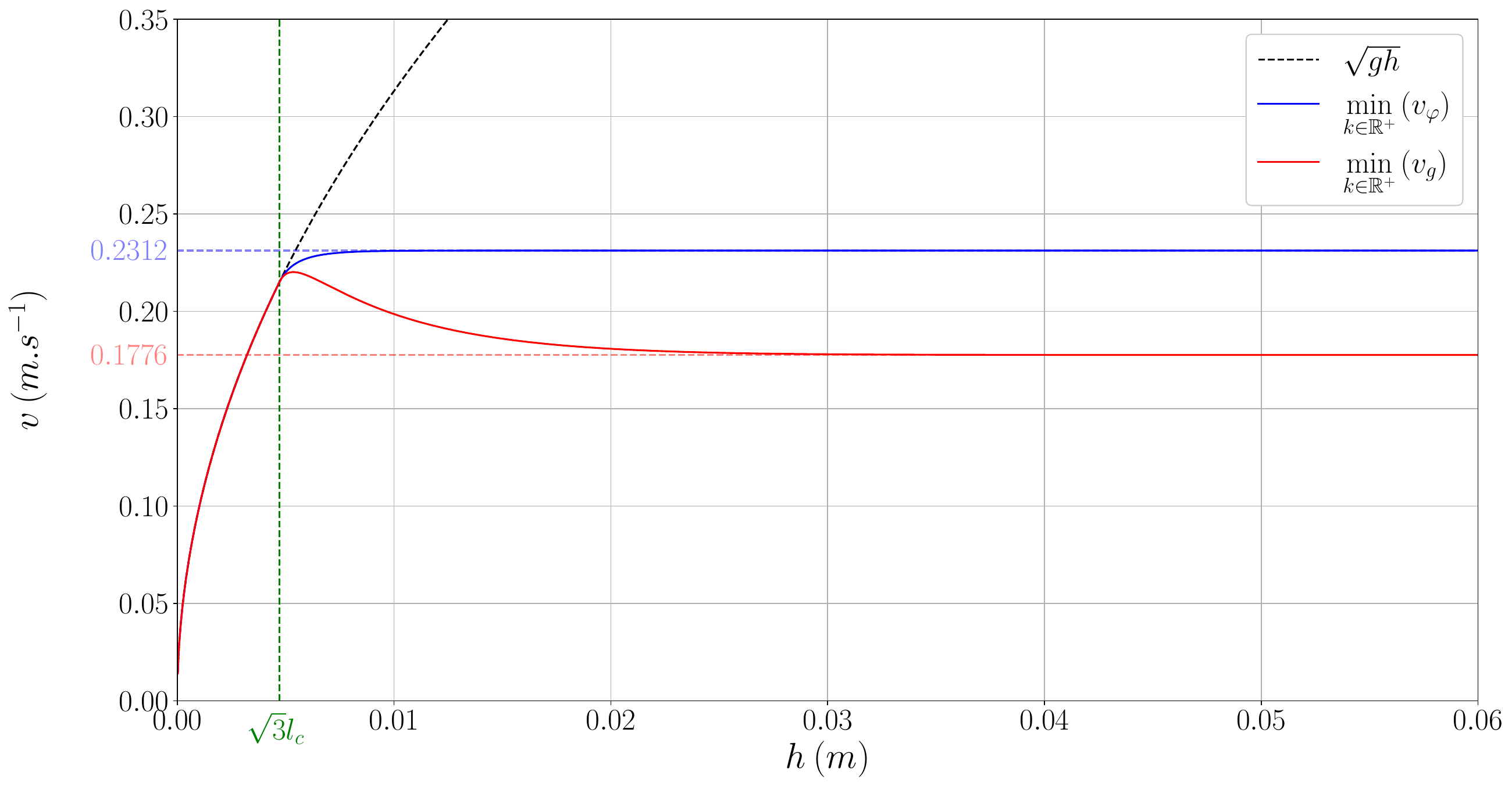}
    \caption{Evolution of minimum group and phase velocity as a function of water height. The graph was constructed for water at 20°C, i.e. with $\rho=1000\:kg.m^{-3}$, $\gamma=0.0728\:kg.s^{-2}$ and $g=9.81\:m.s^{-2}$, in which case the green limit is: $\sqrt{3}l_c=0.004718\: m$}
    \label{v vs h}
\end{figure}

In order to classify the various regimes while taking dispersive effects into account, we will compare the flow velocity with two natural thresholds: the minimum of the co-moving phase velocity, denoted $\underset{k\in \mathbb{R}^+}{\min}\left(v_\varphi\right)$ with $v_\varphi=\omega_\text{cm}/k$, and the minimum of the co-moving group velocity, denoted $\underset{k\in \mathbb{R}^+}{\min}\left(v_g\right)$ with $v_g=\partial\omega_\text{cm}/\partial k$. These two thresholds are important, because $\underset{k\in \mathbb{R}^+}{\min}\left(v_\varphi\right)$ corresponds to the threshold for the appearance of negative-energy waves (such as the partner mode of Hawking radiation that is trapped inside the black hole) \cite{rousseaux2010horizon, Rousseaux-BASICS-2013, rousseaux2020classical}, while $\underset{k\in \mathbb{R}^+}{\min}\left(v_g\right)$ corresponds to the appearance of the first dispersive horizon ({\it i.e.}, a turning point where a wave packet is slowed to a standstill) \cite{nardin2009wave, rousseaux2010horizon, Rousseaux-BASICS-2013, rousseaux2020classical}.
The critical velocities $\underset{k\in \mathbb{R}^+}{\min}\left(v_\varphi\right)$ and $\underset{k\in \mathbb{R}^+}{\min}\left(v_g\right)$ have a non-trivial dependence on the water level, as shown in Figure \ref{v vs h}.  Nevertheless, we can obtain their asymptotic behaviours.
In fact, in the area where capillary effects predominate over gravitational effects, i.e. when the water height is less than $\sqrt{3}l_{c}$ (the so-called thin film limit) \cite{Rousseaux-BASICS-2013}, then since the dispersion is superluminal for all $k$ it follows that the minima of the phase and group velocities both occur at $k = 0$, yielding the long-wavelength limit:
\begin{equation}
    \underset{k\in \mathbb{R}^+}{\min}\left(v_\varphi\right)= \underset{k\in \mathbb{R}^+}{\min}\left(v_g\right)=c=\sqrt{gh}\quad \text{for}\quad h\in[0;\sqrt{3}l_c] \,.
\end{equation}
Once this threshold has been passed ($h\geqslant\sqrt{3}l_c$), the minima of the phase and group velocities separate from the non-dispersive limit, becoming smaller thanks to the subluminal corrections at small $k$. 
As the water depth $h$ is increased, we find that these critical velocities approach the following asymptotic behaviours: 
\begin{gather}
\underset{k\in\mathbb{R}^+}{\min}\left(v_\varphi\right)=\sqrt{2}\sqrt[4]{\frac{\gamma g}{\rho}}\left(1-e^{-2\sqrt{\frac{\rho g}{\gamma}}h}\right)+\underset{h\rightarrow +\infty}{O}\left(e^{-4\sqrt{\frac{\rho g}{\gamma}}h}\right)\label{asympt1}\\
\underset{k\in \mathbb{R}^+}{\min}\left(v_g\right)=\frac{\sqrt{3}}{\sqrt[4]{3+2\sqrt{3}}}\sqrt[4]{\frac{\gamma g}{\rho}}+2\frac{\sqrt[4]{8\sqrt{3}-12}}{\sqrt{3}}\sqrt[4]{\frac{\rho g^3}{\gamma}}he^{-2\sqrt{\frac{2-\sqrt{3}}{\sqrt{3}}}\sqrt{\frac{\rho g}{\gamma}}h}+\underset{h\rightarrow +\infty}{O}\left(e^{-2\sqrt{\frac{2-\sqrt{3}}{\sqrt{3}}}\sqrt{\frac{\rho g}{\gamma}}h}\right)
\label{asympt2}
\end{gather}
These formulae show that the minimum of the phase velocity approaches its asymptotic value from below while the group velocity does so from above; this behaviour can be clearly seen in Figure~\ref{v vs h}. 
When surface tension is negligible, both minima tend towards zero and only one global maximum remains (for both velocities), which is precisely the long-wavelength limit 
$c=\sqrt{gh}$. As the water depth tends to infinity, we recover the limits in deep water:
\begin{gather}
    \underset{k\in \mathbb{R}^+}{\min}\left(v_\varphi\right)\underset{h\rightarrow +\infty}{\sim}\sqrt{2}\sqrt[4]{\frac{\gamma g}{\rho}}\quad\text{noted $|U_\gamma|$ in \cite{rousseaux2010horizon}}\\ \underset{k\in \mathbb{R}^+}{\min}\left(v_g\right)\underset{h\rightarrow +\infty}{\sim}\frac{\sqrt{3}}{\sqrt[4]{3+2\sqrt{3}}}\sqrt[4]{\frac{\gamma g}{\rho}}\quad\text{noted $|U_c|$ in \cite{rousseaux2010horizon}}
\end{gather}
$|U_\gamma|=23.1 \, {\rm cm}/{\rm s}$ in water is the so-called Landau speed \cite{rousseaux2020classical} and is the threshold 
for the existence of a non-trivial zero-frequency mode, the appearance of a stationary undulation on the downstream side of a decelerating flow, 
and the existence of negative-energy waves (Hawking's partners) in deep water \cite{rousseaux2010horizon, Rousseaux-BASICS-2013, coutant2014undulations, euve2016observation}. In deep water with no surface tension, there is no threshold for the appearance of both the undulation and the negative-energy modes \cite{nardin2009wave}, whereas in extremely shallow waters without dispersion (including surface tension effect), there is no undulation and the speed threshold for the appearance of negative-energy modes is $c=\sqrt{gh}$ \cite{Rousseaux-BASICS-2013}, {\it i.e.}, negative-energy waves only exist in the supercritical region (as in General Relativity where the negative partners exist only inside the horizon \cite{Hawking-1975}).

\subsection{Classification of transcritical flows}\label{trans classi}

\begin{figure}[!h]
    \centering
    \includegraphics[scale=0.35]{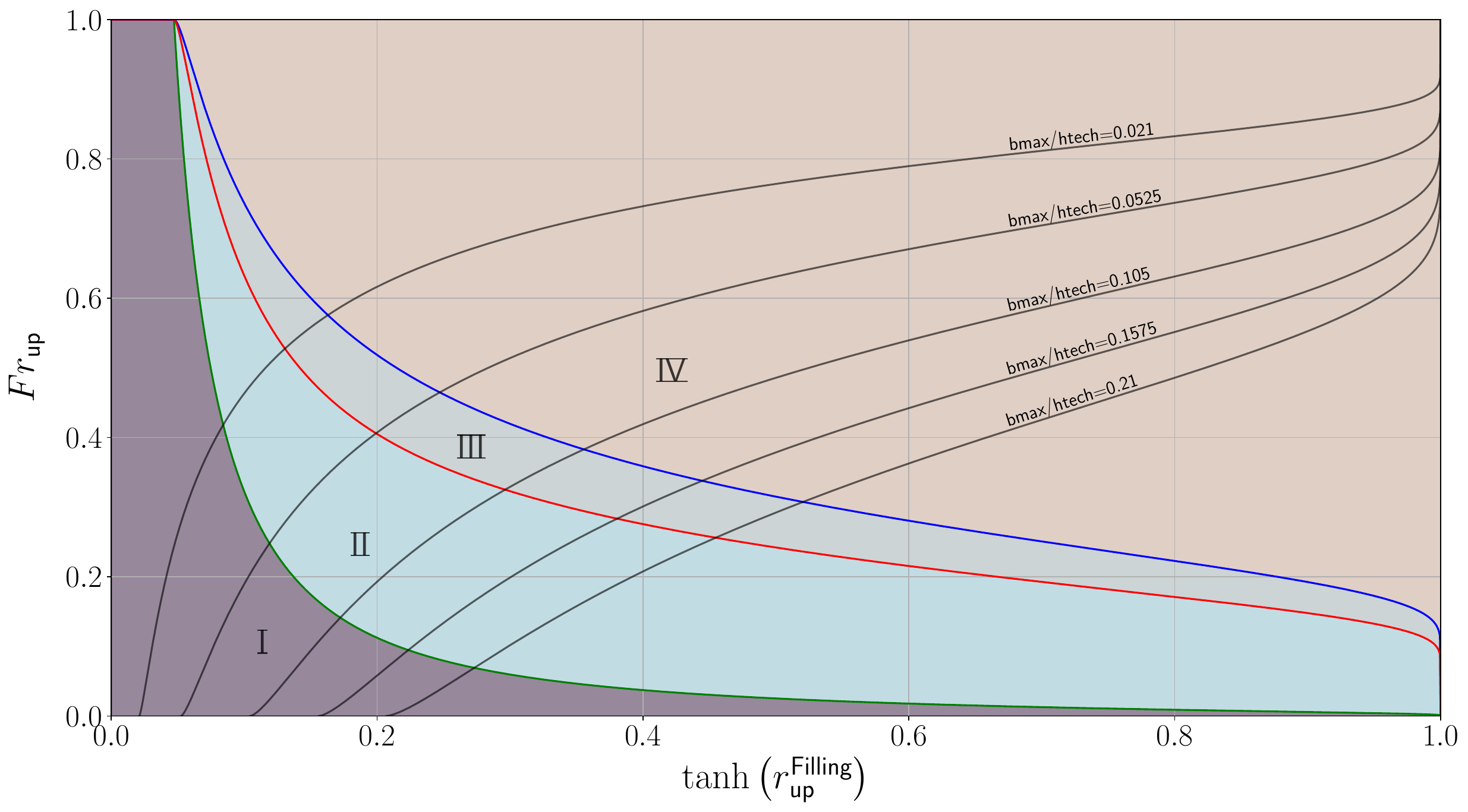}
    \caption{Phase diagram for a transcritical flow regime, taking dispersive and geometrical effects into account. The upstream Froude numbers (group velocity, phase velocity, transitional) are plotted against the hyperbolic tangent of the filling ratio. Here, the filling ratio was constructed with $h_\text{tech}=0.1\:m$.}
    \label{Fr vs tanhh}
\end{figure}

Let us first construct the classification for transcritical regimes, 
keeping Analogue Gravity in mind and taking dispersive effects into account.
Starting from Figure \ref{v vs h} and trying to scale its axes, we were led to introduce the following dimensionless numbers:
\begin{equation}
    Fr_{(X)}^\text{up}=\frac{\underset{k\in \mathbb{R}^+}{\min}\left(v_X^{\rm up}\right)}{\sqrt{gh_\text{up}}}\quad\text{with}\quad X\in\{\varphi , g\} \,,
\end{equation}
where the superscript `up' indicates that the number is to be evaluated in the upstream asymptotic region.
As Figure \ref{v vs h} is valid for any water depth, we're going to fix it with the upstream water depth $h_{\rm up}$, as the historical Long's law \cite{Long, houghton1968nonlinear, bouhadef1988contribution,  Lowery-Liapis-1999, Baines, rousseaux2020classical} was established with the upstream water depth albeit scaled by the maximum height of the obstacle; this procedure will make the comparison easier. Scaling the vertical axis of Figure \ref{v vs h} by dividing by the speed of long gravity waves in the upstream asymptotic region ($c_{\rm up} = \sqrt{g h_{\rm up}}$), we obtain the red and blue curves of Figure \ref{Fr vs tanhh}. Scaling the upstream water level on the horizontal axis of the same plot is not trivial.  We have chosen to scale it, not as before to get an obstruction ratio, by the maximum depth compatible with both the geometry of the water channel and with the measurements methods of the water depth.  To this end, we define the ``technological water depth'' as $h_\text{tech}=\min \left(h_{\rm max}^{\rm channel},h_{\rm maxROI}^{\rm Camera}\right)$, where $h_{\rm max}^{\rm channel}$ is the maximum water depth that can be put in the water channel and where $h_{\rm maxROI}^{\rm Camera}$ is the maximum height measured by the sensor (such as a camera) with region of interest ROI. For the water channel used in this work, we have $h_\text{tech}=0.1\:m$. The resulting dimensionless number is what we call from now on the (liquid) 
``filling ratio'' of the water in the open channel, and is defined explicitly as:
\begin{equation}
    r_\text{up}^\text{Filling}=\frac{h_\text{up}}{h_\text{tech}}=\frac{b_\text{max}}{h_\text{tech}}\frac{h_{up}}{b_{max}}=\frac{b_\text{max}}{h_\text{tech}}\frac{1}{r_\text{up}^\text{Ob}}
\end{equation}
The (solid) obstruction ratio $r_\text{up}^\text{Ob}=b_{\rm max}/h_{\rm up}$, noted $r$ up to now, is the historical parameter \cite{Long, houghton1968nonlinear, bouhadef1988contribution,  Lowery-Liapis-1999, pompee2015modelling, Baines, rousseaux2020classical} that was used so far in the literature to classify the pure hydraulic regimes and that we propose to replace by $r_\text{up}^\text{Filling}$ in our new dispersive and geometrical classification. Of course, the maximum height of the obstacle is smaller than the technological height for obvious reasons. In addition, in order to get a bounded phase diagram with a horizontal range between 0 and 1 (as in the standard ``$Fr$ versus $r$'' diagram), we apply the hyperbolic tangent function to the horizontal axis to highlight the different zones in the diagram of the Figure \ref{Fr vs tanhh}. The scaling height $h_\text{tech}$ has the role of zooming the diagram in or out for the depth range we wish to study.

The transition curve shown in green on Figure \ref{Fr vs tanhh} corresponds to the limit where the minima of the phase and group velocities are no longer equal to $c=\sqrt{gh}$. According to the Figure \ref{v vs h}, $\underset{k\in \mathbb{R}^+}{\min}\left(v_\varphi\right)$ and $\underset{k\in \mathbb{R}^+}{\min}\left(v_g\right)$ are separated from $c$ at a depth of $h=\sqrt{3}l_c$. To display this point of separation in Figure \ref{Fr vs tanhh}, we need to set $h_c=\sqrt{3}l_\text{c}$ in equation \ref{h_c} and inject it into the definition of the upstream Froude number (equation \ref{poly}). We thereby obtain the transitional Froude number:
\begin{equation}
    Fr_\text{trans}^\text{transit}=\left(\frac{h_c}{h_{\rm up}}\right)^{3/2}=\frac{\left(\sqrt{3}\sqrt{\frac{\gamma}{\rho g}}\right)^{\frac{3}{2}}}{h_\text{up}^{\frac{3}{2}}}
\end{equation}
The green curve therefore corresponds to an isoflux curve with a flow rate equal to $q_\text{transit}=\sqrt{gh_c^3}=\sqrt[4]{27}\sqrt{g}l_{c} ^{\frac{3}{2}}$.
To find out the trajectory of a transcritical regime in this phase diagram, we project the reciprocal function of Long's law, namely the equation \ref{Long_explicit_bas}, onto the diagram. Since Long's law is parameterized by the maximum height of the obstacle $b_\text{max}$, we obtain different trajectories for different values of $b_\text{max}$; some examples are shown with solid black lines on Figure \ref{Fr vs tanhh}. As the different curves corresponding to the transcritical regime cross the different zones (delimited by the red, blue and green curves) for different values of the upstream Froude number, this means that there is a scaling effect on the domains of existence of the flow regimes with regard to dispersive effects. It is also for this reason that the upstream water depth cannot be scaled by $b_\text{max}$ as was done in anterior works where dispersion was not taken into account \cite{Long, houghton1968nonlinear, pratt1982dynamics, Pratt1984, lawrence1987steady, bouhadef1988contribution, Lowery-Liapis-1999, Baines}. Experimental realisations of the hydrodynamic regimes 
corresponding to the different zones of the diagram in Figure \ref{Fr vs tanhh} are later shown in Figures \ref{Planches_trans_sans_porte} and \ref{Planches_trans_avec_porte}.

The transcritical case admits four different zones, delimited in Figure \ref{Fr vs tanhh} by the different coloured curves (green, blue, and red). We label the zones using Roman numerals:
\begin{itemize}
    \item[-] Area \Romanbar{1}: if the transcritical regime curve is less than $Fr_\text{trans}^{\text{transit}}$ (the green curve), then negative energy modes exist only in the supercritical zone (downstream from the horizon) because the dispersive speed limits ($\underset{k\in \mathbb{R}^+}{\min}\left(v_\varphi\right)$ and $\underset{k\in \mathbb{R}^+}{\min}\left(v_g\right)$) coincide at the horizon;
    \item[-] Area \Romanbar{2}: if the transcritical regime is between curve $Fr_\text{trans}^{\text{transit}}$ (the green curve) and $Fr_{(g)}$ (the red curve), then there is a water depth $h_g$ and a water depth $h_\varphi$ with $h_g,h_\varphi\in]h_c;h^{\text{up}}_{\rm trans}[$ such that $\frac{q}{h_g}=\underset{k\in \mathbb{R}^+}{\min}\left(v_g\right)$ and $\frac{q}{h_\varphi}=\underset{k\in \mathbb{R}^+}{\min}\left(v_\varphi\right)$ and $h_g>h_\varphi$. This reflects the appearance of a dispersive zone before the horizon. In this dispersive region, negative-energy modes can exist upstream of the horizon, and turning points appear in the subcritical zone \cite{rousseaux2010horizon}, but only in a limited region; 
    \item[-] Area \Romanbar{3}: if the transcritical regime lies between $Fr_{(g)}^\text{up}$ (the red curve) and $Fr_{(\varphi)}^\text{up}$ (the blue curve), then there exists $h_\varphi^\prime$ with $h_\varphi^\prime\in]h_c;h_{\text{trans}}^{\text{up}}[$ such that $\frac{q}{h_\varphi^\prime}=\underset{k\in \mathbb{R}^+}{\min}\left(v_\varphi\right)$. In this zone the minimum of the group velocity has been sent to infinity, that is: $\frac{q}{h_{\text{trans}}^{\text{up}}}\geqslant\underset{k\in \mathbb{R}^+}{\min}\left(v_g\right)$. In this dispersive region, negative-energy modes can exist before the horizon, in the subcritical zone, but only in a limited region. Moreover, there now exist additional (dispersive) positive-energy modes that are incident from infinity;
    \item[-] Area \Romanbar{4}: if the transcritical regime is above $Fr_{(\varphi)}^\text{up}$ (the blue curve): $\frac{q}{h_{\text{trans}}^{\text{up}}}\geqslant\underset{k\in \mathbb{R}^+}{\min}\left(v_\varphi\right)$. According to \cite{rousseaux2010horizon}, negative-energy modes 
    exist everywhere in the subcritical zone. 
\end{itemize}

The graphics in the Figure \ref{Fr vs tanhh} was designed for a transcritical regime, accelerated by the presence of an obstacle (black hole flow type). But the diagram can also work in the case of a decelerating transcritical regime (white fountain type). The difference is that the upstream Froude number is transformed into the downstream Froude number and the upstream height is replaced by the downstream height. If the decelerating transcritical regime is not governed by the flow over an obstacle -- if, say, it is instead driven by dissipation, like an undular hydraulic jump \cite{fourdrinoy2022correlations} -- then the black curves shown in the diagram of Figure \ref{Fr vs tanhh} are no longer valid.

From the point of view of astrophysics, it is difficult to draw a parallel with the diagram in Figure \ref{Fr vs tanhh} because of the current theory available namely General Relativity which  is dispersion-less. In fact, for the graph to be relevant to astrophysics, we would need a non-trivial dispersion model (different from $\omega=\pm ck$) breaking Lorentz invariance with its full physical implication, perhaps provided by a quantum gravity regime that remains elusive so far. If we had to return to a non-dispersive description, the equivalent of a $r^\text{Ob}_\text{up}$ could be encoded either in the mass distribution or the topology inside the horizon...

To conclude this section, the main result is the construction of a phase diagram to take account of dispersive effects for transcritical regimes and therefore for the study of surface waves in a transcritical regime. With this diagram, we have identified 3 threshold curves (which are summarised in the table \ref{resum_trans_disp}), which divide the plane into 4 dispersive zones named with Roman numerals (\Romanbar{1}, \Romanbar{2}, \Romanbar{3} and \Romanbar{4}). We conclude from this that if we want to be as close as possible experimentally to the case in astrophysics, i.e. to be as non-dispersive as possible in the vicinity of the horizon of a black hole, we need to locate ourselves as close as possible to the green line in the diagramm \ref{Fr vs tanhh} (first line in the table \ref{resum_trans_disp}). In the next section, we will also construct a phase diagram for the subcritical regime and identify the different dispersive domains.

\begin{table}[h!]
\begin{tabular}{|c|c|c|c|}
\hline
\begin{tabular}[c]{@{}c@{}}Curve \\ colour\end{tabular} & Equations                        & Physical meaning                                                                                                                                                                                                                                                                                                                                                                                     & Comments                                                   \\ \hline
Green                                                   & $Fr_\text{trans}^\text{transit}$ & \begin{tabular}[c]{@{}c@{}}This isoflux curve corresponds to the flow rate \\ ($q=q_\text{transit}$) for which $\underset{k\in \mathbb{R}^+}{\min}\left(v_g\right)$ and $\underset{k\in \mathbb{R}^+}{\min}\left(v_\varphi\right)$\\  dissociate from the horizon and therefore\\  dispersive sub-domains appear in\\  the sub-critical part of the transcritical regime.\end{tabular}               & $q_\text{transit}=\sqrt[4]{27}\sqrt{g}l_{c} ^{\frac{3}{2}}$ \\ \hline
Red                                                     & $Fr_{(g)}^\text{up}$             & \begin{tabular}[c]{@{}c@{}}On the curve, the upstream flow velocity is equal to $\underset{k\in \mathbb{R}^+}{\min}\left(v_g\right)$.\\  In this case, waves with positive energies can be\\  blocked between upstream ($x=-\infty)$ and the\\  horizon ($x=x_\text{hor}$). As for negative energy\\  waves upstream of the horizon, they are confined\\  to a limited region of space.\end{tabular} &                                                            \\ \hline
Blue                                                    & $Fr_{(\varphi)}^\text{up}$       & \begin{tabular}[c]{@{}c@{}}On the curve, the upstream flow velocity is equal to $\underset{k\in \mathbb{R}^+}{\min}\left(v_\varphi\right)$.\\  In this case, waves with positive energies\\  can be blocked between\\  the upstream ($x=-\infty)$ and the\\  horizon ($x=x_\text{hor}$) and waves with negative\\  energies can reach the upstream ($x=-\infty)$.\end{tabular}                       &                                                            \\ \hline
Black                                                   & $Fr_\text{trans}$                & \begin{tabular}[c]{@{}c@{}}Long's Law is projected on \\ top of this to determine \\ the trajectory of the transcritical \\ regime in this phase space.\\ This black line depends \\ on the size of $b_\text{max}$.\end{tabular}                                                                                                                                                                     & Equation \ref{Long_explicit_bas}                           \\ \hline
\end{tabular}
\caption{Summary table of the curves separating the different dispersive regimes in the case of an accelerating transcritical regime (see figure \ref{Fr vs tanhh}).}
\label{resum_trans_disp}
\end{table}

\subsection{Classification of subcritical flows}

\begin{figure}[!h]
    \centering
    \includegraphics[scale=0.35]{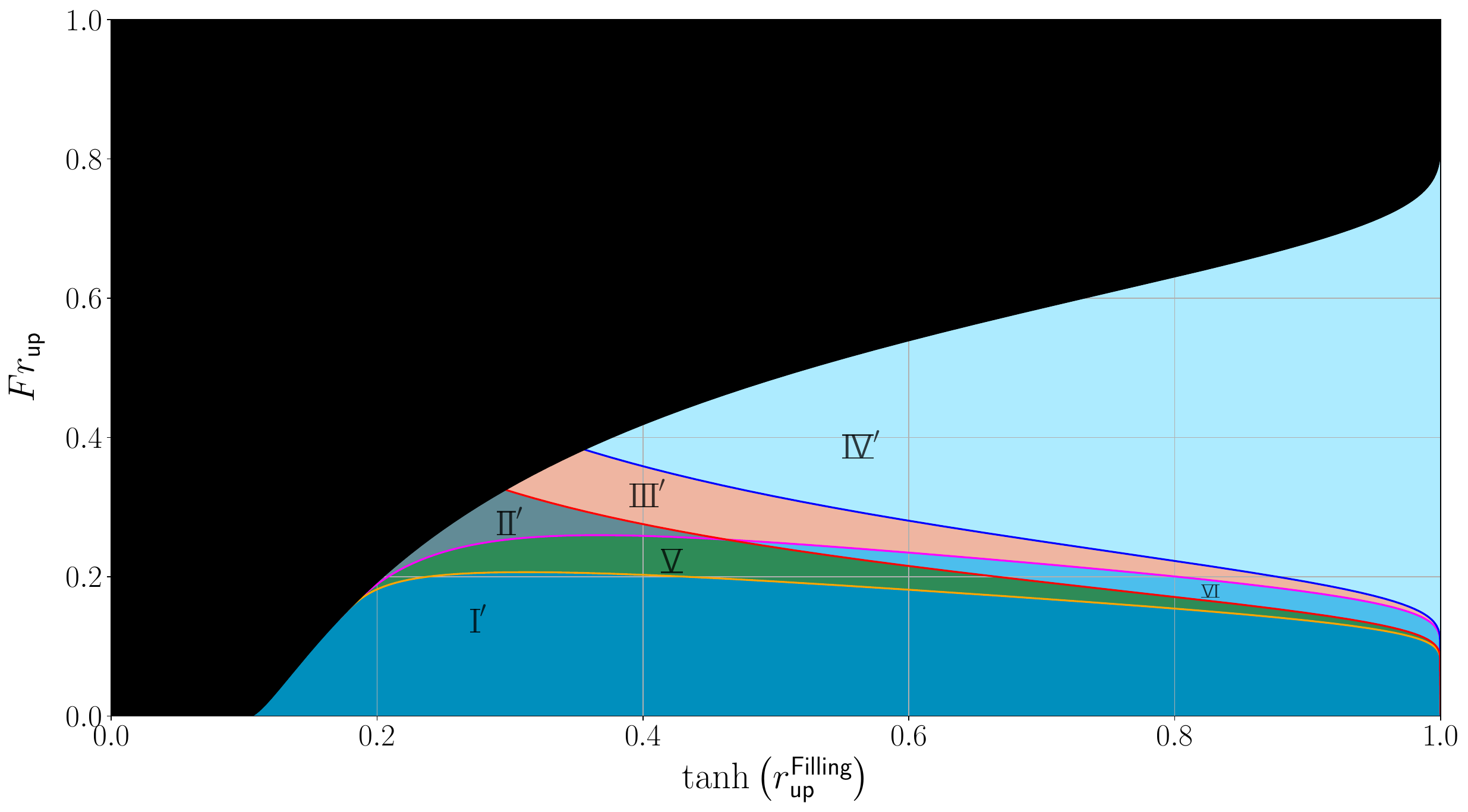}
    \caption{Example of a phase diagram for subcritical flow regimes. This diagram was constructed for an obstacle with a maximum height of $b_\text{max}=0.0105\:m$ and $h_\text{tech}=0.1\:m$.}
    \label{Fr vs tanhh (sous)}
\end{figure}

A phase diagram can also be constructed for subcritical flows, taking dispersive and geometrical effects into account. The starting point for this phase diagram, for subcritical regimes, is the diagram of Figure \ref{v vs h}. The scaling chosen is the same as for Figure \ref{Fr vs tanhh}. 
The $Fr_{(g)}^\text{up}$ and $Fr_{(\varphi)}^\text{up}$ curves continue to be represented in red and blue, respectively, as can be seen in Figure \ref{Fr vs tanhh (sous)}. However, as there is no longer a horizon, the constraint with its corresponding curve $Fr_\text{trans}^\text{transit}$ no longer applies. We therefore need to compare both dispersive limits with the current speed at the minimum height of the subcritical regime. Moreover, as the minimum height is different from $h_c$ (more precisely $h_\text{min}>h_c\:\forall q$) then there is no longer a single horizon appearance height ($h=\sqrt{3}l_{c}$) as in the transcritical case: there are therefore new horizon appearance limits. To find the new limits, we parameterized the upstream water level with the minimum water level for the subcritical regime, which is at $b(x)=b_\text{max}$ (see the hydraulic regime b) in figure \ref{Long_regime}). These conditions are injected into the equation \ref{Bernou}. Thus, we obtain a polynomial of the third order for the upstream water depth:
\begin{equation}
    h_\text{up}^{3}-\left(\frac{\underset{k\in \mathbb{R}^+}{\min}\left(v_X\right)^2}{2g}+h_\text{min}+b_\text{max}\right)h_\text{up}^2+\frac{\underset{k\in \mathbb{R}^+}{\min}\left(v_X\right)^2h_\text{min}^2}{2g}=0\quad\text{with}\quad X\in\{\varphi,g\}
\end{equation}
The flow rate does not appear in this equation, as it is parameterized too by the minimum water depth of the subcritical regime:
$q=\underset{k\in \mathbb{R}^+}{\min}\left(v_X\right)h_\text{min}$. For the latter polynomial, we can extract an exact solution (using again Cardan's formulae) which will be used to plot the desired limits:
\begin{multline}
    h_\text{up,(X)}^{\text{transit,sub}}=\frac{1}{3}\left(\frac{\underset{k\in \mathbb{R}^+}{\min}\left(v_X\right)^2}{2g}+h_\text{min}+b_\text{max}\right)\left[1+2\cos\left(\frac{1}{3}\Arccos\left(1-\frac{27}{4g}\frac{\underset{k\in \mathbb{R}^+}{\min}\left(v_X\right)^2h_\text{min}^2}{\left(\frac{\underset{k\in \mathbb{R}^+}{\min}\left(v_X\right)^2}{2g}+h_\text{min}+b_\text{max}\right)^3}\right)\right)\right]\\
    \quad\text{with}\quad X\in\{\varphi,g\}
    \label{sol}
\end{multline}
with $\underset{k\in \mathbb{R}^+}{\min}\left(v_X\right)=f(h_\text{min})$ according to the graphics in the Figure \ref{v vs h}.
We can construct the upstream Froude number for the transitions of appearance of the minimum of the phase velocity and the minimum of the group velocity in the case of a subcritical regime, which is:
\begin{equation}
    Fr_\text{sub,(X)}^{\text{transit}}=\frac{\underset{k\in \mathbb{R}^+}{\min}\left(v_X\right) h_\text{min}}{\sqrt{g}\left(h_\text{up,(X)}^{\text{transit,sub}}\right)^{\frac{3}{2}}}\quad\text{with}\quad X\in\{\varphi,g\}
    \label{Fr_limit_sous}
\end{equation}

By varying $h_\text{min}$ between $[\sqrt{3}l_c;+\infty[$, we can form the phase diagram for the subcritical flow regimes, which gives the Figure \ref{Fr vs tanhh (sous)}.

 As $h_\text{up,(X)}^{\text{transit,sub}}$ depends on $b_\text{max}$ then the lower limits $Fr_\text{sub,(X)}^{\text{transit}}$ will also depend on $b_\text{max}$. It is for this reason that the diagram shown in figure \ref{Fr vs tanhh (sous)} is only valid for a given value of $b_\text{max}$, i.e. here $b_\text{max}=0.0105\:m$. On the graphics of the Figure \ref{Fr vs tanhh (sous)}, the limit of appearance of the minimum of the group velocity, $Fr_\text{sub,(g)}^{\text{transit}}$, is represented by the orange curve and the limit of appearance of the minimum of the phase velocity, $Fr_\text{sub,($\varphi$)}^{\text{transit}}$, is represented by the magenta curve. Furthermore, as we are studying a subcritical regime, we are interested in a range of Froude numbers that is below the curve corresponding to the Long's law (equation \ref{Long_explicit_bas}); this is why the part of the plane for Froude numbers above the latter curve is a dark area, as it is not of interest for subcritical flow regimes. Using equation \ref{sol}, \ref{Fr_limit_sous}, \ref{asympt1} and \ref{asympt2}, we can determine the asymptotic developments of the limits $Fr_\text{sub,(X)}^{\text{transit}}$ as $h_\text{up}$ tends to infinity, which gives:
\begin{equation}
    Fr_\text{sub,(X)}^{\text{transit}}=Fr_{(X)}-\frac{3}{2}\frac{b_\text{max}}{h_\text{up}}Fr_{(X)}+\underset{h_\text{up}\rightarrow +\infty}{O}\left(Fr_{(X)}^2\right)\quad\text{with}\quad X\in\{\varphi,g\}
\end{equation}
In fact, the $Fr_\text{sub,(X)}^{\text{transit}}$ limits approach $Fr_\text{(X)}$ by lower values, which is logical because the appearance of minima precedes their disappearance, and the way in which $Fr_\text{sub,(X)}^{\text{transit}}$ approaches $Fr_\text{(X)}$ depends on $r_\text{up}^\text{Ob}=b_\text{max}/h_\text{up}$. Finally, as the upstream water level decreases, the limits $Fr_\text{sub,(g)}^{\text{transit}}$ and $Fr_\text{sub,($\varphi$)}^{\text{transit}}$ both tend towards the same point, which is the point of intersection between $Fr_\text{trans}^\text{transit}$ and $Fr_\text{trans}$. By construction, the diagram has six zones whatever the values of $b_\text{max}$, as for example on the Figure \ref{Fr vs tanhh (sous)} diagram for $b_\text{max}=0.0105\:m$. Those six zones are:
\begin{itemize}
    \item[-]Area $\text{\Romanbar{1}}^\prime$: if the subcritical regime has a Froude number less than $Fr_\text{sub,(g)}^{\text{transit}}$, then the regime admits no negative-energy waves, and no wave blocking \cite{nardin2009wave, rousseaux2010horizon, Rousseaux-BASICS-2013, rousseaux2020classical};
    \item[-]Area \Romanbar{5}: if the subcritical regime has a Froude number greater than $Fr_\text{sub,(g)}^{\text{transit}}$ and less than $Fr_\text{sub,($\varphi$)}^{\text{transit}}$ and $Fr_\text{(g)}^\text{up}$, then there is a height $h^a_g\in[h_\text{min};h_\text{up}[$ such that $\frac{q}{h^a_g}=\underset{k\in \mathbb{R}^+}{\min}\left(v_g\right)$. There is therefore a zone above the obstacle where height fluctuations could be blocked;
    \item[-]Area $\text{\Romanbar{2}}^\prime$: if the subcritical regime has a Froude number greater than $Fr_\text{sub,($\varphi$)}^{\text{transit}}$ and less than $Fr_\text{(g)}^\text{up}$, then there is $h^b_g\in]h_\text{min};h_\text{up}[$ and $h^b_\varphi\in[h_\text{min};h^b_g[$ such that $\frac{q}{h^b_g}=\underset{k\in \mathbb{R}^+}{\min}\left(v_g\right)$ and $\frac{q}{h^b_\varphi}=\underset{k\in \mathbb{R}^+}{\min}\left(v_\varphi\right)$. There is a blocked zone and a zone above the obstacle where waves with negative energies can exist but they are confined \cite{rousseaux2010horizon};
    \item[-]Area \Romanbar{6}: if the subcritical regime has a Froude number greater than $Fr_\text{(g)}^\text{up}$ and less than $Fr_\text{sub,($\varphi$)}^{\text{transit}}$, then $\frac{q}{h_\text{up}}\geqslant\underset{k\in \mathbb{R}^+}{\min}\left(v_g\right)$ means that height fluctuations can be blocked everywhere;
    \item[-]Area $\text{\Romanbar{3}}^\prime$: if the subcritical regime has a Froude number greater than $Fr_\text{(g)}^\text{up}$ and $Fr_\text{sub,($\varphi$)}^{\text{transit}}$ and less than $Fr_\text{($\varphi$)}^\text{up}$, then $\frac{q}{h_\text{up}}\geqslant\underset{k\in \mathbb{R}^+}{\min}\left(v_g\right)$ and there is $h^c_\varphi\in[h_\text{min};h_\text{up}[$ such that $\frac{q}{h^c_\varphi}=\underset{k\in \mathbb{R}^+}{\min}\left(v_\varphi\right)$ .This means that height fluctuations can be blocked everywhere and there is a zone above the obstacle where waves with negative energies can exist but they are confined \cite{rousseaux2010horizon};
    \item[-]Area $\text{\Romanbar{4}}^\prime$: if the subcritical regime has a Froude number greater than $Fr_\text{($\varphi$)}^\text{up}$, then $\frac{q}{h_\text{up}}\geqslant\underset{k\in \mathbb{R}^+}{\min}\left(v_\varphi\right)$ means that waves with negative energies can exist everywhere and they are not confined \cite{nardin2009wave, rousseaux2010horizon, Rousseaux-BASICS-2013, rousseaux2020classical}.
\end{itemize}
Roman numerals with $\prime$ are used to distinguish the dispersive domains common to the transcritical regime. Although they represent the same type of hydro-dispersive regime, they do not occupy the same area in the diagram -$Fr_\text{up}$ versus $\tanh\left(r_\text{up}^\text{Filling}\right)$-. In addition, as the $\Romanbar{5}$ and $\Romanbar{6}$ regimes do not appear in the transcritical phase diagram, they do not appear without $\prime$. In addition, hydro-dispersive regime \Romanbar{6} is a new regime that has never been identified before. So in the case of subcritical flow regimes and within the framework of Analogue Gravity, one usually look for flow regimes that allow the existence of negative norm modes or negative energy waves to produce the Hawking amplification process \cite{nardin2009wave, rousseaux2010horizon, Rousseaux-BASICS-2013, rousseaux2020classical}. To do this, all we need is, for the subcritical regime, to be above the magenta curve, i.e. above $Fr_\text{sub,($\varphi$)}^{\text{transit}}$.

To conclude this section, the main result is the construction of a phase diagram to take into account dispersive effects for subcritical regimes and therefore for the study of surface waves in a subcritical regime.  This diagram enabled us to identify 4 threshold curves (summarised in table \ref{resum_sous}), which divide the plane into 6 dispersive zones named by Roman numerals ($\text{\Romanbar{1}}^\prime$, $\text{\Romanbar{2}}^\prime$, $\text{\Romanbar{3}}^\prime$, $\text{\Romanbar{4}}^\prime$, \Romanbar{5} and \Romanbar{6}).  We conclude that if we want to obtain waves with negative energies, we must place ourselves in a region above the magenta curve (see the second line of the summary table \ref{resum_sous}). In what follows, we will use the results of this and the previous section to propose a classification of the experimental regimes.

\begin{table}[h!]
\begin{tabular}{|c|c|c|c|}
\hline
\begin{tabular}[c]{@{}c@{}}Curve \\ colour\end{tabular} & Equations                                    & Physical meaning                                                                                                                                                                                                                                                                                                                                                                                                 & Comments                                                                                                                                                                                                         \\ \hline
Orange                                                  & $Fr_\text{sub,(g)}^{\text{transit}}$         & \begin{tabular}[c]{@{}c@{}}This curve corresponds to the fact that \\ the maximum flow velocity is equal to $\underset{k\in \mathbb{R}^+}{\min}\left(v_g\right)$. \\ A zone therefore appears where waves \\ with positive energies can be blocked. \\ This is the $\underset{k\in \mathbb{R}^+}{\min}\left(v_g\right)$ threshold curve for \\ a subcritical regime.\\ This curve depends on \\ $b_\text{max}$.\end{tabular}                                        & \begin{tabular}[c]{@{}c@{}}The maximum velocity \\ of the subcritical \\ flow occurs when \\ $b(x)=b_\text{max}$,\\  i.e. when $h(x)=h_\text{min}$ \\ (see case b) in \\ figure \ref{Long_regime}).\end{tabular} \\ \hline
Magenta                                                 & $Fr_\text{sub,($\varphi$)}^{\text{transit}}$ & \begin{tabular}[c]{@{}c@{}}This curve corresponds to the fact that the \\ maximum speed of the flow is equal to $\underset{k\in \mathbb{R}^+}{\min}\left(v_\varphi\right)$. \\ A zone therefore appears where waves \\ with negative energies can exist and be \\ blocked. This is the $\underset{k\in \mathbb{R}^+}{\min}\left(v_\varphi\right)$ threshold curve \\ for a subcritical regime.\\ This curve depends on \\ $b_\text{max}$.\end{tabular}                    & \begin{tabular}[c]{@{}c@{}}The maximum velocity \\ of the subcritical flow \\ occurs when \\ $b(x)=b_\text{max}$, \\ i.e. when $h(x)=h_\text{min}$ \\ (see case b) in \\ figure \ref{Long_regime}).\end{tabular} \\ \hline
Red                                                     & $Fr_{(g)}^\text{up}$                         & \begin{tabular}[c]{@{}c@{}}On the curve, the  upstream flow velocity is \\ equal to $\underset{k\in \mathbb{R}^+}{\min}\left(v_g\right)$. In this case, waves with positive \\ energies can be blocked between $x=-\infty$ or \\ $x=+\infty$ and the position $x=0$ where \\ $h(x)=h_\text{min}$. Negative energy waves \\ are confined to a limited region \\ of space around $h(x)=h_\text{min}$.\end{tabular} &                                                                                                                                                                                                                  \\ \hline
Blue                                                    & $Fr_{(\varphi)}^\text{up}$                   & \begin{tabular}[c]{@{}c@{}}On the curve, the upstream flow velocity is \\ equal to $\underset{k\in \mathbb{R}^+}{\min}\left(v_\varphi\right)$. In this case, waves with positive \\ energies can be blocked between $x=-\infty$ or \\ $x=+\infty$ and the position where \\ $h(x)=h_\text{min}$ and waves with \\ negative energies can \\ reach $x=-\infty$ or $x=+\infty$.\end{tabular}                        &                                                                                                                                                                                                                  \\ \hline
Black                                                   & $Fr_\text{trans}$                            & \begin{tabular}[c]{@{}c@{}}The black area corresponds to the\\  zone that is not accessible for a \\ sub-critical regime. It is delimited by \\ the reciprocal function of Long's \\ law. This boundary depends \\ on the size of $b_\text{max}$.\end{tabular}                                                                                                                                                   & Equation \ref{Long_explicit_bas}                                                                                                                                                                                 \\ \hline
\end{tabular}
\caption{Summary table of the curves separating the different dispersive regimes in the case of a subcritical regime (see figure \ref{Fr vs tanhh (sous)}).}
\label{resum_sous}
\end{table}

\section{Classification of the experimental flow regimes including the effects of dispersion and geometry}
\label{Exp-disp}

We may be able to build a classification of the hydrodynamic flow regimes realised in our experiments. 
As a general remark, we note that even though our flows are turbulent (as confirmed by the value of the Reynolds number $Re=Uh/\nu=q/\nu$ which is in the range $Re\in[600,11 500]$), our classification assumes that the flow is uniform over the entire depth (a ``plug'' flow), with a negligible boundary layer and no vorticity. We neglect also both the viscous damping \cite{robertson2017viscous} and the turbulent head losses in our classification; these would gradually deform the flow and could be described using an effective obstacle.  Nevertheless, even with these strong assumptions, we believe our classification to be robust. This strong assumption will be tested numerically (the numerical results are presented in the appendices) since particle velocity measurements are impossible with our experimental setup since the bottom of the channel is opaque to a LASER sheet.

\subsection{A quantification of the Depression regime}
\label{Flat}

\begin{figure}[!h]
    \centering
    \includegraphics[scale=0.35]{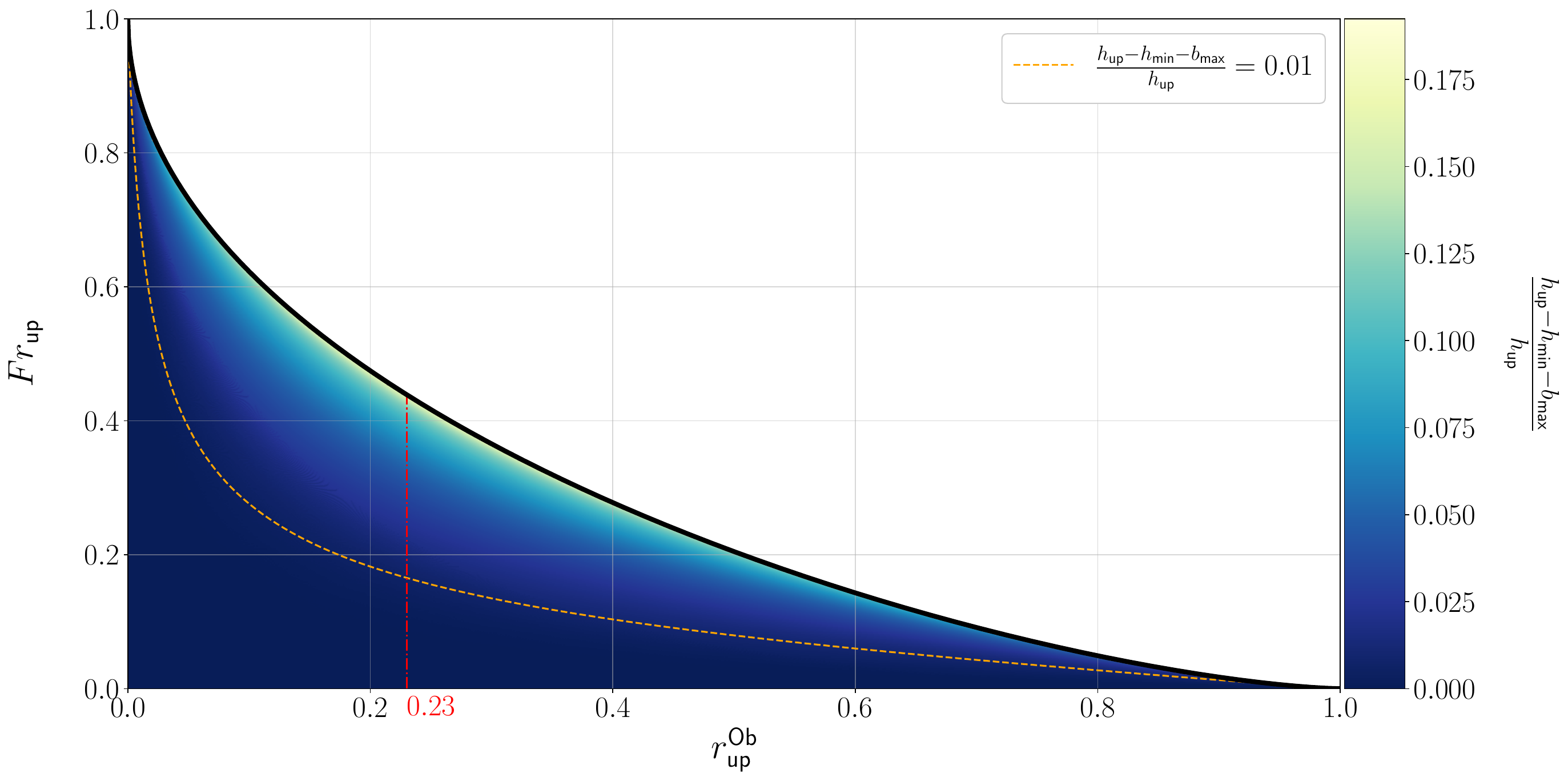}
    \caption{Hydraulic diagram -$Fr_\text{up}$ versus $r_\text{up}^\text{Ob}$- with the frontier for the Depression zone appearance. Below the corresponding dotted orange curve (obtained with the equation \ref{flat_formula} for $C_\text{flat}=0.01$.), the subcritical flow regimes satisfy the condition \ref{Cond} with $C_\text{flat}=0.01$. These regimes are considered to be Flat (subcritical).}
    \label{seuil_plat}
\end{figure}

Firstly, we propose to differentiate between a subcritical {\it depression} regime and a subcritical {\it flat} regime, as would be observed by the naive observer along the river (see the classification based on the free surface discussed earlier in the paper).  In practice, for a subcritical regime, it can be difficult to identify a depression, as it can be extremely small visually. This is why we introduced the notion of a flat regime inspired by \cite{rousseaux2020classical}. As an explicit definition of this regime, we use the following quantitative criterion:
\begin{equation}
    C = \frac{h_\text{up}-h_\text{min}-b_\text{max}}{h_\text{up}} < C_{\rm flat} \,.
    \label{Cond}
\end{equation}
The number $C$, which must lie between 0 and 1, characterises the deflection of the depression in the subcritical regime in relation to the upstream water depth. 
Condition \ref{Cond} defines in turn a critical upstream Froude number $Fr_{C}$ where $C = C_{\rm flat}$:
\begin{equation}
    Fr_C=\sqrt{2}\left(1-C_{\rm flat}-r_\text{up}^\text{Ob}\right)\sqrt{\frac{C_{\rm flat}}{1-\left(1-C_{\rm flat}-r_\text{up}^\text{Ob}\right)^2}} \,,
    \label{flat_formula}
\end{equation}
where we have used equation \ref{Bernou}. This critical Froude number marks the boundary between the flat and depression regimes.
In the remainder of this paper, to define a Flat subcritical regime, we choose 
we take $C_{\rm flat}=0.01$; the corresponding threshold is shown in the Figure \ref{seuil_plat}.

While we know $C < 1$, we can derive a stricter upper bound by recalling that $h_{\rm min} \geq h_{c}$.
We arrive at the subsequent inequalities:
\begin{equation}
    r_\text{up}^\text{Ob}+2\sin\left(\frac{1}{3}\Arcsin\left(1-r_\text{up}^\text{Ob}\right)\right)<1-C<\frac{h_\text{min}}{h_\text{up}}+r_\text{up}^\text{Ob}<1
\end{equation}
By studying the function of $r_\text{up}^\text{Ob}+2\sin\left(\frac{1}{3}\Arcsin\left(1-r_\text{up}^\text{Ob}\right)\right)-1+C$, we can see that the function is no longer negative for :
\begin{equation}
    C_\text{max}\approx 0.19245\quad\text{for}\quad r_\text{up}^\text{Ob}\approx 0.23
\end{equation}
This means that the subcritical regime with the greatest subcritical depression is located in the - Fr versus r- diagram close to $r_\text{up}^\text{Ob}\approx 0.23$ and just below the Long transcritical boundary \cite{pompee2015modelling}. In addition, the maximum depression is:
\begin{equation}
    \underset{r,Fr\in[0,1]}{\min}\left(\frac{h_\text{min}+b_\text{max}}{h_\text{up}}\right)=1-C_\text{max}\approx 0.80755
\end{equation}
and $C\in]0;C_\text{max}[$. The Flat subcritical domain can also be placed in the diagram -Fr versus $\tanh(r_\text{up}^\text{Filling})$- (see section \ref{Flat+disp} in the appendix).

\subsection{Introducing the nomenclature for the hydrodynamics regimes}

As a shorthand for characterizing the experimental regimes, we have settled on 
the following nomenclature:
\begin{equation}
    {\huge X}^{U,B,\ldots}_{{\small\Romanbar{1}/\Romanbar{2}/}\cdots}
    \label{nomenclature}
\end{equation}
We now explain the meaning of the symbols in the expression \ref{nomenclature}.

\begin{figure}[h!]
    \centering
    \includegraphics[scale=0.35]{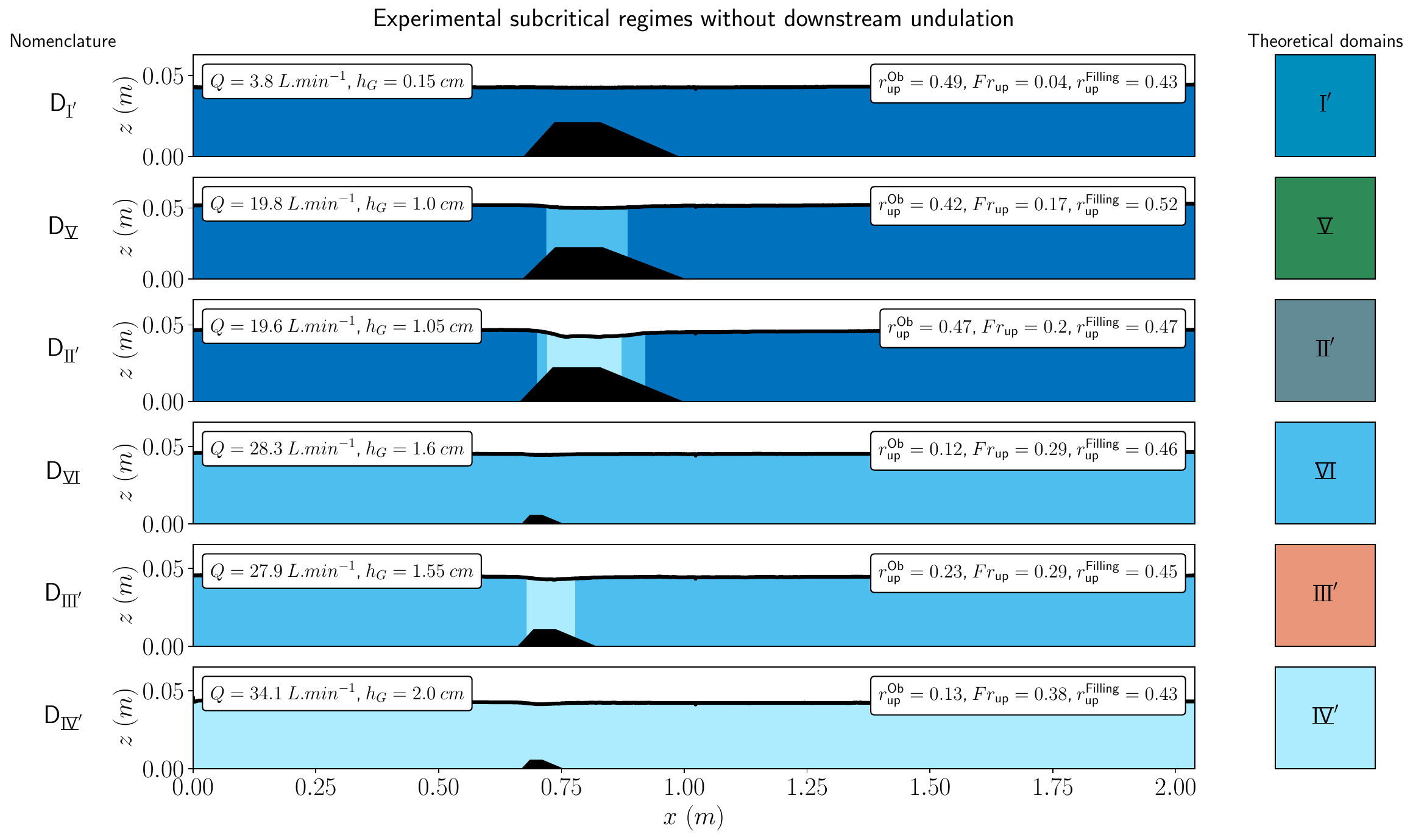}
    \caption{Experimental subcritical flow regimes without undulation. These regimes are accompanied by their nomenclature, on the left, and by the name of the theoretical regime and the colour of the corresponding domain, on the right.}
    \label{Planches_sous_sans_ondu}
\end{figure}

The letter $X$ refers to the type of hydraulic regime.
It takes one among five values:
\begin{itemize}
\item $\text{T}^{\nearrow}$ for the accelerating transcritical regime with a waterfall (a descending jump);
\item $\text{T}^{\searrow}$ for the decelerating transcritical regime with a cataract (a climbing jump);
\item $\text{D}$ for the subcritical regime with a depression above the obstacle and flat asymptotic regions; 
\item $\text{F}$ for the subcritical regime with a flat surface 
everywhere;
\item $\text{S}$ for the supercritical regime.
\end{itemize}

\begin{table}[h!]
\begin{tabular}{|cl|c|c|cc|}
\hline
\multicolumn{2}{|c|}{}                                                                              &                                                                                  &                                                                                      & \multicolumn{2}{c|}{\begin{tabular}[c]{@{}c@{}}Historical\\ experiment\end{tabular}}                                                     \\ \cline{5-6} 
\multicolumn{2}{|c|}{\multirow{-2}{*}{\begin{tabular}[c]{@{}c@{}}Theoretical\\ color\end{tabular}}} & \multirow{-2}{*}{\begin{tabular}[c]{@{}c@{}}Nomenclature\\ regimes\end{tabular}} & \multirow{-2}{*}{\begin{tabular}[c]{@{}c@{}}Experimental\\ realisation\end{tabular}} & \multicolumn{1}{c|}{\begin{tabular}[c]{@{}c@{}}Subluminal\\ LASER effect?\end{tabular}} & \begin{tabular}[c]{@{}c@{}}Hawking\\ effect?\end{tabular} \\ \hline
\multicolumn{2}{|c|}{\cellcolor[HTML]{008FBD}}                                                      & $\text{D}_{\text{\Romanbar{1}}^\prime}$                                          & \textcolor{orange}{$\star$}                                                          & \notableentry                                             & \notableentry                                             \\ \hline
\multicolumn{2}{|c|}{\cellcolor[HTML]{2E8B57}}                                                      & $\text{D}_\text{\Romanbar{5}}$                                                   & \textcolor{green}{\checkmark}                                                        & \notableentry                                             & \notableentry                                         \\ \hline
\multicolumn{2}{|c|}{\cellcolor[HTML]{628B96}}                                                      & $\text{D}_{\text{\Romanbar{2}}^\prime}$                                          & \textcolor{green}{\checkmark}                                                        & \multicolumn{1}{c|}{\textcolor{red}{x}}                                      & 
\begin{tabular}[c]{@{}c@{}}Rousseaux et al.(2008)\cite{rousseaux2008observation} \\ Chaline et al.(2013) \cite{chaline2013some}\end{tabular}
\\ \hline
\multicolumn{2}{|c|}{\cellcolor[HTML]{4DBEED}}                                                      & $\text{D}_\text{\Romanbar{6}}$                                                   & \textcolor{green}{\checkmark}                                                        & \notableentry                                             & \notableentry                                             \\ \hline
\multicolumn{2}{|c|}{\cellcolor[HTML]{F0B6A1}}                                                      & $\text{D}_{\text{\Romanbar{3}}^\prime}$                                          & \textcolor{green}{\checkmark}                                                        & \multicolumn{1}{c|}{\textcolor{red}{x}}                                      & Rousseaux et al.(2008)\cite{rousseaux2008observation}                                       \\ \hline
\multicolumn{2}{|c|}{\cellcolor[HTML]{ADEBFF}}                                                      & $\text{D}_{\text{\Romanbar{4}}^\prime}$                                          & \textcolor{green}{\checkmark}                                                        & \notableentry                                        & Rousseaux et al.(2008)\cite{rousseaux2008observation}                                        \\ \hline
\end{tabular}
\caption{Summary table of subcritical flow regimes with depression and their importance for the Analogue Gravity domain.}
\label{Tab D}
\end{table}

The subscript on expression \ref{nomenclature} designates the dispersive domain of the system. This dispersive domain is identified by the Roman numerals in Figures \ref{Fr vs tanhh} and \ref{Fr vs tanhh (sous)}.

The superscript of the nomenclature designates the visual form of the free surface. We have identified the following possibilities:
\begin{itemize}
\item U when there is an undulation (a train of stationary secondary waves known as whelps) downstream of a depression on the top of the obstacle;
\item B when the interface is breaking somewhere ({\it e.g.}, at a turbulent jump); \item M when there is a modulation of the changing average speed due to the presence of the undulation whelps whose amplitudes is varying due to the succession of extrema (crests and troughs).
\end{itemize}
Combinations of letters can be used ({\it e.g.}, we may write $UB$ when there are breaking waves on the undulation).

\subsection{Experimental realisations}

We now present a series of experimental realisations to indicate the various flow regimes described above.  In the appendix, we revisit historical experiments in Analogue Gravity and interpret them in the light of our proposed classification.  This article is also accompanied by a ZIP file containing animations describing the positions of the different dispersive regimes in a new phase diagram.

\subsubsection{Subcritical flows}

Figure \ref{Planches_sous_sans_ondu} shows the different experimental subcritical flow regimes that show a depression on top of the obstacle.  The theoretical dispersive domain associated with each regime is shown on the right, and the corresponding label (defined according to the nomenclature outlined above) is shown on the left. The obstacle used is shown in black, and the free surface is symbolised by a thick dark line. Three colours (dark blue, light blue and cyan) are used to visualise the dispersive nature of the flows; these correspond in Fig.~\ref{Fr vs tanhh (sous)} to the $\text{\Romanbar{1}}^\prime$, $\text{\Romanbar{4}}^\prime$ and \Romanbar{6} domains, respectively. These colours are used to describe the appearance of the phase velocity minimum and the group velocity minimum locally. This gives us an idea of the size of the dispersive cavities. Moreover, as can be seen again in figure \ref{Planches_sous_sans_ondu} and also in Figures \ref{Planches_trans_sans_porte} and \ref{Planches_trans_avec_porte}, the size of the obstacle is {\it not} the same for all regimes.  This is an important point: because of the scaling effect (discussed in Section \ref{Th-disp}), the size of the obstacle must be reduced to reach the highest dispersive regimes in terms of water height (such as the \Romanbar{6} regime). This imposes a constraint for the experimentalists, and in effect it means that, for given technological limitations (such as the strength of the pump or the height of the channel), the desired flow and/or dispersive regime determines the size of the obstacle that must be used.

Table \ref{Tab D} shows the experimental regimes in Figure \ref{Planches_sous_sans_ondu} and compares them with the regimes studied in the literature. Under ``Experimental realisation'', a green check mark means that the flow regime has been realised in an experiment, while an orange star means that we have not realised the regime but we know that it exists. Under ``Historical experiment", we use a red cross to show that the regime is interesting from an Analogue Gravity perspective, in that it allows either for an analogue of the Hawking effect (in the sense of scattering between positive- and negative-energy waves), or for a variant with an amplifying cavity known as the black hole LASER effect~\footnote{The black hole LASER may manifest itself, provided viscous dissipation is not too high, either in a subcritical cavity with a subluminal dispersion correction or in a supercritical cavity with a superluminal correction to the acoustic metric associated to a non-dispersive dispersion relation with its Doppler shift \cite{peloquin2016analog, robertson2017viscous}.} \cite{peloquin2016analog}. If such a regime has already been realised in an Analogue Gravity experiment, a reference to that experiment is placed in the cell instead. A black cross across the whole cell means that the regime is not interesting from an Analogue Gravity perspective because it does not allow negative-energy modes.  

\begin{figure}[!h]
    \centering
    \includegraphics[scale=0.35]{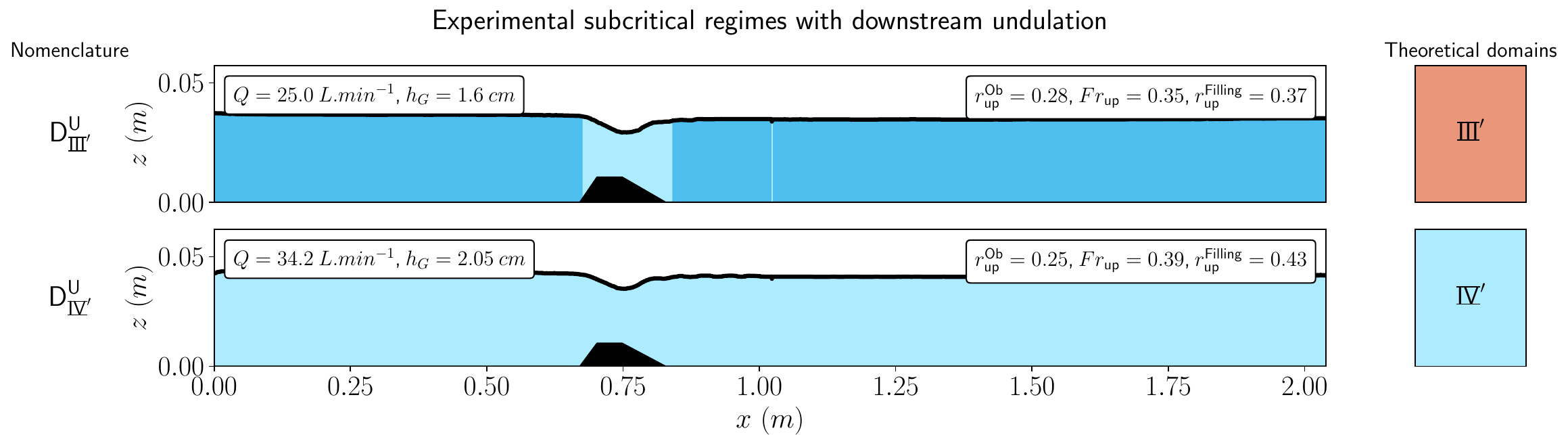}
    \caption{Experimental subcritical flow regimes with an undulation. These regimes are accompanied by their nomenclature, on the left, and by the name of the theoretical regime and the colour of the corresponding domain, on the right.}
    \label{Planches_sous_avec_ondu}
\end{figure}

The regimes in the Figure \ref{Planches_sous_avec_ondu} are characterized by the presence of an undulation (which was also visible in the historical experiments reported in Table \ref{tab DU} and in the appendices, section \ref{histoire}).

\begin{table}[h!]
\begin{tabular}{|cl|c|c|cc|}
\hline
\multicolumn{2}{|c|}{\multirow{2}{*}{\begin{tabular}[c]{@{}c@{}}Theoretical\\ color\end{tabular}}} & \multirow{2}{*}{\begin{tabular}[c]{@{}c@{}}Nomenclature\\ regimes\end{tabular}} & \multirow{2}{*}{\begin{tabular}[c]{@{}c@{}}Experimental\\ realisation\end{tabular}} & \multicolumn{2}{c|}{\begin{tabular}[c]{@{}c@{}}Historical\\ experiment\end{tabular}}                                                     \\ \cline{5-6} 
\multicolumn{2}{|c|}{}                                                                             &                                                                                 &                                                                                     & \multicolumn{1}{c|}{\begin{tabular}[c]{@{}c@{}}Subluminal\\ LASER effect?\end{tabular}} & \begin{tabular}[c]{@{}c@{}}Hawking\\ effect?\end{tabular} \\ \hline
\multicolumn{2}{|c|}{\cellcolor[HTML]{628B96}}                                                     & $\text{D}^{\text{U}}_{\text{\Romanbar{2}}^\prime}$                              & \textcolor{orange}{$\star$}                                                         & \multicolumn{1}{c|}{\textcolor{red}{x}}                                      & \textcolor{red}{x}                                        \\ \hline
\multicolumn{2}{|c|}{\cellcolor[HTML]{F0B6A1}}                                                     & $\text{D}^{\text{U}}_{\text{\Romanbar{3}}^\prime}$                              & \textcolor{green}{\checkmark}                                                       & \multicolumn{1}{c|}{\textcolor{red}{x}}                                      & 
\begin{tabular}[c]{@{}c@{}} Weinfurtner et al.(2011)\cite{weinfurtner2011measurement} \\ Euvé and Rousseaux (2021) \cite{euve2021non}\end{tabular}
                                            \\ \hline
\multicolumn{2}{|c|}{\cellcolor[HTML]{ADEBFF}}                                                     & $\text{D}^{\text{U}}_{\text{\Romanbar{4}}^\prime}$                              & \textcolor{green}{\checkmark}                                                       & \notableentry                                          & Euvé et al.(2016)\cite{euve2016observation}                                             \\ \hline
\end{tabular}
\caption{Summary table of subcritical flow regimes with depression and downstream undulation and their importance in Analogue Gravity.}
\label{tab DU}
\end{table}

\subsubsection{Transcritical flows}

\begin{figure}[h!]
    \centering
    \includegraphics[scale=0.35]{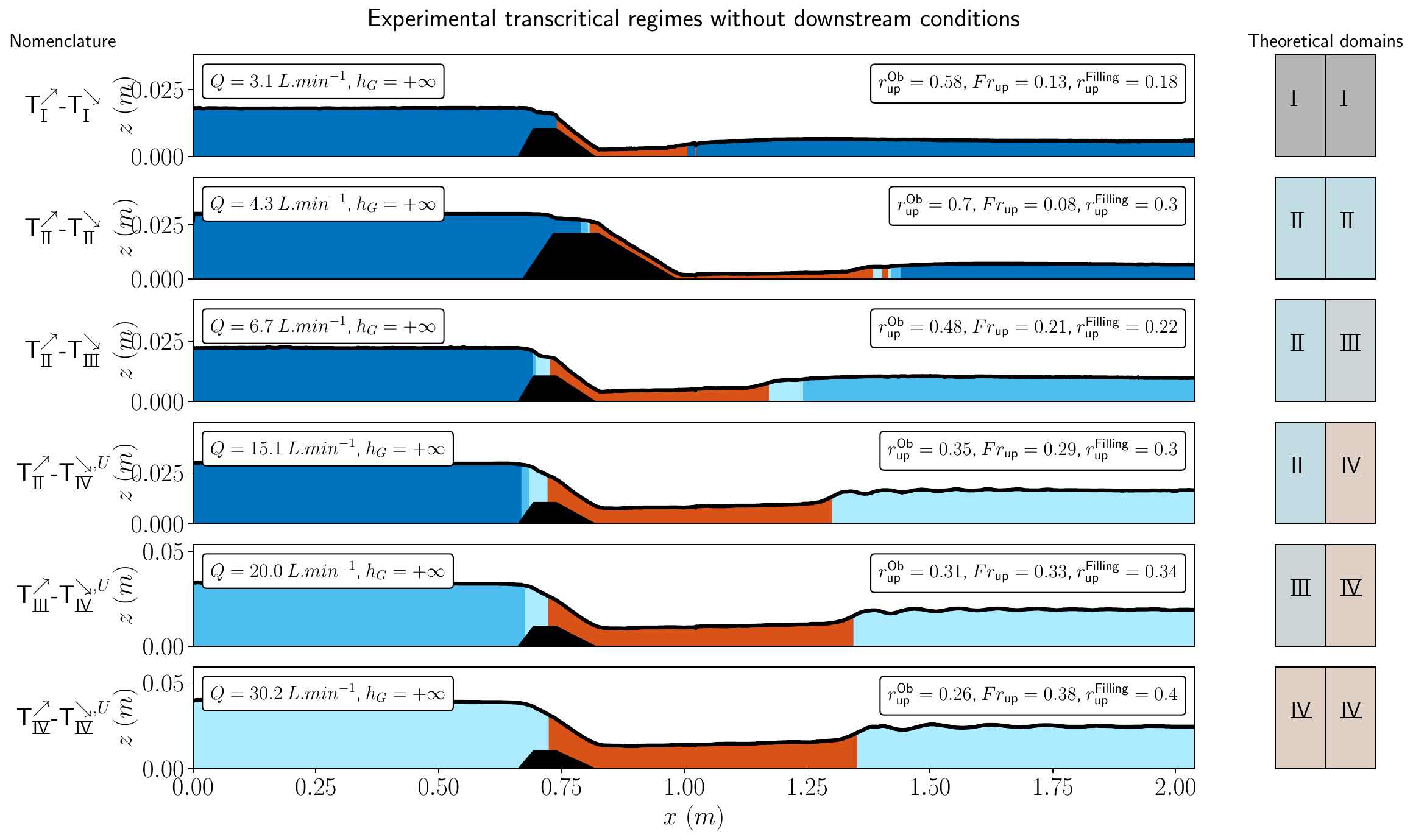}
    \caption{Experimental transcritical regimes without downstream condition. These regimes are accompanied by their nomenclature, on the left, and by the name of the theoretical regime and the colour of the corresponding domain, on the right. The theoretical domains are separated into two to designate respectively the theoretical domain for the accelerating transcritical part and the decelerating transcritical part.}
    \label{Planches_trans_sans_porte}
\end{figure}

The same approach can be taken for transcritical systems. For transcritical regimes that can be considered as free, {\it i.e.}, without any downstream boundary conditions, we obtain Figure \ref{Planches_trans_sans_porte}. For the nomenclature of the regimes, we employ two ordered symbols. The first symbol corresponds to the accelerating part where the flow first becomes supercritical, while the second corresponds to the downstream part where the flow decelerates to become subcritical again.
This deceleration occurs via a hydraulic jump, which occurs some distance downstream from the obstacle and is governed by dissipation~\cite{fourdrinoy2022correlations}). As soon as $\text{T}^{\searrow}$ is in the dispersive domain \Romanbar{4}, a spontaneous undulation is observed downstream from the jump (hence the U in the superscript). 
Moreover, when the hydraulic jump is in the $\text{T}^{\searrow}_{\text{\Romanbar{1}}}$ or $\text{T}^{\searrow}_{\text{\Romanbar{2}}}$ regime, it becomes unsteady and emits waves downstream: this is how the instability predicted in General Relativity is "solved" with the emission of waves by the white horizon where energy is reputed to be concentrated. The emission of waves is a original way to regularize the energy caustics in this flow regime not anticipated in General Relativity.

\begin{table}[h!]
\begin{tabular}{|cc|c|c|cc|}
\hline
\multicolumn{2}{|c|}{\multirow{2}{*}{\begin{tabular}[c]{@{}c@{}}Theoretical\\ color\end{tabular}}} & \multirow{2}{*}{\begin{tabular}[c]{@{}c@{}}Nomenclature\\ regimes\end{tabular}}             & \multirow{2}{*}{\begin{tabular}[c]{@{}c@{}}Experimental\\ realisation\end{tabular}} & \multicolumn{2}{c|}{\begin{tabular}[c]{@{}c@{}}Historical\\ experiment\end{tabular}}                                                               \\ \cline{5-6} 
\multicolumn{2}{|c|}{}                                                                             &                                                                                             &                                                                                     & \multicolumn{1}{c|}{\begin{tabular}[c]{@{}c@{}}Superluminal LASER\\ effect?\end{tabular}} & \begin{tabular}[c]{@{}c@{}}Hawking\\ effect?\end{tabular}           \\ \hline
\multicolumn{1}{|c|}{\cellcolor[HTML]{7D6B80}\phantom{111.}}              & \cellcolor[HTML]{7D6B80}             & $\text{T}^{\nearrow}_{\text{\Romanbar{1}}}-\text{T}^{\searrow}_{\text{\Romanbar{1}}}$        & \textcolor{green}{\checkmark}                                                       & \multicolumn{1}{c|}{\textcolor{red}{x}}                                      & \textcolor{red}{x}                                                  \\ \hline
\multicolumn{1}{|c|}{\cellcolor[HTML]{C0D9E3}}              & \cellcolor[HTML]{C0D9E3}             & $\text{T}^\nearrow_{\text{\Romanbar{2}}}-\text{T}^\searrow_{\text{\Romanbar{2}}}$            & \textcolor{green}{\checkmark}                                                       & \multicolumn{1}{c|}{\textcolor{red}{x}}                                      & \textcolor{red}{x}                                                  \\ \hline
\multicolumn{1}{|c|}{\cellcolor[HTML]{C0D9E3}}              & \cellcolor[HTML]{C8CECF}             & $\text{T}^\nearrow_{\text{\Romanbar{2}}}-\text{T}^\searrow_{\text{\Romanbar{3}}}$            & \textcolor{green}{\checkmark}                                                       & \multicolumn{1}{c|}{\textcolor{red}{x}}                                      & \textcolor{red}{x}                                                  \\ \hline
\multicolumn{1}{|c|}{\cellcolor[HTML]{C0D9E3}}              & \cellcolor[HTML]{DDCABE}             & $\text{T}^\nearrow_{\text{\Romanbar{2}}}-\text{T}^{\searrow,\text{U}}_{\text{\Romanbar{4}}}$ & \textcolor{green}{\checkmark}                                                       & \multicolumn{1}{c|}{\textcolor{red}{x}}                                      & \textcolor{red}{x}                                                  \\ \hline
\multicolumn{1}{|c|}{\cellcolor[HTML]{C8CECF}}              & \cellcolor[HTML]{DDCABE}             & $\text{T}^\nearrow_{\text{\Romanbar{3}}}-\text{T}^{\searrow,\text{U}}_{\text{\Romanbar{4}}}$ & \textcolor{green}{\checkmark}                                                       & \multicolumn{1}{c|}{\textcolor{red}{x}}                                      & \begin{tabular}[c]{@{}c@{}}Fourdrinoy \\ et al. (2022)\cite{fourdrinoy2022correlations}\end{tabular} \\ \hline
\multicolumn{1}{|c|}{\cellcolor[HTML]{DDCABE}}              & \cellcolor[HTML]{DDCABE}             & $\text{T}^\nearrow_{\text{\Romanbar{4}}}-\text{T}^{\searrow,\text{U}}_{\text{\Romanbar{4}}}$ & \textcolor{green}{\checkmark}                                                       & \multicolumn{1}{c|}{\textcolor{red}{x}}                                      & Euvé et al. (2020)\cite{euve2021non}                                                  \\ \hline
\end{tabular}
\caption{Summary table of transcritical regimes without downstream conditions and their importance for the Analogue Gravity domain.}
\label{Tab TNG}
\end{table}

Table \ref{Tab TNG} summarises the free transcritical regimes observed experimentally and the importance of these regimes for Analogue Gravity, in particular, dissipation controls the slowing of the supercritical flow downstream of the obstacle, an undular jump is thus produced that is stabilized by both the effect of dispersion and non-linearity that saturates its amplitude (another way to regularize the caustics). 
These transcritical regimes are of interest for the analogue Hawking effect, both on the accelerating and decelerating sides. 
These regimes are also interesting for the capillary/superluminal black hole LASER effect \cite{peloquin2016analog} in the supercritical region (shown in red in Figure \ref{Planches_trans_sans_porte}).

\begin{figure}[h!]
    \centering
    \includegraphics[scale=0.35]{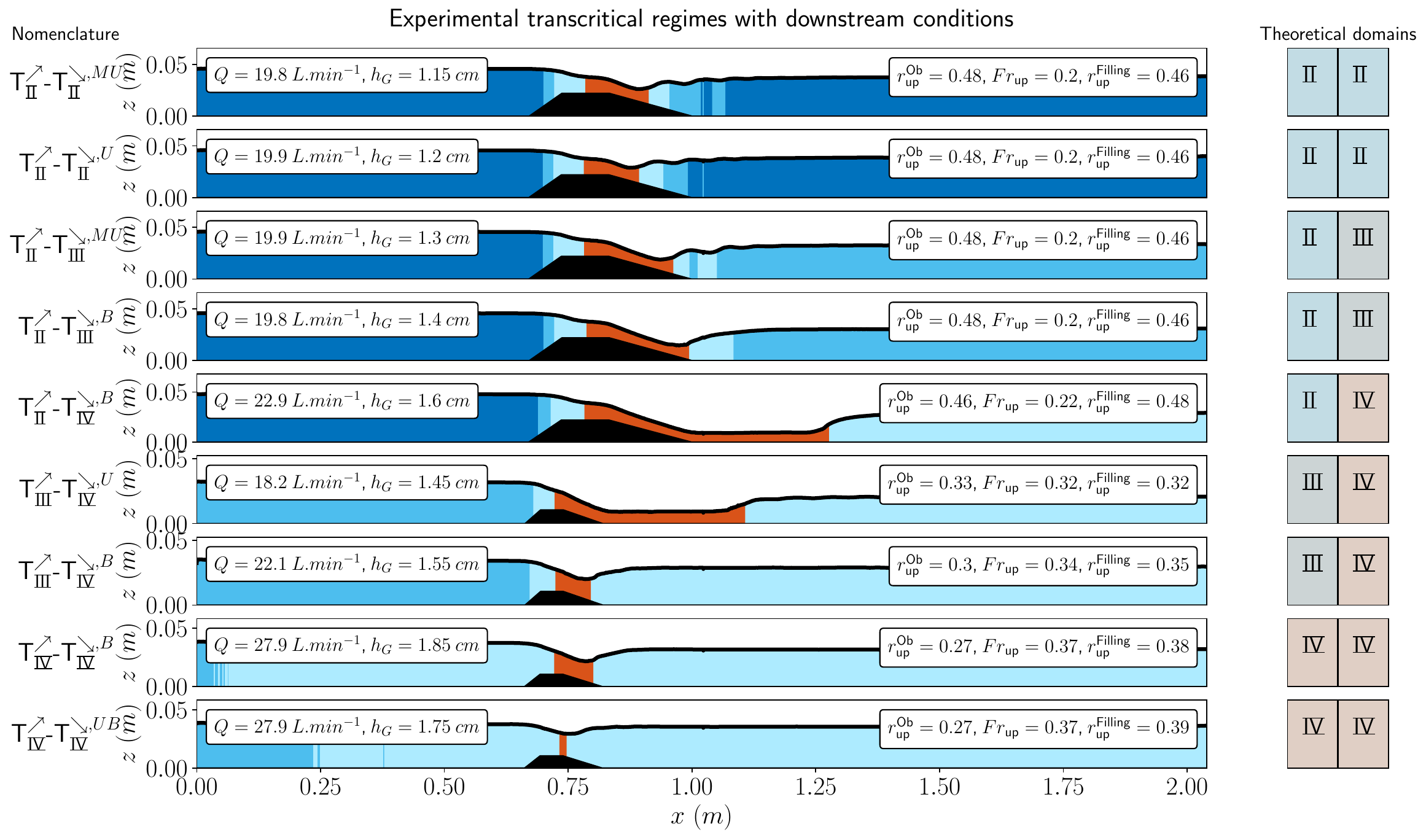}
    \caption{Experimental transcritical regimes with downstream condition. These regimes are accompanied by their nomenclature, on the left, and by the name of the theoretical regime and the colour of the corresponding domain, on the right. The theoretical domains are separated into two to designate respectively the theoretical domain for the accelerating transcritical part and the decelerating transcritical part.}
    \label{Planches_trans_avec_porte}
\end{figure}

Adding a downstream boundary condition by partially lowering a gate at the end of the channel, we find that the position and nature of the hydraulic jump is modified (this was previously observed in~\cite{fourdrinoy2022correlations}). Some such experiments are shown in Figure \ref{Planches_trans_avec_porte}. 
In the presence of the gate, the jump tends to occur much closer to the obstacle, and sometimes we find that it tends to break. For breaking jumps, the turbulence is a non-linear and time-dependent way to dissipate the energy accumulated at the white fountain horizon: again, fluid mechanics solves the difficulty with the assumed instability of white fountain in General Relativity \cite{leonhardt2002intrinsic} by condensed-matter scenarii and purely classical processes (either wave emission, dispersive shock wave saturated by non-linearity or wave breaking).

Table \ref{Tab TrG} summarises these cases.  It shows that these regimes are also interesting for the black hole LASER effect \cite{peloquin2016analog}, because by bringing the hydraulic jump (the stabilized white fountain) closer to the black horizon it reduces the size of the LASER cavity, which helps to fight against viscous damping \cite{robertson2017viscous}.

\begin{table}[h!]
\begin{tabular}{|cc|c|c|cc|}
\hline
\multicolumn{2}{|c|}{\multirow{2}{*}{\begin{tabular}[c]{@{}c@{}}Theoretical\\ color\end{tabular}}} & \multirow{2}{*}{\begin{tabular}[c]{@{}c@{}}Nomenclature\\ regimes\end{tabular}}               & \multirow{2}{*}{\begin{tabular}[c]{@{}c@{}}Experimental\\ realisation\end{tabular}} & \multicolumn{2}{c|}{\begin{tabular}[c]{@{}c@{}}Historical\\ experiment\end{tabular}}                                                              \\ \cline{5-6} 
\multicolumn{2}{|c|}{}                                                                             &                                                                                               &                                                                                     & \multicolumn{1}{c|}{\begin{tabular}[c]{@{}c@{}}Superluminal LASER\\ effect?\end{tabular}} & \begin{tabular}[c]{@{}c@{}}Hawking\\ effect?\end{tabular}          \\ \hline
\multicolumn{1}{|c|}{\cellcolor[HTML]{C0D9E3}\phantom{111.}}              & \cellcolor[HTML]{C0D9E3}             & $\text{T}^\nearrow_{\text{\Romanbar{2}}}-\text{T}^{\searrow,\text{MU}}_{\text{\Romanbar{2}}}$ & \textcolor{green}{\checkmark}                                                       & \multicolumn{1}{c|}{\textcolor{red}{x}}                                      & \textcolor{red}{x}                                                 \\ \hline
\multicolumn{1}{|c|}{\cellcolor[HTML]{C0D9E3}}              & \cellcolor[HTML]{C0D9E3}             & $\text{T}^\nearrow_{\text{\Romanbar{2}}}-\text{T}^{\searrow,\text{U}}_{\text{\Romanbar{2}}}$  & \textcolor{green}{\checkmark}                                                       & \multicolumn{1}{c|}{\textcolor{red}{x}}                                      & \begin{tabular}[c]{@{}c@{}}Fourdrinoy\\ et al. (2022)\cite{fourdrinoy2022correlations}\end{tabular} \\ \hline
\multicolumn{1}{|c|}{\cellcolor[HTML]{C0D9E3}}              & \cellcolor[HTML]{C8CECF}             & $\text{T}^\nearrow_{\text{\Romanbar{2}}}-\text{T}^{\searrow,\text{MU}}_{\text{\Romanbar{3}}}$ & \textcolor{green}{\checkmark}                                                       & \multicolumn{1}{c|}{\textcolor{red}{x}}                                      & \textcolor{red}{x}                                                 \\ \hline
\multicolumn{1}{|c|}{\cellcolor[HTML]{C0D9E3}}              & \cellcolor[HTML]{C8CECF}             & $\text{T}^\nearrow_{\text{\Romanbar{2}}}-\text{T}^{\searrow,\text{B}}_{\text{\Romanbar{3}}}$  & \textcolor{green}{\checkmark}                                                       & \multicolumn{1}{c|}{\textcolor{red}{x}}                                      & \textcolor{red}{x}                                                 \\ \hline
\multicolumn{1}{|c|}{\cellcolor[HTML]{C0D9E3}}              & \cellcolor[HTML]{DDCABE}             & $\text{T}^\nearrow_{\text{\Romanbar{2}}}-\text{T}^{\searrow,\text{B}}_{\text{\Romanbar{4}}}$  & \textcolor{green}{\checkmark}                                                       & \multicolumn{1}{c|}{\textcolor{red}{x}}                                      & \textcolor{red}{x}                                                 \\ \hline
\multicolumn{1}{|c|}{\cellcolor[HTML]{C8CECF}}              & \cellcolor[HTML]{DDCABE}             & $\text{T}^\nearrow_{\text{\Romanbar{3}}}-\text{T}^{\searrow\text{U}}_{\text{\Romanbar{4}}}$   & \textcolor{green}{\checkmark}                                                       & \multicolumn{1}{c|}{\textcolor{red}{x}}                                      & \textcolor{red}{x}                                                 \\ \hline
\multicolumn{1}{|c|}{\cellcolor[HTML]{C8CECF}}              & \cellcolor[HTML]{DDCABE}             & $\text{T}^\nearrow_{\text{\Romanbar{3}}}-\text{T}^{\searrow\text{B}}_{\text{\Romanbar{4}}}$   & \textcolor{green}{\checkmark}                                                       & \multicolumn{1}{c|}{\textcolor{red}{x}}                                      & \textcolor{red}{x}                                                 \\ \hline
\multicolumn{1}{|c|}{\cellcolor[HTML]{DDCABE}}              & \cellcolor[HTML]{DDCABE}             & $\text{T}^\nearrow_{\text{\Romanbar{4}}}-\text{T}^{\searrow\text{B}}_{\text{\Romanbar{4}}}$   & \textcolor{green}{\checkmark}                                                       & \multicolumn{1}{c|}{\textcolor{red}{x}}                                      & \textcolor{red}{x}                                                 \\ \hline
\multicolumn{1}{|c|}{\cellcolor[HTML]{DDCABE}}              & \cellcolor[HTML]{DDCABE}             & $\text{T}^\nearrow_{\text{\Romanbar{4}}}-\text{T}^{\searrow\text{BU}}_{\text{\Romanbar{4}}}$  & \textcolor{green}{\checkmark}                                                       & \multicolumn{1}{c|}{\textcolor{red}{x}}                                      & \textcolor{red}{x}                                                 \\ \hline
\end{tabular}
\caption{Summary table of transcritical flow regimes with downstream conditions and their importance in Analogue Gravity.}
\label{Tab TrG}
\end{table}

\section{Conclusion}
\label{sec:conclusion}
To sum up, we have placed ourselves in the theoretical framework of Analogue Gravity in interfacial hydrodynamics, i.e. with a perfect fluid, an irrotational flow and in the non-dispersive limit, to find a hydraulic classification of free surface flows. The hydraulic regimes include the transcritical regime, the subcritical regime and the supercritical regime. These regimes are controlled by two dimensionless numbers: the upstream obstruction ratio $r_\text{up}^\text{Ob}=b_\text{max}/h_\text{up}$ and the upstream Froude number $Fr_\text{up}=U_\text{up}/\sqrt{gh_\text{up}}$, with $U_\text{up}=q/h_\text{up}$. The -Fr versus r- diagram distinguishes between these three hydraulic regimes, in which Long's law is also projected (relation between the upstream quantities $r_\text{up}^\text{Ob}$ and the Froude number $Fr_\text{up}$ to obtain a transcritical regime). However, in the context of Analogue Gravity, a purely hydraulic description is not enough and we need to add a new ingredient: linear dispersion. Starting from characteristic velocities of the dispersion relation, such as the minimum of the phase velocity, which is the threshold for the appearance of waves with negative energies, and the minimum of the group velocity, which corresponds to the first dispersive horizon, we have obtained a new phase diagram that counts the different dispersive domains. This diagram is based on upstream quantities such as the upstream Froude number and the upstream water level. However, we noticed that if we tried to project Long's law (the law governing transcritical regimes) in the -Fr versus h- diagram, we found that for two different obstacles, i.e. for two different $b_\text{max}$, the dispersive domains are crossed for different Froude numbers, which is therefore incompatible with a purely hydraulic law such as Long's law. There is therefore a size effect. It is therefore no longer interesting to use the -Fr versus r- diagram to describe hydro-dispersive regimes. We have therefore chosen to scale the upstream water depth by a quantity related to the material used in the experiment, called $h_\text{tech}$. This type of scaling allows us to identify the experimentally accessible hydro-dispersive regimes. If we take the inverse problem, we can design an experiment, or change the size of the obstacle to observe certain types of hydro-dispersive regime. In addition, we have given names to the different areas of the diagrams to construct a nomenclature for the hydro-dispersive regimes. This nomenclature provides information on the hydraulic regime, the dispersive range of the regime and the visual appearance of the free surface. Finally, we have sorted out the historical Analogue Gravity experiments in interfacial hydrodynamics to find out which hydro-dispersive regimes are interesting to study (for the analogue Hawking effect or the analogue LASER effect). According to our study, the transcritical regime is of interest for the analogue superluminal LASER effect, i.e. an amplification of the Hawking effect by the overproduction of its partner in the supercritical zone, which in our case corresponds to the overproduction of the capillary waves in the supercritical zone between the accelerating transcritical horizon and the decelerating transcritical horizon. The subluminal LASER effect is also of interest in the case of subcritical regimes, where the Hawking effect is amplified between two horizons of dispersive origin. Concerning some perspectives of this work, our approach relies on two simplifications: we neglected the effect of viscosity whereas it has an obvious influence, say on the appearance of the downstream transcritical ondulation for instance, and we assume that the bottom obstacle has an aspect ratio (length to height) which is big compared to one. An interesting generalisation is the possibility to have small aspect ratio of the obstacle. Coupled to the effect of viscosity, flow recirculations certainly are generated both in front and in the lee side of the geometrical obstacle like in weir fishways \cite{ead2004flow}. It is probable that the flow generates a kind of effective obstacle (space-time) that smooths the sum of the real geometry with its associated spurious recirculations. The inclusion of viscosity in a future generalization of our work should rely on the seminal work by Pratt in 1986 where he showed that the position of the critical point (the horizon in Analogue Gravity) lies where the friction drag coefficient is the negative of the bottom slope $db/dx|_{x=x_\text{hor}}=-C_d$ \cite{pratt1986hydraulic}: a viscous hydro-dispersive theory including all the dispersive scenarii (subluminal, superluminal as in the BEC used by Steinhauer \cite{steinhauer2016observation}, and subluminal followed by a superluminal correction as in this work) for any aspect ratio that includes the prediction of the undulation in Analogue Gravity is still lacking but we believe our work to be an important milestone in its History...

\section*{Acknowledgements}

We are indebted to the Professor E. Lamballais from the Poitiers University for the lending of the pedagogical water channel in between lectures for a few months each year since 2020.  This work pertains (namely is not funded but enters in the corresponding scientific perimeter) to the French government program ``Investissements d'Avenir'' (LABEX INTERACTIFS, reference ANR-11-LABX-0017-01 and EUR INTREE, reference ANR-18-EURE-0010). S. Robertson is funded through the CNRS Chair in Physical Hydrodynamics (reference ANR-22-CPJ2-0039-01). This work was not funded by the french national research agency ANR through its annual call since 2015: this constant lack of funding for interdisciplinary research was a great chance since it pushed the authors to use a small pedagogical water channel that allowed us to discover the effect of size on the regimes observed, an unexpected result that would have been missed with larger channels that required (and still require) more funding to run (renting, costs, sophisticated metrology) and test the classification introduced in this work...
\bibliographystyle{crunsrt}

\nocite{*}

\bibliography{samplebib}

\section{Appendix}

\subsection{A (false) classification that excludes the surface tension effect}

\begin{figure}[!h]
    \centering
    \includegraphics[scale=0.35]{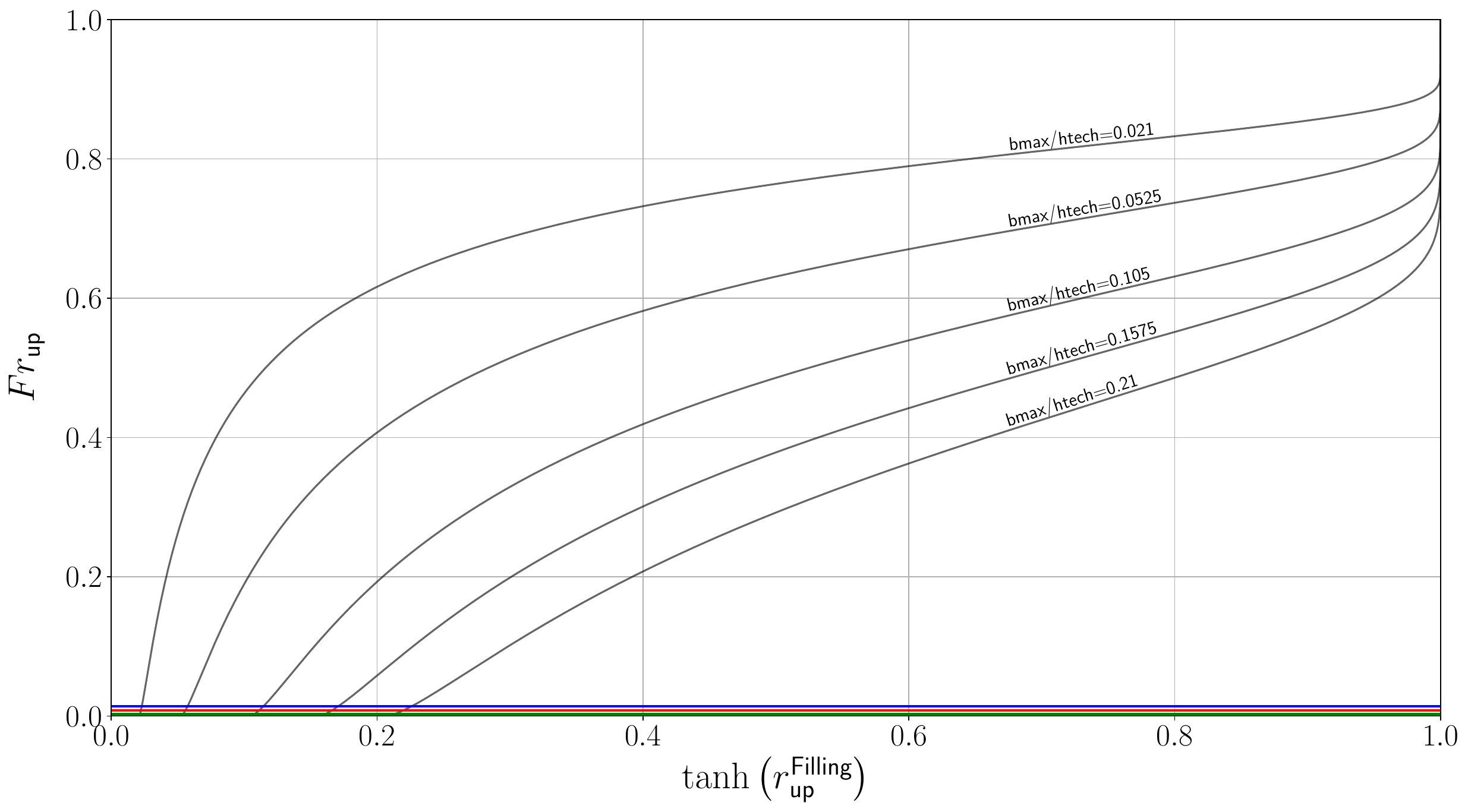}
    \caption{Phase diagram for transcritical regimes where the value of the surface tension is null. This graph was constructed with $h_\text{tech}=0.1\:m$.}
    \label{trans gamma0}
\end{figure}

\begin{figure}[!h]
    \centering
    \includegraphics[scale=0.35]{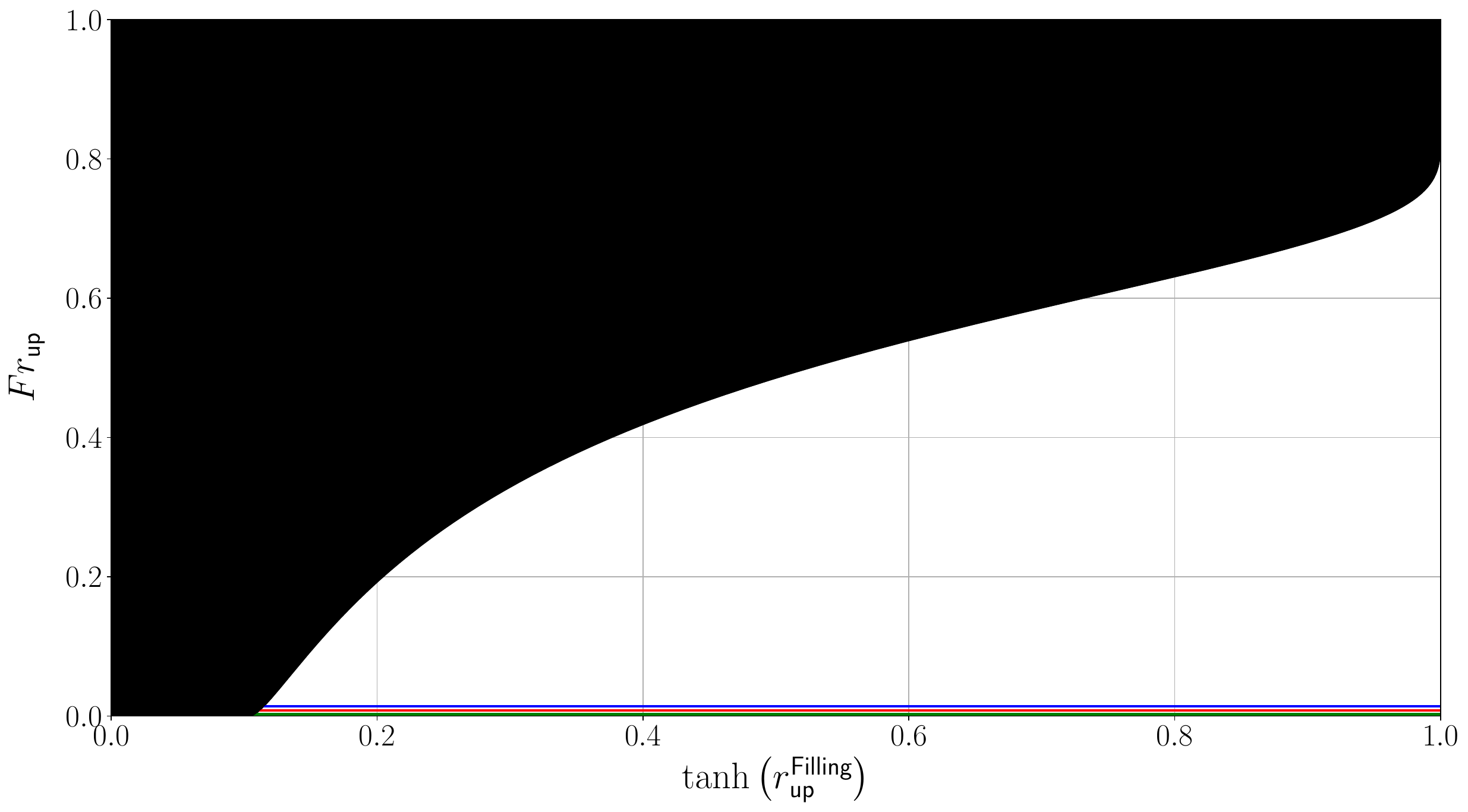}
    \caption{Phase diagram for subcritical regimes where the value of the surface tension is null. This graph was constructed with $b_\text{max}=0.0105\: m$ and $h_\text{tech}=0.1\:m$.}
    \label{Sous gamma0}
\end{figure}

For completeness, we can look at the limit of no surface tension $\gamma=0$. Indeed, it is the natural theoretical regime where a hydraulic metric is defined since the pioneering work \cite{Schuetzhold-Unruh-2002}. In this case, we obtain the Figures \ref{trans gamma0} and \ref{Sous gamma0}. The Figure \ref{trans gamma0} repeats the phase diagram \ref{Fr vs tanhh}, however with $\gamma=0$: the different limits expressed in the section \ref{trans classi}, i.e $Fr_\text{trans}^{\text{transit}}$ (the green curve), $Fr_{(g)}^\text{up}$ (the red curve) and $Fr_{(\varphi)}^\text{up}$ (the blue curve) are now superimposed with $Fr=0$. Similarly, for the subcritical phase diagram \ref{Sous gamma0}, the limits $Fr_\text{sub,(g)}^{\text{transit}}$ (the orange curve), $Fr_\text{sub,($\varphi$)}^{\text{transit}}$ (the magenta curve), $Fr_{(g)}^\text{up}$ (the red curve) and $Fr_{(\varphi)}^\text{up}$ (the blue curve) are now superimposed with $Fr=0$. In the context of Analogue Gravity, we no longer have a threshold for the appearance of negative-energy waves (the partner of analogue Hawking radiation), an astounding departure from the prediction of General Relativity: these negative-energy waves are therefore present at all positions and for all flow rates in both the transcritical and subcritical regimes. This conclusion is similar to the one reached in \cite{nardin2009wave} that was sooner generalized to include the necessary effect of surface tension \cite{rousseaux2010horizon} as discussed in the main text of this paper since the seminal experiments \cite{rousseaux2008observation} pointed to the existence of an experimental threshold (whose origin was unkownn by the authors at that time) for the appearance of negative norm modes that we have finally demonstrated in the present study for all possible flow regimes: surface tension is the true explanation for the threshold of appearance of both the negative norm modes and the undulation which is the same for the confinement of Hawking radiation by the dispersive blue horizon \cite{rousseaux2010horizon, Rousseaux-BASICS-2013, chaline2013some, Peloquin, Euve-Rousseaux-2017}, a fact overlooked by most of the theoretical works \cite{Schuetzhold-Unruh-2002, nardin2009wave, weinfurtner2011measurement, coutant2014undulations, Michel-Parentani-2014, robertson2016scattering, Michel-et-al-2018, Coutant-Weinfurtner-2018}. The so-called robustness of Hawking prediction \cite{gryb2021universality} is challenged in presence of both subluminal and superluminal dispersion corrections due to the water depth and the surface tension since Hawking radiation can be emitted by a horizon but it may also not reach the asymptotic observer because of the existence of the dispersive blue horizon with a finite speed threshold. This conclusion is halfway between the one of General Relativity (the speed at the horizon is the appearance threshold of the negative norm modes which are trapped inside the horizon and in absence of dispersion) and the pure dispersive gravity regime (no speed threshold) of the Figures \ref{trans gamma0} and \ref{Sous gamma0} \cite{Rousseaux-BASICS-2013}.

\subsection{Location of the flat subcritical regime in the -Fr versus $\tanh(r_\text{up}^\text{Filling})$- diagram}\label{Flat+disp}
\begin{figure}[!h]
    \centering
    \includegraphics[scale=0.35]{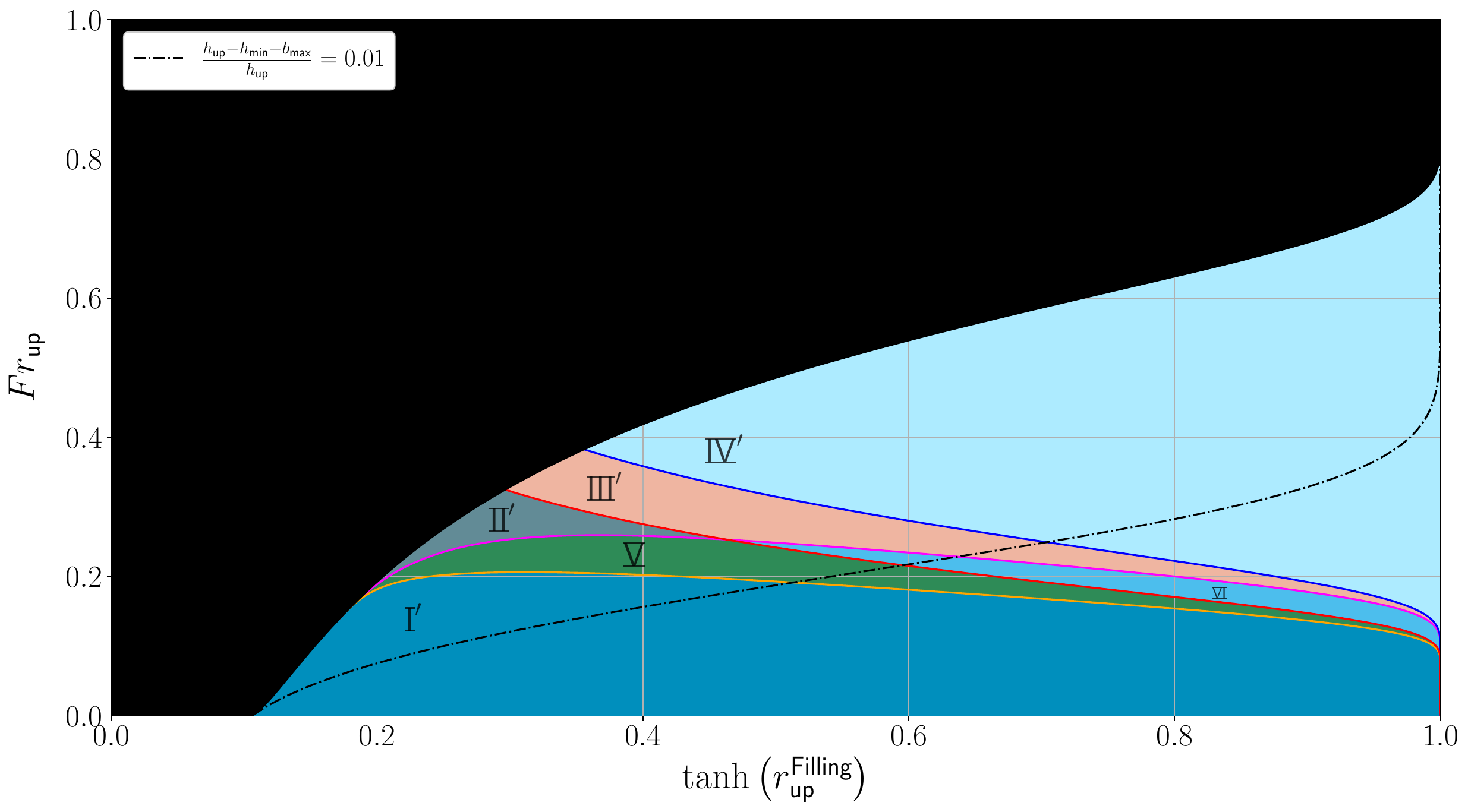}
    \caption{Phase diagram for subcritical regimes taking dispersion into account. This graph was constructed with $b_\text{max}=0.0105\: m$, $h_\text{tech}=0.1\: m$, $b_\text{max}<b_\text{max}^\text{F}$ and $b_\text{max}^\text{F}\approx 0.02479\: m$.}
    \label{Flat inf}
\end{figure}

\begin{figure}[!h]
    \centering
    \includegraphics[scale=0.35]{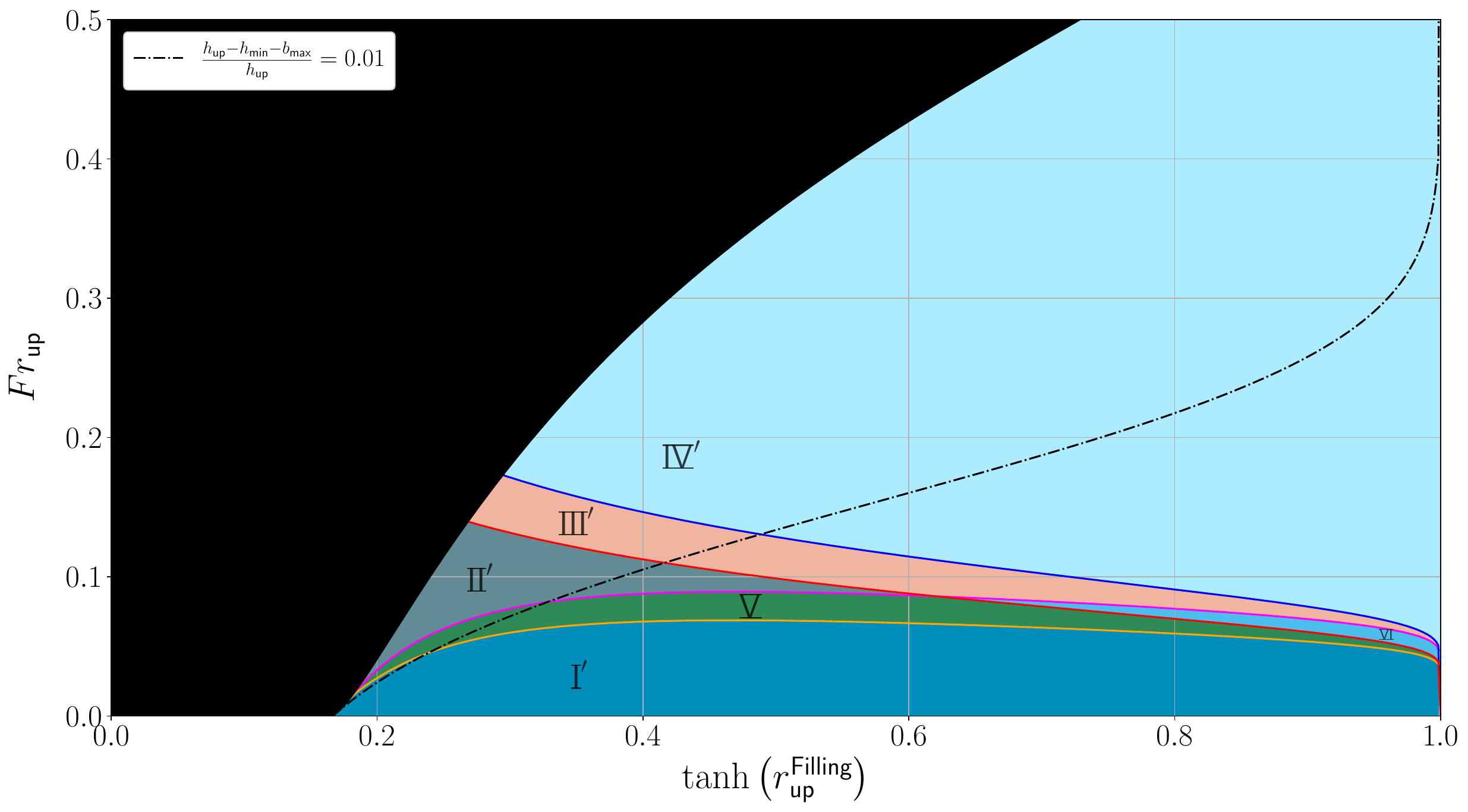}
    \caption{Phase diagram for subcritical regimes taking dispersion into account. This graph was constructed with $b_\text{max}=0.1\: m$, $h_\text{tech}=0.6\: m$, $b_\text{max}>b_\text{max}^\text{F}$ and $b_\text{max}^\text{F}\approx 0.02479\: m$.}
    \label{Flat Sup}
\end{figure}

Using the definition of a flat subcritical regime given in the section \ref{Flat}, we can use the diagram -Fr versus $\tanh(r_\text{up}^\text{Filling})$- to visualise the dispersive domains in which the so-called flat regimes are found, giving the figures \ref{Flat inf} and \ref{Flat Sup}. In these two figures, everything below the dotted line corresponds to the flat regime. We can also see where the flat regimes reach all the dispersive domains. There is therefore a maximum obstacle height, noted $b_\text{max}^\text{F}$ such that:
\begin{itemize}
    \item if $b_\text{max}<b_\text{max}^\text{F}$, then there are 5 dispersive regimes: $\text{F}_{\text{\Romanbar{1}}^\prime}$, $\text{F}_{\text{\Romanbar{3}}^\prime}$, $\text{F}_{\text{\Romanbar{4}}^\prime}$, $\text{F}_{\text{\Romanbar{5}}}$ and $\text{F}_{\text{\Romanbar{6}}}$. The dispersive regime $\text{F}_{\text{\Romanbar{2}}^\prime}$ is not reached (see Figure \ref{Flat inf} constructed for $b_\text{max}=0.0105\: m$);
    \item if $b_\text{max}>b_\text{max}^\text{F}$, the 6 dispersive regimes are reached, i.e. $\text{F}_{\text{\Romanbar{1}}^\prime}$, $\text{F}_{\text{\Romanbar{2}}^\prime}$, $\text{F}_{\text{\Romanbar{3}}^\prime}$, $\text{F}_{\text{\Romanbar{4}}^\prime}$, $\text{F}_{\text{\Romanbar{5}}}$ and $\text{F}_{\text{\Romanbar{6}}}$ (see Figure \ref{Flat Sup} constructed for $b_\text{max}=0.1\: m$).
\end{itemize}
Numerically, we find that $b_\text{max}^\text{F}\approx 0.02479\: m$. Most of the historical experiments in the early years where such that $b_{max}>b_\text{max}^\text{F}$ except \cite{euve2016observation}, \cite{euve2020scattering},  \cite{fourdrinoy2022correlations} and the present work.

\subsection{A resume of the historical experiments in Analogue Gravity with open channel flows}\label{histoire}

Historically, one can classify the experiments in Analogue Gravity in interfacial hydrodynamics according to the following criteria:
\begin{itemize}

\item{\it Transcritical flows and non-linear regimes}: a hydraulic turbulent and aerated jump was observed for high flow rates on the top of the obstacle with the ACRI 2008 geometry in the Nice experiments with dimensions of order 10 m \cite{rousseaux2008observation}. This black hole flow type regime with a turbulent jump as a model of a classical and time-dependent central singularity was dismissed because of the impossibility to stimulate the dispersive white hole horizon with incoming waves from a piston wave-maker placed downstream of the obstacle in presence of a static water depth superior to the maximum height of the obstacle: the waves would have gone through the jump. This case was not reported in \cite{rousseaux2008observation} since it was replaced by the next regimes.\\

\item {\it Subcritical flows and non-linear regimes}: Flat and Depression 
regimes were observed in the Nice 2008 experiments \cite{rousseaux2008observation} where a dispersive white fountain flow was studied on the decelerating side of a bottom obstacle (initially submerged by a static water depth superior to the maximum height of the obstacle). Counter-propagating stimulating waves were sent from the downstream piston wave-maker. Multiple mode conversions were observed but in a non-linear regime with partial wave breaking and harmonic generation (as acknowledged in the conclusions of \cite{rousseaux2008observation}); the generated harmonics were later recognized as free harmonics \cite{euve2021non}, which are themselves solutions of the dispersion relation. Wave blocking was definitely observed \cite{rousseaux2010horizon} for high flow rates with mode conversion towards the so-called ``blue-shifted'' modes 
\cite{nardin2009wave}. Shorter co-propagating modes were reported whose origin is still unknown but which were interpreted, at the time, as indications of the presence of negative-energy modes \cite{rousseaux2008observation}. 
Wave blocking was understood as not necessarily synonymous with mode conversion since some conversion was reported without wave blocking \cite{rousseaux2008observation}. Visual observations did not allow the distinguishing of linearly converted blue-shifted modes \cite{nardin2009wave} from non-linearly excited harmonics of either the incoming or converted modes \cite{euve2021non}. 

Improved geometries were used subsequently by the Nice \cite{chaline2013some} and Vancouver \cite{weinfurtner2011measurement} experiments to minimize flow recirculation on the descending slope of the obstacle \cite{weinfurtner2013classical}. The Vancouver geometry \cite{weinfurtner2011measurement} improved the ACRI 2008 geometry \cite{rousseaux2008observation, rousseaux2010horizon} by smoothing it with the help of a NACA wing-like profile. The Vancouver experiment included a long flat part on top of the obstacle (similar to the ACRI 2008 geometry) which avoided a tunneling effect over 
the obstacle. Moreover, the Vancouver obstacle had a reduced descending slope thanks to the insertion of an inclined flat plate that was glued to the wing profile in order to avoid flow recirculation 
\cite{weinfurtner2013classical}. Simultaneously, the downstream part of the ACRI 2008 geometry was replaced by a flat plate with smaller slope to get the ACRI 2010 geometry with the same slope, so as to avoid flow recirculation \cite{chaline2013some}. In Vancouver \cite{weinfurtner2011measurement, weinfurtner2013classical}, the amplitudes of the stimulating waves were reduced by a factor of 25 (1-2 mm) compared to the 2008 Nice experiments \cite{rousseaux2008observation, rousseaux2010horizon} (2.5-5 cm of physical amplitudes a or 5-10 cm of coastal engineering amplitudes $H=2a$). Mode conversion towards blue-shifted modes and negative-energy modes was reported in the Vancouver experiment \cite{weinfurtner2011measurement, weinfurtner2013classical}.  Contrary to the Nice experiments \cite{rousseaux2008observation, rousseaux2010horizon}, non-linear conversion towards harmonics likely affected only the blue-shifted modes that were themselves the results of a linear conversion \cite{euve2021non}. 

The 2016 Poitiers experiments \cite{euve2016observation} reduced the amplitude of the stimulating waves by a factor of $10$ compared to the 2011 Vancouver experiments \cite{weinfurtner2011measurement, weinfurtner2013classical}.  This allowed all waves to propagate linearly, including the blue-shifted waves. 
Both the 2011 Vancouver 
\cite{weinfurtner2011measurement, weinfurtner2013classical} and the 2016 Poitiers 
experiments \cite{euve2016observation} featured a stationary undulation \cite{coutant2014undulations}. The 2011 Vancouver depth-averaged flow speeds \cite{weinfurtner2011measurement, weinfurtner2013classical} were below the minimum of the group velocity (the Landau speed \cite{rousseaux2020classical}, 23.1 cm/s in water) in both the upstream and downstream regions, whereas the surface speed was above the Landau speed on the entire free surface. 
\cite{euve2017interactions, euve2021non}. It is unclear whether the converted modes (if they would have been linear) would have reached the asymptotic region downstream of the obstacle since the average speed was below the Landau threshold with its corresponding blue horizon \cite{rousseaux2010horizon} far from the obstacle or if the fact that the surface speed was above the same threshold would allow the propagation of Hawking radiation towards the asymptotic region, a matter of current research on the effect of vorticity on wave propagation since the Vancouver flow profile was not plug-like \cite{euve2017interactions, euve2021non}. In any case, the linear Hawking radiation produced in the Vancouver experiments did not reach the asymptotic observer because of harmonics generation \cite{euve2021non}.\\

\item {\it Subcritical flows and linear regimes}:
The 2016 Poitiers flow speed was superior to the Landau speed everywhere \cite{euve2016observation}: linear mode conversions, hence stimulated Hawking radiation, were observed in the asymptotic region downstream of the obstacle contrarily to the 2011 Vancouver experiments \cite{weinfurtner2011measurement, weinfurtner2013classical} where both blue and negative were observed on the top of the obstacle before disappearing because of the free harmonics generation during the propagation away from the conversion zone localized on the top of the obstacle \cite{euve2017interactions, euve2021non}. The 2016 Poitiers experiments \cite{euve2016observation} reported also for the first time the linear Hawking radiation in the asymptotic region stimulated by the free surface noise due to the wave turbulence induced by the bulk flow turbulence. Neither the 2011 Vancouver experiments \cite{weinfurtner2011measurement, weinfurtner2013classical} nor the 2016 Poitiers experiments \cite{euve2016observation} reported a divergence of the scattering coefficients of both the blue (amplified Hawking modes) and negative modes for vanishing frequency because of the disappearance of the dispersive group velocity horizon (hence wave blocking). Hence, the predicted thermal spectrum of Hawking \cite{Hawking-1975} with its divergence of the scattering coefficient at small frequency is still not demonstrated in experiments with the classical platform using water waves and open water channels for subcritical flows in presence of an undulation. The latter enhances mode conversion since the measured scattering coefficients in the 2016 Poitiers experiments \cite{euve2016observation} were much higher than theoretically expected without the undulation. \\

\item {\it Transcritical flows and linear regimes}: The 2020 Poitiers experiments \cite{euve2020scattering} moved on to an accelerating transcritical and non-dispersive regime of a black hole flow type. Linear negative mode conversion (by conformal coupling and not Hawking process) was observed starting with stimulating waves plunging into the waterfall (their amplitudes were between $0.25-0.5-0.75-1mm$) and the hydraulic metric was validated for the first time with a classical platform by a measurement of the non-dispersive dispersion relations in both the subcritical and supercritical regions. The 2022 Poitiers experiments \cite{fourdrinoy2022correlations} used the wave turbulence on a decelerating transcritical flow of a white hole type with its dispersive undular jump located on the descending slope of the obstacle and in presence of a downstream gate that reflected the co-current noise due to the free-surface turbulence and that stimulated a non-dispersive white hole horizon contrary to all previous experiments with dispersive group velocity horizon \cite{rousseaux2008observation, rousseaux2010horizon, weinfurtner2011measurement, weinfurtner2013classical, euve2016observation} or sometimes even without the latter \cite{euve2015wave}. Hawking correlations were observed \cite{fourdrinoy2022correlations} on a region close to the obstacle unlike in the 2016 Poitiers experiments where it was observed in the asymptotic region far from the obstacle \cite{euve2016observation}.
\end{itemize}

\subsection{An \emph{a posteriori} hydrodynamics classification of the historical experiments}

Now, as we have anticipated in the main part of the article, it is possible to classify the historical experiments thanks to the theory introduced in this work that combines for the first time both hydraulics and dispersive aspects, a hydro-dispersive theory, albeit without being able to predict the existence zone of the subcritical undulation that we know to be within the existence zone of the subcritical depression regime as a peculiar case and close to the transcritical boundary à la Long \cite{rousseaux2020classical} in the historical -Froude versus obstruction ratio- diagram for a given obstacle (the U(ndulation) regime of \cite{rousseaux2020classical}).

\begin{figure}[!h]
    \centering
    \includegraphics[scale=0.35]{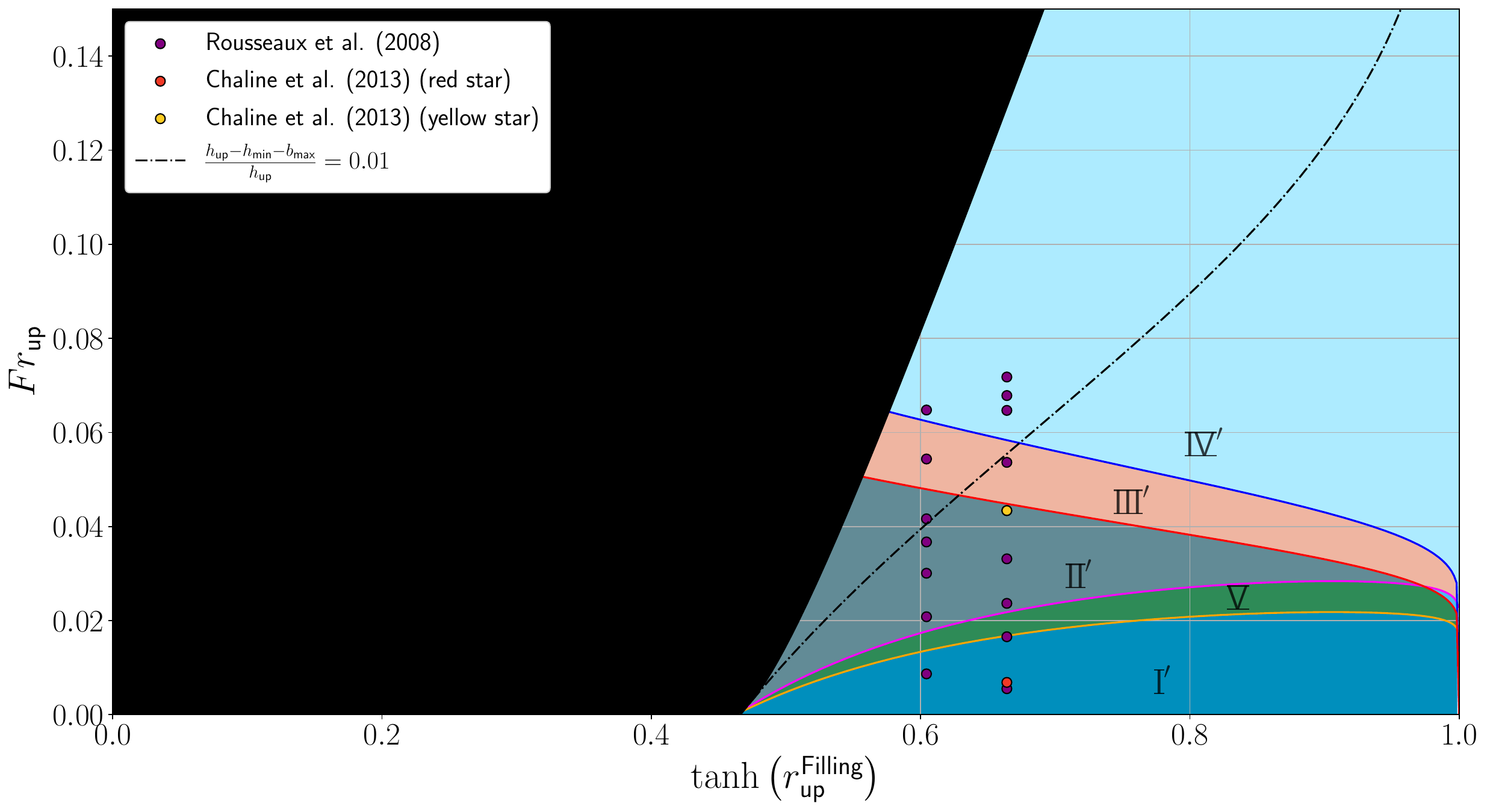}
    \caption{Phase diagram for subcritical regimes present in the literature of Analogue Gravity in hydrodynamics. This diagram was constructed for an obstacle with a maximum height of $b_\text{max}=1.1\:m$ and $h_\text{tech}=2\:m$. For this diagram, the experimental point (i.e. ($\tanh\left(r_\text{up}^\text{Filling}\right),Fr_\text{up}$))) used are: (0.664,0.0055), (0.604,0.0087), (0.664,0.017), (0.664,0.024), (0.604,0.021), (0.664,0.033), (0.604,0.030), (0.664,0.043), (0.604,0.037), (0.604,0.042), (0.604,0.054), (0.604,0.065), (0.664,0.065), (0.664,0.068), (0.664,0.072), (0.664,0.054) in the case of Rousseaux et al.(2008)\cite{rousseaux2008observation} and (0.664,0.0067), (0.664,0.043) in the case of Chaline et al.(2013)\cite{chaline2013some}.}
    \label{Fr_vs_tanhh_histo_(sous)_1.1}
\end{figure}

An illuminating conclusion can be reached already: the experiments in the first years of Analogue Gravity with free surface flows were not done in the canonical regime that was envisaged by the theoretical studies initially \cite{Schuetzhold-Unruh-2002, LivingReview, barcelo2019analogue}, namely, transcritical black hole flows with only one single criterion based on the local Froude number as a function of the position. Somehow, the dynamics was hidden in the theory to the detriment of the wave kinematics analogy. However, as we have seen in this work, the dynamics is important to settle the type of Analogue flows. No theoretical guidelines were available when the first experiments in Analogue Gravity were performed using water waves and open channel flows except that the flow has to be transcritical somewhere to get the analogue of a black hole event horizon in General Relativity. Hence, it is natural that the first experiments performed with subcritical flows regimes of a  white fountain type instead of a transcritical flow regime of a black hole type met strong skepticism, if not critics \cite{rousseaux2008observation, rousseaux2010horizon, weinfurtner2011measurement, weinfurtner2013classical}. As we have seen, technological constraints have been the essential guidelines for the experimentalists: the size of the open channel flow available ($30m$ in Nice, $7m$ in Vancouver and Poitiers and now $3m$); the range of the flow rates available for each channel and the maximal water depths; the metrological measurements method (resistive sensor and a side camera looking at farthest side meniscus, acoustic sensor(s) side camera(s) looking at the fluorescent signal of a LASER sheet impacting a free surface, acoustic sensor(s) and side camera(s) looking a the closest side meniscus) ; the size of the region of interest (ROI) of the cameras (2-3m typically).
The latter size ($30m$ then $7m$) was at the origin of the focus on white fountain type flow instead of black hole flow in the seminal years: indeed, it was easier to send long waves and to measure short converted modes contrary to Hawking's original prediction where short quantum noise redshifts when escaping towards the asymptotic observer being amplified by the trapping of negative norm modes partners. Then, partial dispersive theories \cite{nardin2009wave, rousseaux2010horizon, Rousseaux-BASICS-2013, coutant2014undulations, Michel-Parentani-2014, robertson2016scattering, Michel-et-al-2018, Coutant-Weinfurtner-2018} were developed that now culminate with the hydro-dispersive theory of the present work.

\begin{figure}[!h]
    \centering
    \includegraphics[scale=0.35]{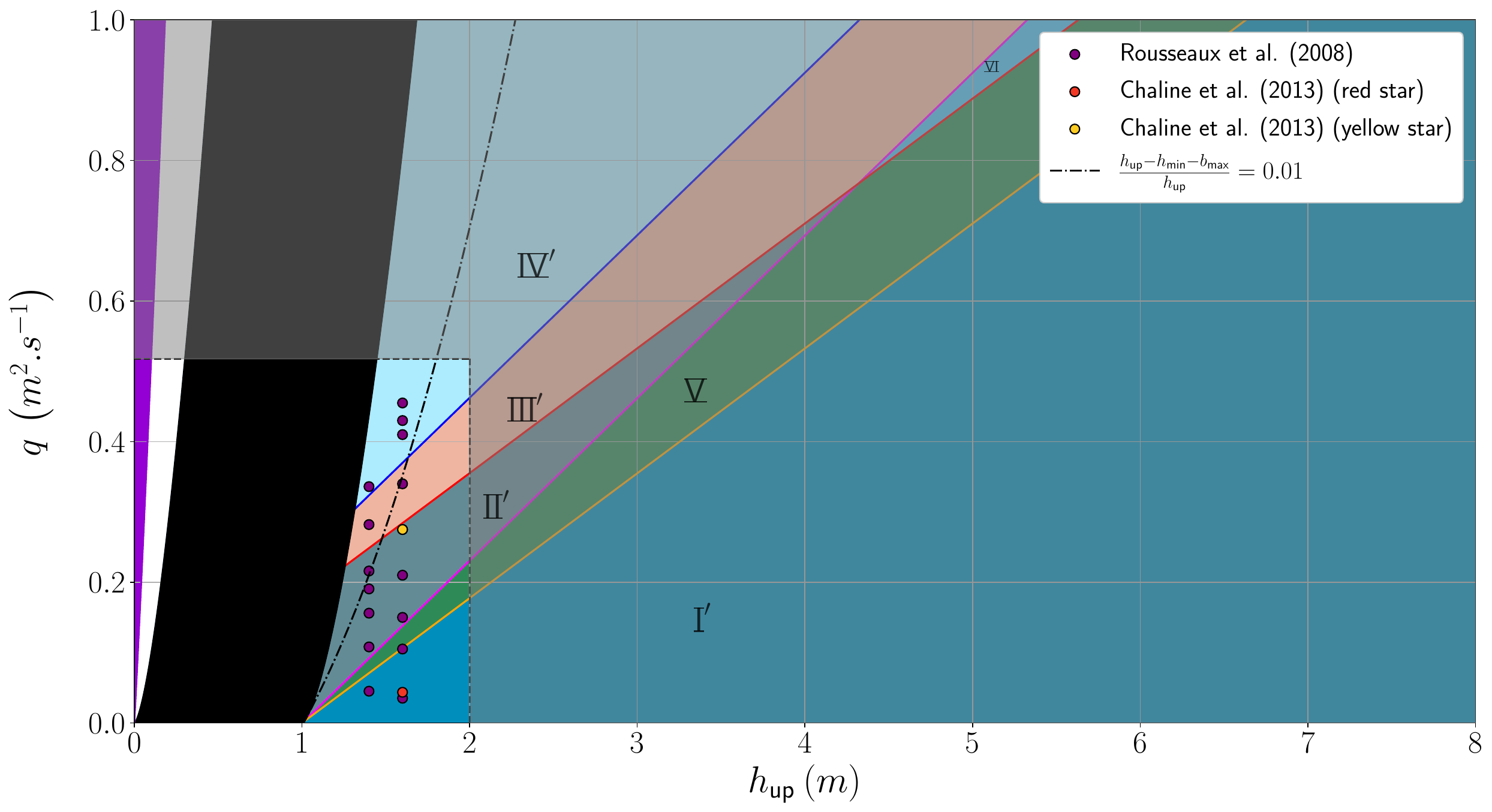}
    \caption{Phase diagram for subcritical regimes present in the literature of Analogue Gravity in hydrodynamics. This diagram was constructed for an obstacle with a maximum height of $b_\text{max}=1.1\:m$ and $h_\text{tech}=2\:m$. For this diagram, the experimental point (i.e. ($h_\text{up},q$)) used are: ($1.6\:  m\:\text{,}\:0.035\: m^2.s^{-1}$), ($1.4\: m\:\text{,}\:0.045\: m^2.s^{-1}$), ($1.6\: m\:\text{,}\:0.105\: m^2.s^{-1}$), ($1.6\: m\:\text{,}\:0.15\: m^2.s^{-1}$), ($1.4\: m\:\text{,}\:0.108\: m^2.s^{-1}$), ($1.6\: m\:\text{,}\:0.21\: m^2.s^{-1}$), ($1.4\: m\:\text{,}\:0.156\: m^2.s^{-1}$), ($1.6\: m\:\text{,}\:0.275\: m^2.s^{-1}$), ($1.4\: m\:\text{,}\:0.1904\: m^2.s^{-1}$), ($1.4\: m\:\text{,}\:0.216\: m^2.s^{-1}$), ($1.4\: m\:\text{,}\:0.282\: m^2.s^{-1}$), ($1.4\: m\:\text{,}\:0.336\: m^2.s^{-1}$), ($1.6\: m\:\text{,}\:0.41\: m^2.s^{-1}$), ($1.6\: m\:\text{,}\:0.43\: m^2.s^{-1}$), ($1.6\: m\:\text{,}\:0.455\: m^2.s^{-1}$), ($1.6\: m\:\text{,}\:0.34\: m^2.s^{-1}$) in the case of Rousseaux et al.(2008)\cite{rousseaux2008observation} and ($1.6\: m\:\text{,}\:0.0435\: m^2.s^{-1}$), ($1.6\: m\:\text{,}\:0.275\: m^2.s^{-1}$) in the case of Chaline et al.(2013)\cite{chaline2013some}. The grey zone corresponds to inaccessible flow parameters of the experimental setup.}
    \label{q_vs_h_histo_(sous)_1.1}
\end{figure}

In this appendix, the dimensionless plane $(\tanh\left(r_\text{up}^\text{Ob}\right), Fr_\text{up})$ is plotted first for each historical experiments in Analogue Gravity in interfacial hydrodynamics and then is supplemented by another plane with axes keeping their dimensions into the plane $\left(h_\text{up},q\right)$ that is as a function of two of the three control parameters imposed by an experimentalist, the third control parameter $b_{max}$ being fixed otherwise. 

\begin{figure}[!h]
    \centering
    \includegraphics[scale=0.35]{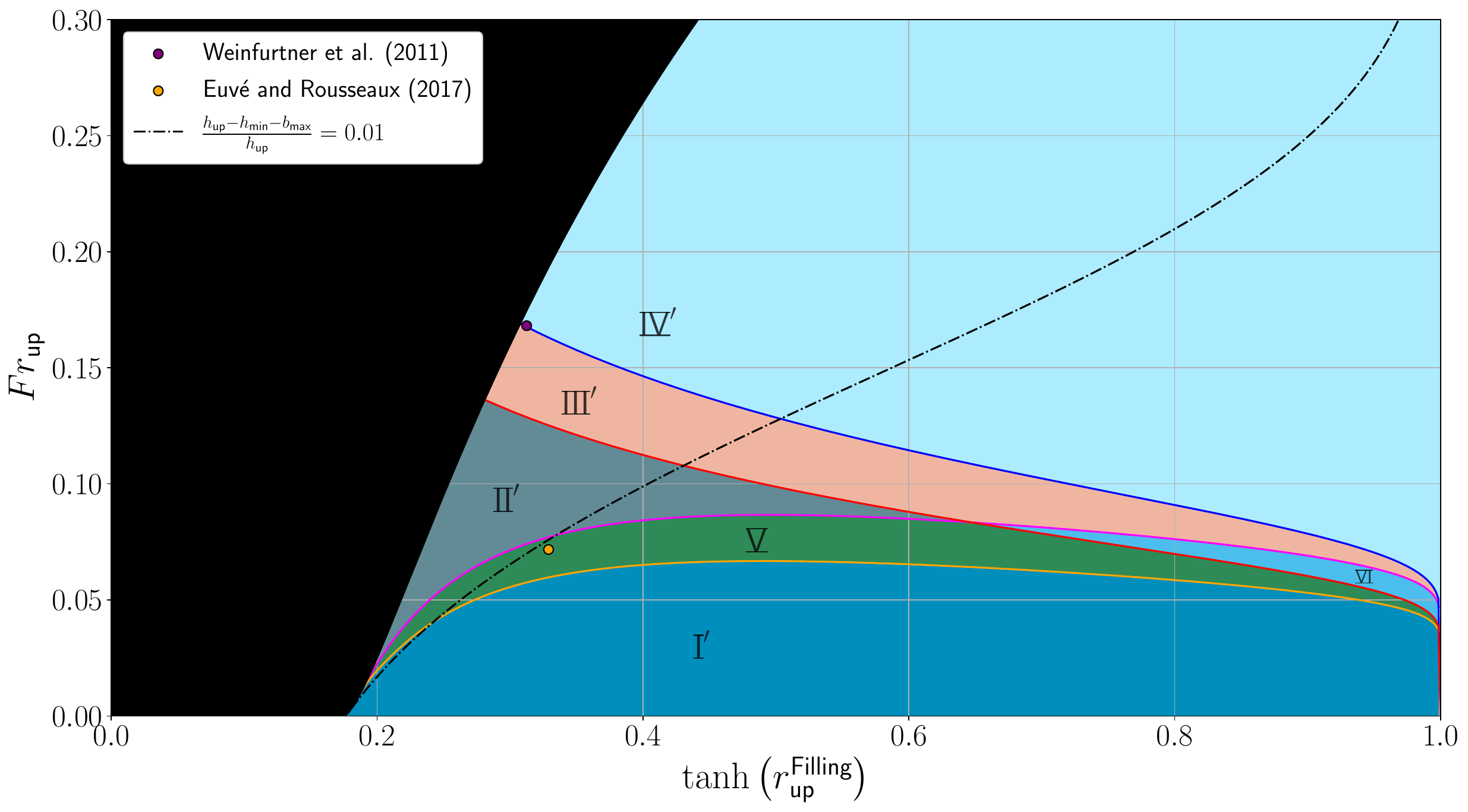}
    \caption{Phase diagram for subcritical regimes. This diagram was constructed for an obstacle with a maximum height of $b_\text{max}=0.106\:m$ and $h_\text{tech}=0.6\:m$. For this diagram with the same Vancouver 2011 geometry, the experimental points (i.e. ($\tanh\left(r_\text{up}^\text{Filling}\right),Fr_\text{up}$)) used are: (0.313,0.168) in the case of Weinfurtner et al. (2011) \cite{weinfurtner2011measurement} and (0.33,0.07) in the case of Euvé \& Rousseaux (2017) \cite{Euve-Rousseaux-2017}.}
    \label{Fr vs tanhh histo (sous) 10.6}
\end{figure}

\begin{figure}[!h]
    \centering
    \includegraphics[scale=0.35]{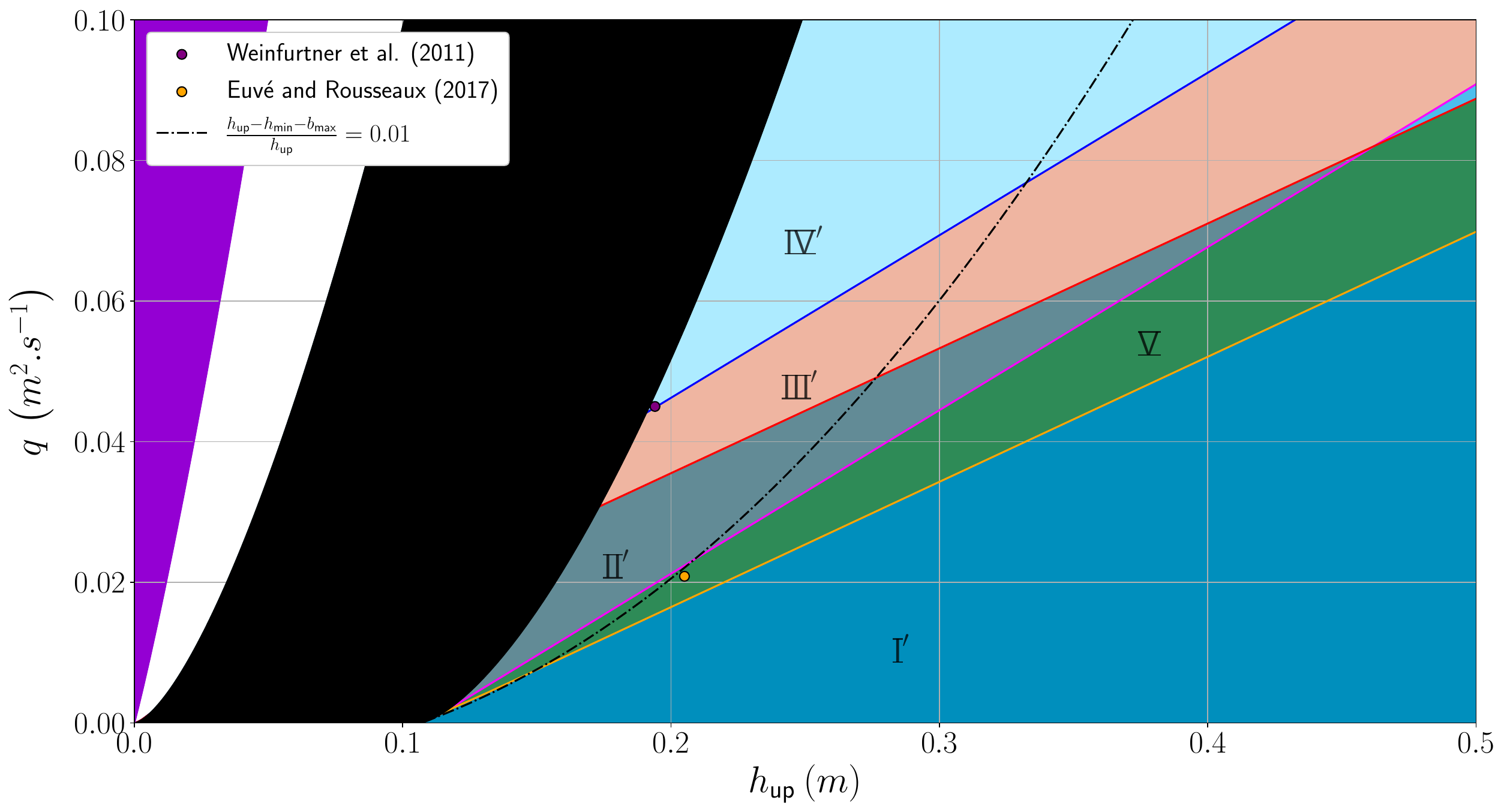}
    \caption{Phase diagram for subcritical regimes. This diagram was constructed for an obstacle with a maximum height of $b_\text{max}=0.106\:m$. For this diagram with the same Vancouver 2011 geometry, the experimental points (i.e. ($h_\text{up},q$)) used are: ($0.194\: m\:\text{,}\:0.045\: m^2.s^{-1}$) in the case of Weinfurtner et al. (2011) \cite{weinfurtner2011measurement} and $0.205\: m\:\text{,}\:0.0208\: m^2.s^{-1}$) in the case of Euvé \& Rousseaux (2017) \cite{Euve-Rousseaux-2017}. The maximum flow rate of the pump, for the experiment by Euvé and Rousseaux (2017)\cite{Euve-Rousseaux-2017}, is $q_\text{max}=0.1538\:m^2.s^{-1}$.}
    \label{q vs h (sous) 10.6}
\end{figure}

In the ACRI 2008 Nice experiments \cite{rousseaux2008observation, rousseaux2010horizon}, we understand now thanks to the dimensionless Figure  \ref{Fr_vs_tanhh_histo_(sous)_1.1} and its associated Figure \ref{q_vs_h_histo_(sous)_1.1} with dimensions for the experimentalists that the theoretical domains $\text{\Romanbar{1}}^\prime$, \Romanbar{5}, $\text{\Romanbar{2}}^\prime$, $\text{\Romanbar{3}}^\prime$ and $\text{\Romanbar{4}}^\prime$ where probed experimentally without the authors' knowledge at that time. The regime \Romanbar{6} could not be explored (the grey zone in the Figure \ref{q_vs_h_histo_(sous)_1.1} for values of the parameters unreachable in the experimental setup): it is fortunate since there is no negative norm mode conversion possible in this regime so no Hawking radiation. The authors of \cite{rousseaux2008observation, rousseaux2010horizon} reported no wave blocking for a range of frequencies and small speeds (black circles in the phase diagram of the Figure 7 of \cite{rousseaux2008observation}): this is consistent with the points in the zone $\text{\Romanbar{1}}^\prime$ of our new diagram were wave blocking is impossible whatever the incoming frequency for all speeds in the very zone. Moreover, the authors of \cite{rousseaux2008observation, rousseaux2010horizon} clearly showed the existence the existence of a threshold with intermediate speeds in their phase diagram (the green line in the Figure 7 of \cite{rousseaux2008observation}) for the existence of what had been interpreted as negative norm modes: here again, the magenta line between zone \Romanbar{5} and $\text{\Romanbar{2}}^\prime$ in the Figures \ref{Fr_vs_tanhh_histo_(sous)_1.1} and \ref{q_vs_h_histo_(sous)_1.1} is compatible qualitatively with this observation. For higher speeds, the authors of \cite{rousseaux2008observation, rousseaux2010horizon} reported the existence of a dispersive white hole horizon (the merging place of the incoming and blue-shifted modes \cite{nardin2009wave}) since its position changed with the incoming periods probed (the purple line in the Figure 7 of \cite{rousseaux2008observation}): once again, this last observation is consistent with the new diagrams of the Figures \ref{Fr_vs_tanhh_histo_(sous)_1.1} and \ref{q_vs_h_histo_(sous)_1.1} since all the regimes above the magenta line corresponds to possible wave blocking for short periods. The authors of \cite{rousseaux2008observation, rousseaux2010horizon} reported as well indications of negative norm modes existence, albeit with the wrong frequencies indicating non-linear effects possibly due to harmonics generation or partial wave breaking as clearly stated in their conclusion. These modes were observed not only within the existence zone of the white hole horizon namely with wave blocking (above the purple line of Figure 7 of \cite{rousseaux2008observation}) but also, to the authors surprise, in a non-blocking zone and above a speed threshold (in between the purple and green line of the Figure 7 of \cite{rousseaux2008observation}): this is again fully compatible with our new theoretical prediction since the zones $\Romanbar{2}^\prime$, $\Romanbar{3}^\prime$ and $\text{\Romanbar{4}}^\prime$ do feature negative norm modes. For the case $\Romanbar{2}^\prime$, negative norm modes can be produced but they do not reach the asymptotic observer as the blue-shifted modes that become Hawking radiation modes if they are amplified by the very appearance of the negative norm modes. Hawking radiation would be then localized here in a subcritical zone over the obstacle because it would be blocked at the blue horizon if no harmonics generation is produced in the mean time. As the camera used by the authors was placed in the descending slope of the obstacle, they could not watch an asymptotic propagation. For the case $\text{\Romanbar{3}}^\prime$, only the negative norm modes would be localized on the top of the obstacle since they would be blocked at the negative horizon described in \cite{rousseaux2010horizon, Rousseaux-BASICS-2013, rousseaux2020classical} contrary to the blue-shifted modes who could reach the asymptotic observer. For the case $\text{\Romanbar{4}}^\prime$, both blue-shifted and negative norm modes would espace at infinity. The authors of \cite{rousseaux2008observation} could not distinguish between cases $\text{\Romanbar{2}}^\prime$, $\text{\Romanbar{3}}^\prime$ and $\text{\Romanbar{4}}^\prime$ at high speeds since the obstacle was very long and the camera was placed such as to observe part of the flat part and the downstream slope of the ACRI 2008 obstacle so as to check the existence of wave transmission or not. They did not report space-time diagrams outside of the descending slope which was their main region of interest. So the authors of \cite{rousseaux2008observation} could not report that Hawking radiation if it had taken place would have reach the asymptotic observer. It's a spurious effect of all big size experiments \cite{rousseaux2008observation, rousseaux2010horizon, chaline2013some}, because they have a finite length ($2-3m$) of the observation windows located in the descending slope that does not extend up to the asymptotics region because of the size of the obstacle (decametric in \cite{rousseaux2008observation, rousseaux2010horizon, chaline2013some}).

In the ACRI2010 Nice experiments \cite{chaline2013some}, only 4 experimental flow regimes were reported (using the terminology blue, green, red and yellow stars in the terminology of \cite{chaline2013some}) to test specific questions that were raised by the ACRI2008 experiments \cite{rousseaux2008observation} in the light of the pure dispersive theories (without hydraulics as here) developed in between \cite{nardin2009wave, rousseaux2010horizon}. As discussed in the main part of the current paper, the geometry was changed from ACRI 2008 to ACRI 2010 because the duration of the experiments was longer and it allowed to have a longer descending slope with a length to height ratio of the obstacle similar to the present work. Moreover, flow recirculations are known to be suppressed empirically for a slope angle below $7^\circ$. The new experiments were characterized for half of them (blue and green stars) with a period of the incoming stimulating waves such that it was inferior to the so-called cusp period $T_c=0.425s$ in water \cite{rousseaux2010horizon} where the white fountain horizon disappear by being merged with the blue horizon controlled by surface tension effect: we will not discuss them here since they were done just to check that no wave blocking is possible for periods inferior to the cusp one, a striking dispersive effect in wave-current interaction. For the other half (red and yellow stars discussed here), the stimulated period was superior to the cusp one \cite{rousseaux2010horizon} on the contrary: a white fountain horizon could be observed for the short periods of the wave-maker. All the cases for the 2008 ACRI Nice experiments were such that the period was superior to the cusp one (between $0.6s$ and $2.5s$), a fact unknown to the authors at that time. For the red star regime of \cite{chaline2013some} in 2010, the incoming train of sinusoidal waves were reported not to be blocked which is now confirmed since it corresponds to the regime $\text{\Romanbar{1}}^\prime$ of the present hydro-dispersive theory where waves cannot be blocked whatever their period since the speed range of the red star regime ($0.074-0.087m/s$) was below the cusp speed $U_c=0.178m/s$, the minimum of the group velocity for gravity waves in deep water, a fact known to the authors because of the dispersive theory introduced in between both ACRI 2008 and 2010 experiments \cite{chaline2013some}. For the yellow star regime of \cite{chaline2013some}, the incoming waves were reported to be blocked for the same chosen period ($T=1s$): it corresponds to the regimes $\text{\Romanbar{2}}^\prime$ of this work where waves can be blocked or not depending on the wave period and mode conversion towards both blue-shifted mode à la Hawking and negative norm mode may occur. The new information is that if Hawking radiation would have been produced in the ACRI 2010 experiment in the yellow star regime, it could not have reached the asymptotic observer (as we will see shortly for the Vancouver 2011 experiments \cite{weinfurtner2011measurement, weinfurtner2013classical} as well). Indeed, the blue shifted modes would have been blocked at the blue horizon downstream the obstacle. Here, an important remark is worthy of attention: because of the size of the experimental setup (decametric), visual inspection can lead wrongly the observer to conclude that the converted waves at the horizon, essentially the blue-shifted modes with opposite group and phase velocity extend up to infinity which is wrong because the flow velocity would drop below the threshold for the appearance of the group velocity blue horizon and then linear Hawking radiation would have been blocked and confined in a region above the bottom obstacle similarly to our recent observation of blocked Hawking radiation \cite{fourdrinoy2022correlations} in the vicinity of the obstacle by the blue horizon this time for a transcritical flow in the regime $\text{T}^\nearrow_{\text{\Romanbar{2}}}-\text{T}^{\searrow,\text{U}}_{\text{\Romanbar{2}}}$. In the yellow star regime of \cite{chaline2013some}, only wave blocking at the white horizon in the speed range $0.45-0.55m/s$ was reported with a pixel accuracy detection method based on the contrast of the side meniscus on the opposite wall of the channel as observed by a lateral camera. This observation is again fully compatible with the regime $\text{\Romanbar{2}}^\prime$ of the present theory as one can see on the Figures \ref{Fr_vs_tanhh_histo_(sous)_1.1} and \ref{q_vs_h_histo_(sous)_1.1}. As a final remark, in both ACRI 2008 and ACRI 2010, experiments \cite{rousseaux2008observation, rousseaux2010horizon, chaline2013some}, the free surface was either Flat or with a Depression and its vertical extension was between $1$ to $2cm$. The position of the theoretical frontier (black dotted curve) between the Flat region and the Depression region in the Figures \ref{Fr_vs_tanhh_histo_(sous)_1.1} and \ref{q_vs_h_histo_(sous)_1.1} fully confirmed this simple observation. This depression observed in the experiments \cite{rousseaux2008observation, rousseaux2010horizon, chaline2013some} can be predicted theoretically, a posteriori, by implementing the equation \ref{sous formule} in a computer code and assuming, for example, $h_\text{sub}=1.6\: m$ and a flow rate $q\in[0.035;0.282]\:m^2.s^{-1}$ to measure $h_\text{sub}-h_\text{min}-b_\text{max}$. An order of magnitude for the height of the depression can be calculated using the condition $(h_\text{up}-h_\text{min}-b_\text{max})/h_\text{up}=0.01$. In this case, $(h_\text{min}+b_\text{max})/h_\text{up}=0.99$; so if $h_\text{up}=1.6\:m$ then the depression at the limit of the flat case is $h_\text{up}-h_\text{min}-b_\text{max}=1.6\:cm$, very close to the experimental measured ones.

\begin{figure}[!h]
    \centering
    \includegraphics[scale=0.35]{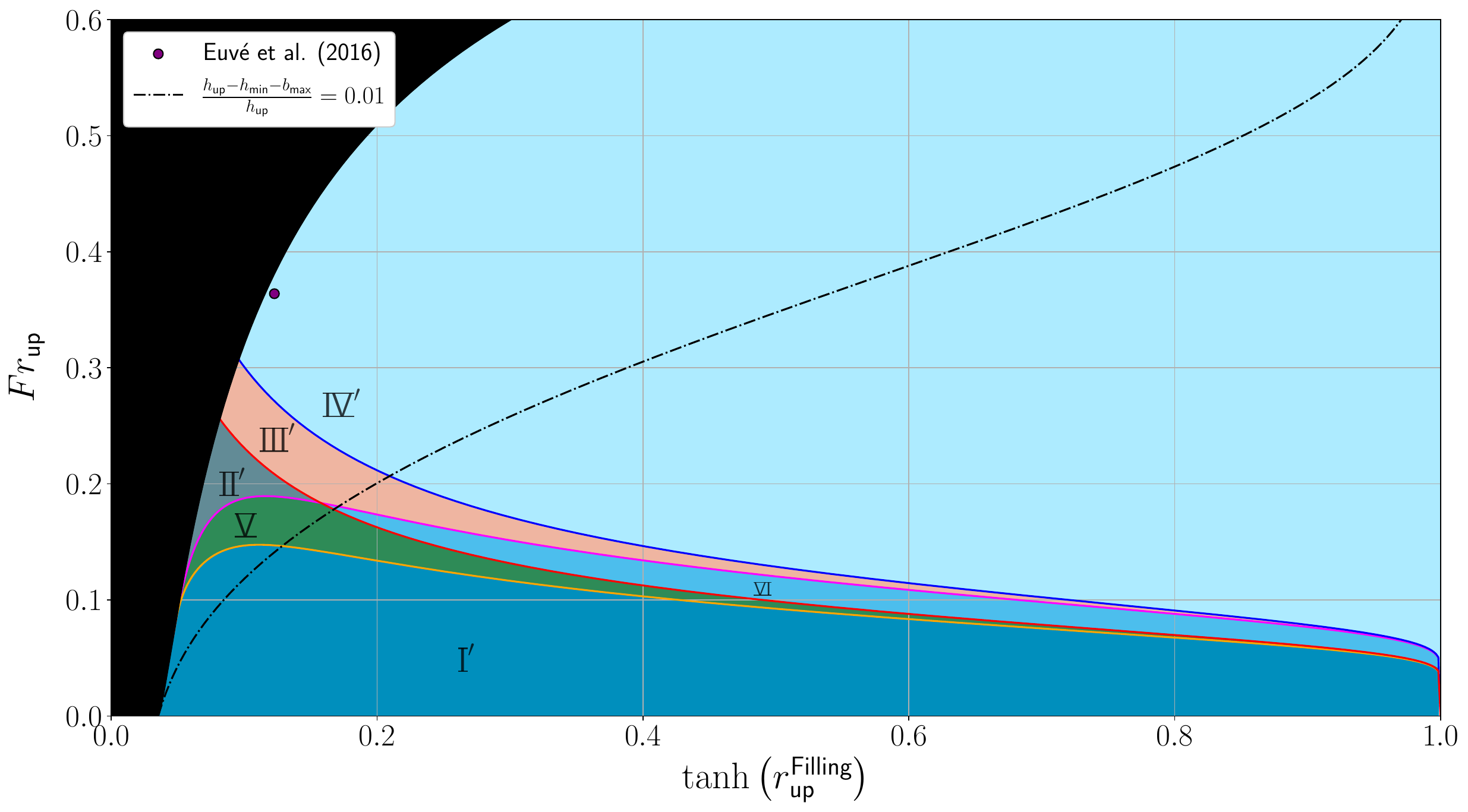}
    \caption{Phase diagram for subcritical regimes. This diagram was constructed for an obstacle with a maximum height of $b_\text{max}=0.022\:m$ and $h_\text{tech}=0.6\:m$.  For this diagram, the experimental points (i.e. ($\tanh\left(r_\text{up}^\text{Filling}\right),Fr_\text{up}$)) used is: (0.1227,0.364) in the case of Euvé et al. (2016) \cite{euve2016observation}.}
    \label{Fr_vs_tanhh_histo_(sous)_2.2}
\end{figure}

In the Vancouver 2011 experiments \cite{weinfurtner2011measurement, weinfurtner2013classical}, stimulated modes conversions were reported for blue-shifted and negative norm modes albeit in the non-linear regime of harmonics generation \cite{euve2021non} hence not in the asymptotics region that would be the linear prediction inspired from Hawking's work \cite{Hawking-1975}. Here, again, the present theory shows that the Vancouver regime was precisely at the frontier between the $\text{\Romanbar{3}}^\prime$ and $\text{\Romanbar{4}}^\prime$ regions as one can see on both Figures \ref{Fr vs tanhh histo (sous) 10.6} and \ref{q vs h (sous) 10.6}. Moreover, it was also closed to the Long transcritical frontier \ref{Long}, at the limit of wave breaking of the undulation itself or of the sum of the undulation and incoming modes. Hence, despite the fact that the stimulated amplitudes of the Vancouver experiments \cite{weinfurtner2011measurement, weinfurtner2013classical} would have been lowered as in its Poitiers reproduction \cite{euve2021non}, the linear converted modes would have not reached the asymptotic observer because of the Landau speed threshold \cite{rousseaux2010horizon, Rousseaux-BASICS-2013, rousseaux2020classical}. It is the origin of the change of variable used by the Vancouver team \cite{weinfurtner2011measurement, weinfurtner2013classical} to take into account the flow speed changes on the descending slope of the bottom obstacle . Indeed, they performed their measurements in a non-asymptotic region filtering their results at the frequency of the wave-maker \cite{weinfurtner2011measurement, weinfurtner2013classical} hence missing the harmonics generation because of too high stimulated amplitudes as discussed in \cite{euve2021non}. Transposed in General Relativity, it is as if the Hawking temperature seen by the asymptotic observer would have been measured close to the black hole horizon with a space dependency of the temperature value as in the Tolman's law where the local temperature varies with the position \cite{santiago2019tolman}.

\begin{figure}[!h]
    \centering
    \includegraphics[scale=0.35]{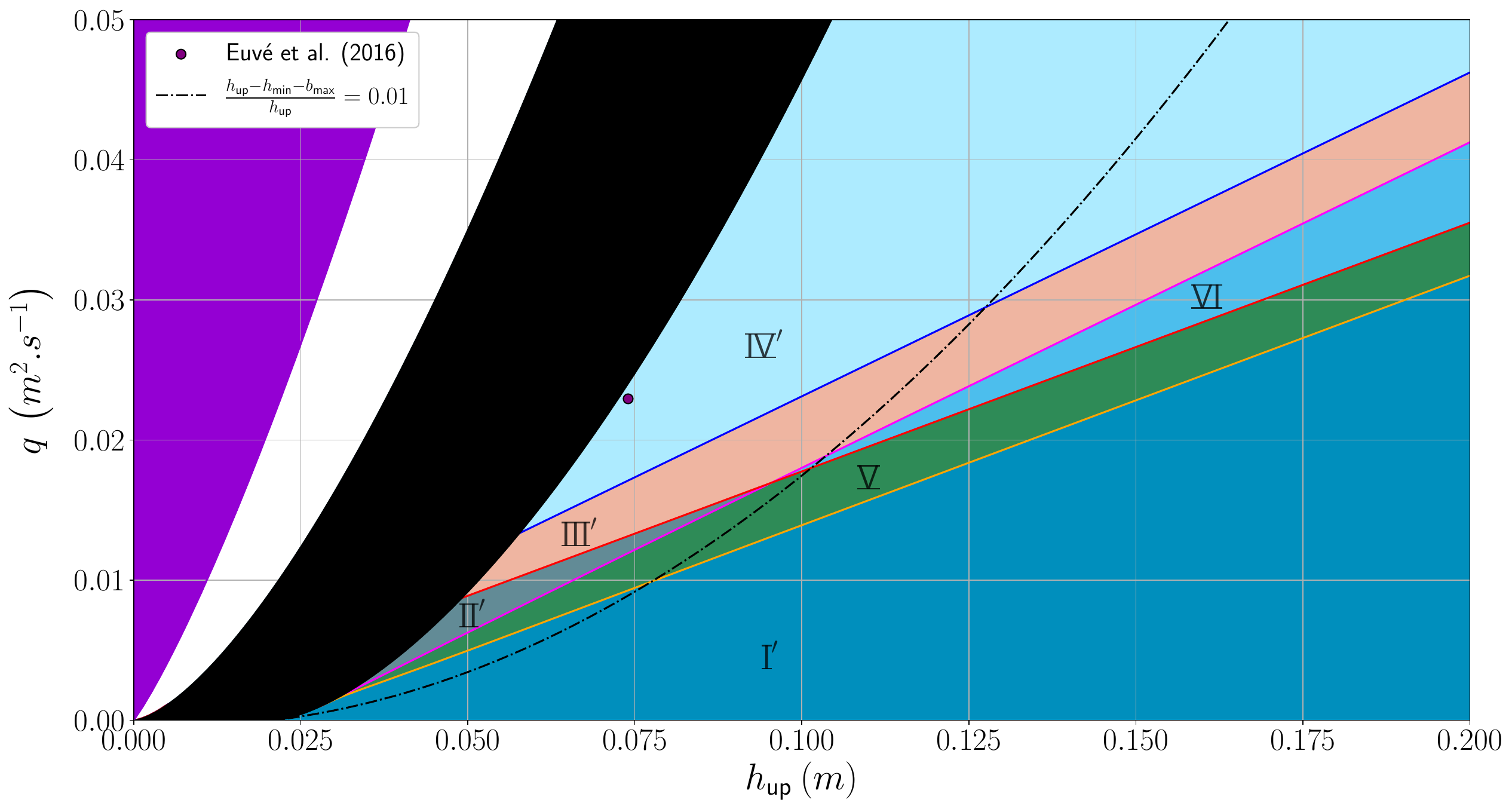}
    \caption{Phase diagram for subcritical regimes. This diagram was constructed for an obstacle with a maximum height of $b_\text{max}=0.022\:m$. For this diagram, the experimental point (i.e. ($h_\text{up},q$)) used is: ($0.074\: m\:\text{,}\:0.023\: m^2.s^{-1}$) in the case of Euvé et al. (2016) \cite{euve2016observation}. The maximum flow rate of the pump, for the experiment by Euvé et al. (2016)\cite{euve2016observation}, is $q_\text{max}=0.1538\:m^2.s^{-1}$.}
    \label{q_vs_h_histo_(sous)_2.2}
\end{figure}

In the Poitiers 2016 experiments \cite{euve2016observation}, a new obstacle avoiding flow recirculations and allowing an increase of the upstream Froude number was used in addition to smaller amplitudes and with an observation in the asymptotic region. The present theory confirmed the choice of the flow regime and parameters which correspond to the zone $\text{\Romanbar{4}}^\prime$ of both Figures \ref{Fr_vs_tanhh_histo_(sous)_2.2} and \ref{q_vs_h_histo_(sous)_2.2}. It is interesting to notice that the Poitiers 2016 flow regime \cite{euve2016observation} (like the Vancouver 2011 regime \cite{weinfurtner2011measurement, weinfurtner2013classical}) is very close to the transcritical boundary given by Long's law \ref{Long} where the similarity with Hawking's predictions is the better as discussed in the theoretical works \cite{Michel-Parentani-2014, robertson2016scattering, Michel-et-al-2018}. Hence, to choose a regime in the vicinity of the limit of appearance of wave breaking on the undulation seems to be an important experimental reference in order to observe linear Hawking radiation of a white fountain for subcritical flows with subluminal dispersion as in \cite{euve2016observation}.

\begin{figure}[!h]
    \centering
    \includegraphics[scale=0.35]{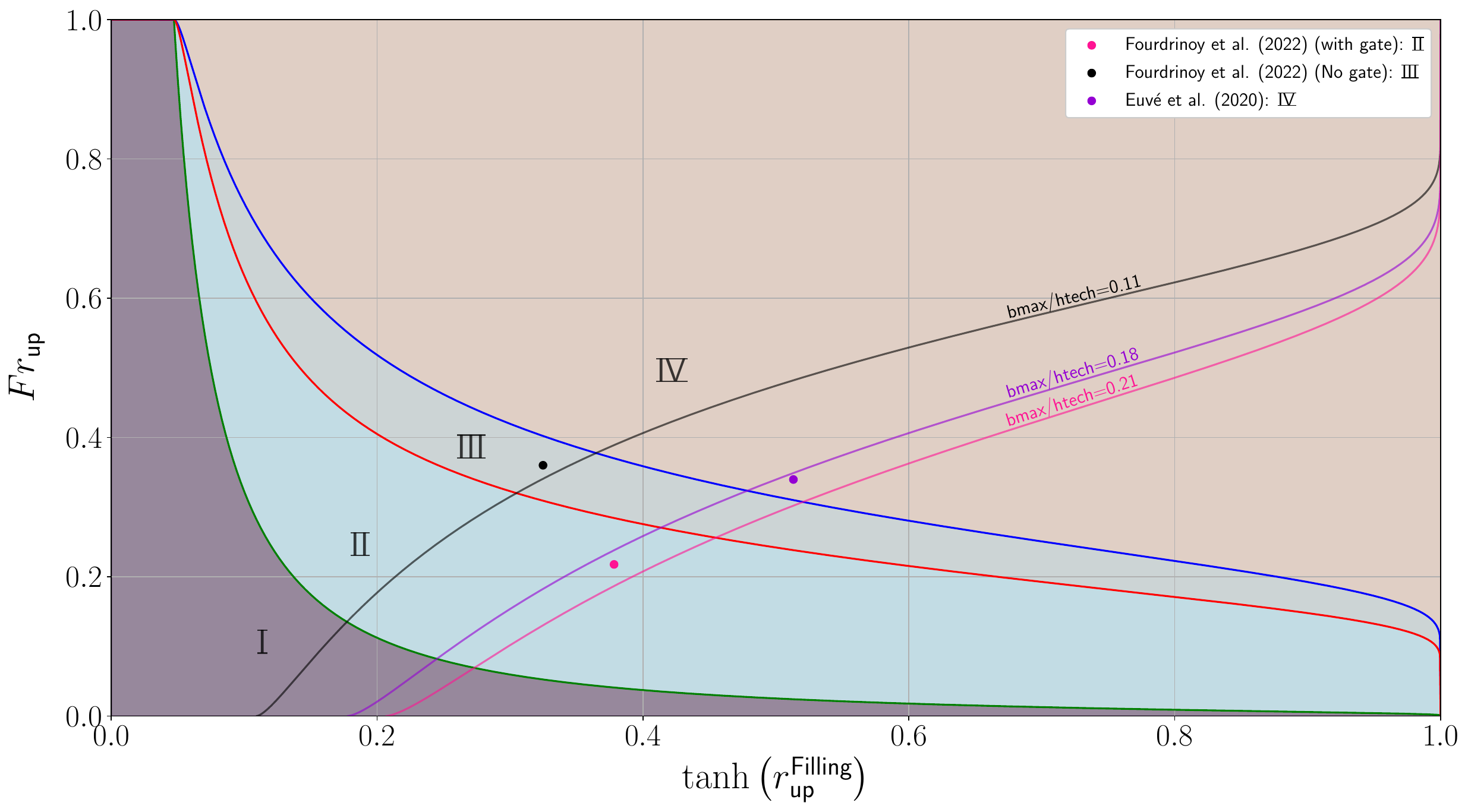}
    \caption{Phase diagram for a transcritical regime, taking dispersive effects into account. The three historical experiments with transcritical horizons \cite{euve2020scattering, fourdrinoy2022correlations} were added in the form of each Long's law for the given bottom obstacle maximum height with the corresponding flow regimes as a point for each case. For this diagram, the experimental points (i.e. ($\tanh\left(r_\text{up}^\text{Filling}\right),Fr_\text{up}$)) used are: (0.513,0.34) in the case of Euvé et al. (2020)\cite{euve2020scattering}, (0.325,0.36) in the case of Fourdrinoy et al. (2022) (without an exit gate) \cite{fourdrinoy2022correlations} and (0.378,0.218) in the case of Fourdrinoy et al. (2022) (with an exit gate) \cite{fourdrinoy2022correlations}. }
    \label{Fr vs tanhh histo}
\end{figure}

In the 2017 reverse-interstellar travel experiments \cite{Euve-Rousseaux-2017}, the flow parameters were chosen so as to avoid negative norm mode conversion on the one hand and on the other hand to generate the so-called double bouncing scenario \cite{rousseaux2010horizon, Peloquin} where incoming gravity waves are blocked at the white hole horizon then converted towards blue-shifted mode only (that are not amplified since no negative norm modes are produced because the maximum speed in the channel is inferior to the Landau speed) that are also blocked at the blue horizon. Then, the blue-shifted are converted into capillary waves that enter the white hole horizon contrary to the gravity waves and the former go through the wormhole in between both horizons before escaping from the black horizon, a forbidden scenario which becomes possible in fluid mechanics since capillary waves are superluminal \cite{Rousseaux-BASICS-2013}. Keeping in mind the enormous damping due to the viscosity \cite{robertson2017viscous}, the authors were forced to use a metric obstacle and a decimetric cavity by tuning the flow parameters correctly by errors and trials \cite{Euve-Rousseaux-2017}. Otherwise, if the reverse interstellar travel experiment would have been performed in a longer tank, the cavity size would be order of the meter and the capillary waves would have been killed by viscous damping inside the wormhole as in the experiments of Badulin et al. of 1983 \cite{badulin1983laboratory}. One checks that the reverse interstellar travel is in the zone \Romanbar{5} of both Figures \ref{Fr vs tanhh histo (sous) 10.6} and \ref{q vs h (sous) 10.6} that are the ones used for the Vancouver 2011 experiments since the geometry was exactly the same: no negative norm modes can be created and the free surface is flat as observed in the 2017 experiments.

\begin{figure}[!h]
    \centering
    \includegraphics[scale=0.35]{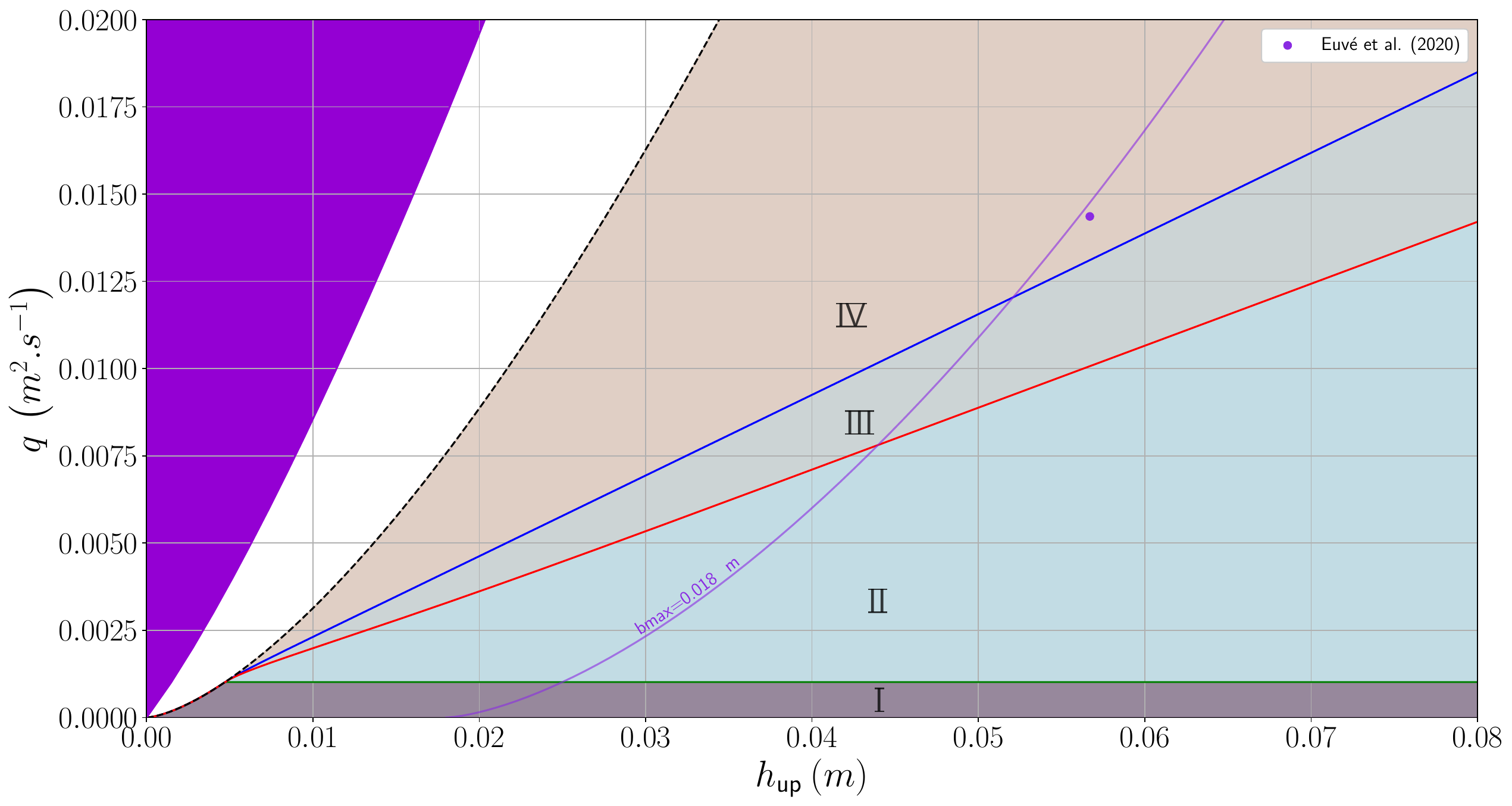}
    \caption{Phase diagram for a transcritical regime, taking dispersive effects into account. For this diagram, the experimental point (i.e. ($h_\text{up},q$)) used is: ($0.0567\: m\:\text{,}\:0.014\: m^2.s^{-1}$) in the case of Euvé et al. (2020) \cite{euve2020scattering}.  The maximum flow rate of the pump, for the experiment by Euvé et al. (2020)\cite{euve2020scattering}, is $q_\text{max}=0.1538\:m^2.s^{-1}$.}
    \label{q vs h histo trans 1.8}
\end{figure}

The same procedure gives the graphics of the dimensionless Figure \ref{Fr vs tanhh histo} for the three historical experiments featuring each time not only a transcritical black hole horizon controlled by their respective bottom obstacle geometries \cite{euve2020scattering} but also a transcritical white fountain horizons (with or without a gate at the exit of the channel), whose position (downstream of the obstacle or attached to it) is controlled by a combination of both turbulent head losses, viscous friction on the boundaries of the channel and bulk viscous dissipation \cite{fourdrinoy2022correlations}, other ways to avoid the assumed instability of white fountains \cite{leonhardt2002intrinsic}. An additional domain has been introduced (the purple domain). This domain corresponds to the dimensional form of the formula \ref{Long_explicit_haut} (the Long's law for supercritical regime): i.e. the regimes which are in the purple zone i.e. above \ref{Long_explicit_haut} are considered to be of a supercritical type. The dotted curve corresponds to the condition $Fr_\text{up}=1$. Finally, the white zone corresponds to the fact that the regime is partially blocked in upstream region according to \cite{Baines}.

We have identified an error in the article \cite{fourdrinoy2022correlations} concerning the flow rate values ($q=4.7\times10^{-3}\:m^2. s^{-1}$ for the regime $\text{T}^\nearrow_{\text{\Romanbar{3}}}-\text{T}^{\searrow,\text{U}}_{\text{\Romanbar{4}}}$ and  $q=3.1\times10^{-3}\:m^2. s^{-1}$ for the regime  $\text{T}^\nearrow_{\text{\Romanbar{2}}}-\text{T}^{\searrow,\text{U}}_{\text{\Romanbar{2}}}$) without any consequence on its conclusions, probably due to a flow meter error, the latter has been changed in the current experiments. Using the data from the experiments described in the article \cite{fourdrinoy2022correlations}, we correct in this work the flow rate values using the measurements of the fluctuations of the free surface fitting the dispersion relation (not reported here), optimising the values of the speed seen by the waves as in \cite{euve2016observation, euve2020scattering}. The new flow rate values are therefore: $q=6.98\times10^{-3}\:m^2.s^{-1}$ for the regime  $\text{T}^\nearrow_{\text{\Romanbar{3}}}-\text{T}^{\searrow,\text{U}}_{\text{\Romanbar{4}}}$ and $q=5. 41\times10^{-3}\:m^2.s^{-1}$ for the regime $\text{T}^\nearrow_{\text{\Romanbar{2}}}-\text{T}^{\searrow,\text{U}}_{\text{\Romanbar{2}}}$.

\begin{figure}[!h]
    \centering
    \includegraphics[scale=0.35]{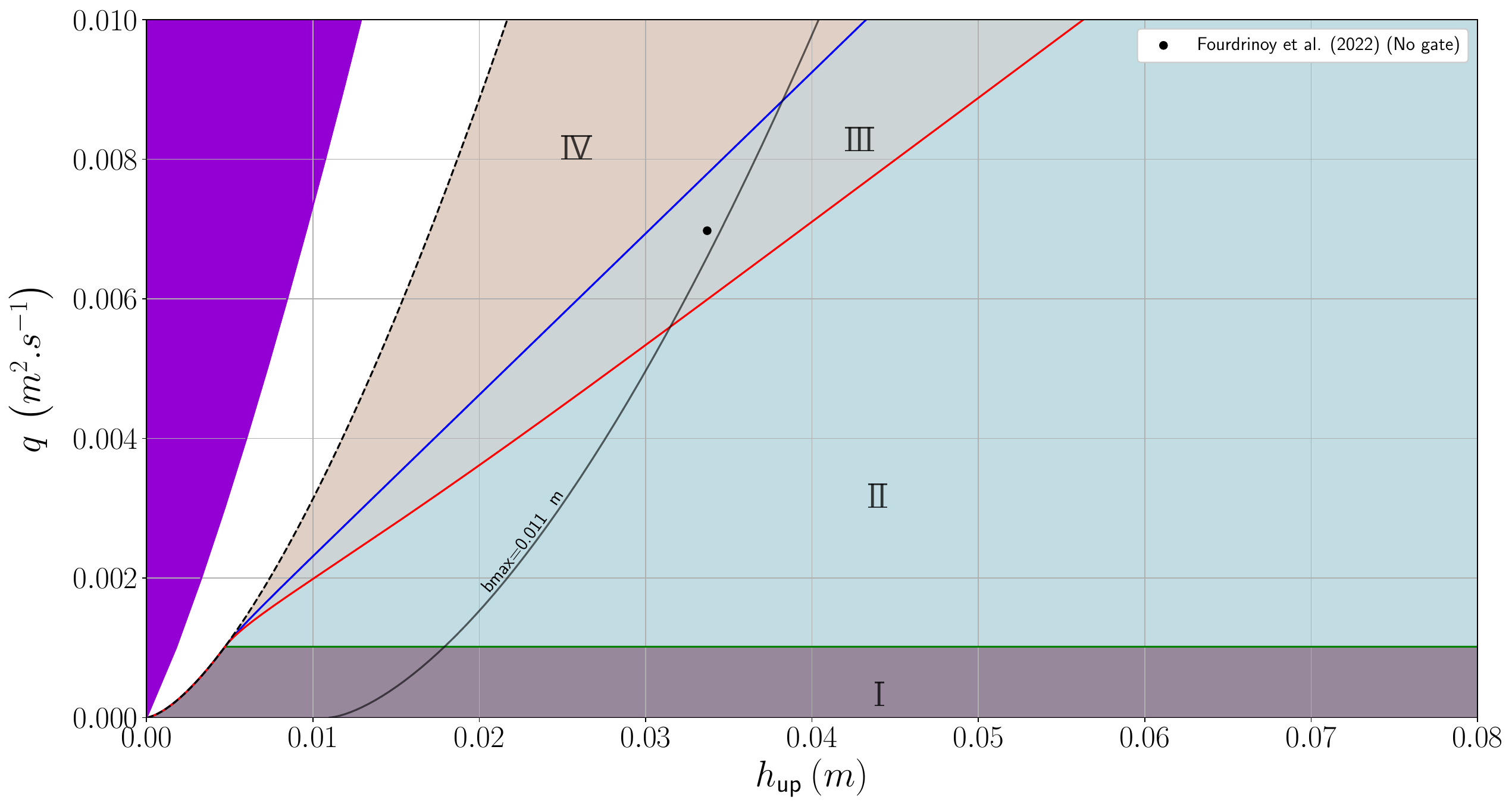}
    \caption{Phase diagram for a transcritical regime, taking dispersive effects into account. For this diagram, the experimental point (i.e. ($h_\text{up},q$)) used is: ($0.0337\: m\:\text{,}\:0.007\: m^2.s^{-1}$) in the case of Fourdrinoy et al. (2022) (without an exit gate) \cite{fourdrinoy2022correlations}.  The maximum flow rate of the pump, for the experiment by Fourdrinoy et al. (2022)\cite{fourdrinoy2022correlations}, is $q_\text{max}=0.0119\:m^2.s^{-1}$.}
    \label{q vs h histo trans 1.1}
\end{figure}

\begin{figure}[!h]
    \centering
    \includegraphics[scale=0.35]{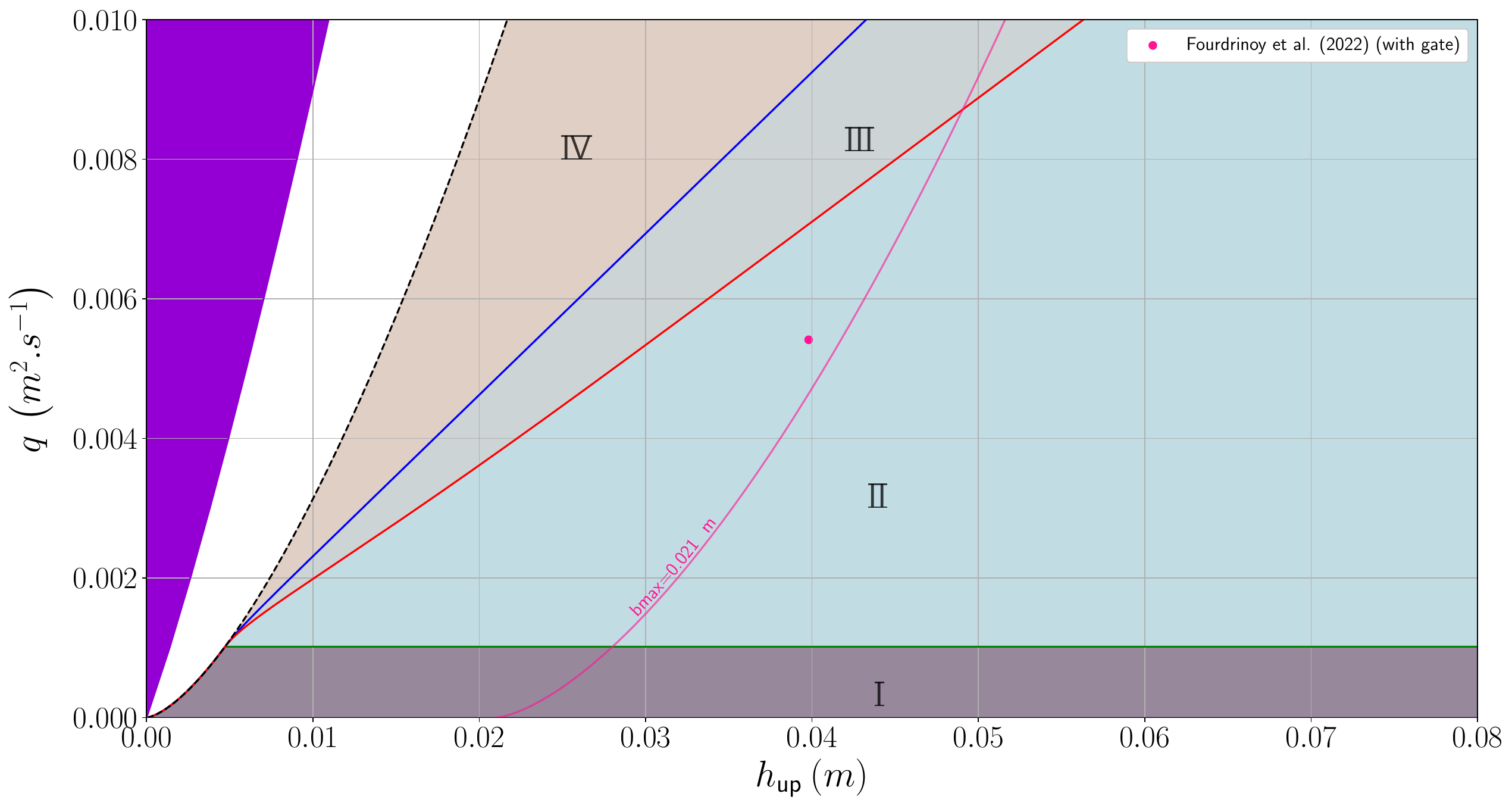}
    \caption{Phase diagram for a transcritical regime, taking dispersive effects into account. For this diagram, the experimental point (i.e. ($h_\text{up},q$)) used is: ($0.0398\: m\:\text{,}\:0.0054\: m^2.s^{-1}$) in the case of Fourdrinoy et al. (2022) (with an exit gate) \cite{fourdrinoy2022correlations}. The maximum flow rate of the pump, for the experiment by Fourdrinoy et al. (2022)\cite{fourdrinoy2022correlations}, is $q_\text{max}=0.0119\:m^2.s^{-1}$.}
    \label{q vs h histo trans 2.1}
\end{figure}

For all the Figures \ref{Fr vs tanhh histo}, \ref{q vs h histo trans 1.8}, \ref{q vs h histo trans 1.1} and \ref{q vs h histo trans 2.1}, it is obvious to confirm that the flow regimes were all transcritical and of a black hole flow type but, more importantly, the experimental points are very close to the theoretical transcritical frontiers of Long, equation \ref{Long} parameterized by each $b_{max}$.

The transcritical flow in presence of the exit gate reported in \cite{fourdrinoy2022correlations} is very interesting since it showed that Hawking process took place as confirmed by the measurements of correlations but it was confined in the vicinity of the obstacle by the blue-shifted horizon since the flow speed drops below the Landau speed in the descending slope. Once again, this scenario was not envisaged in General Relativity since it depends on the existence of a small dispersive scale (here the capillary length). Our future work will be to confirm (or not) the confined Hawking radiation with a subcritical flow by revisiting the 2008 and 2020 Nice experiments \cite{rousseaux2008observation, rousseaux2010horizon, chaline2013some}...

\subsection{Numerical Simulations}

\subsubsection{Description of the code}

The present numerical simulations are achieved using a 2D LevelSet/Cut-cell method (see \cite{bossard2023create} and \cite{BDJ2012} for details) which gives us an approximation of the free surface, but not only that, it also provides velocity and pressure field that improve our understanding of the flow. We are particularly interested in the thickness of the velocity boundary layer and the presence of recirculation zone behind the obstacle. Indeed, these phenomena could explain the difference observed between experimental and theoretical results.

\subsubsection{Numerical Validation}

These simulations will make it possible to validate an important hypothesis, on which all the above is based, i.e. that the longitudinal velocity is of the type $v(x)=v_x(x,z)=q/h(x)\:\forall z$. This hypothesis will mainly be verified for the flow upstream of the obstacle, as it is from the conditions upstream of the obstacle that the phase diagrams and therefore the classification depend. To verify this hypothesis, two simulations were carried out: one for a transcritical regime and one for a subcritical regime.

\begin{figure}[h!]
    \centering
    \includegraphics[scale=0.35]{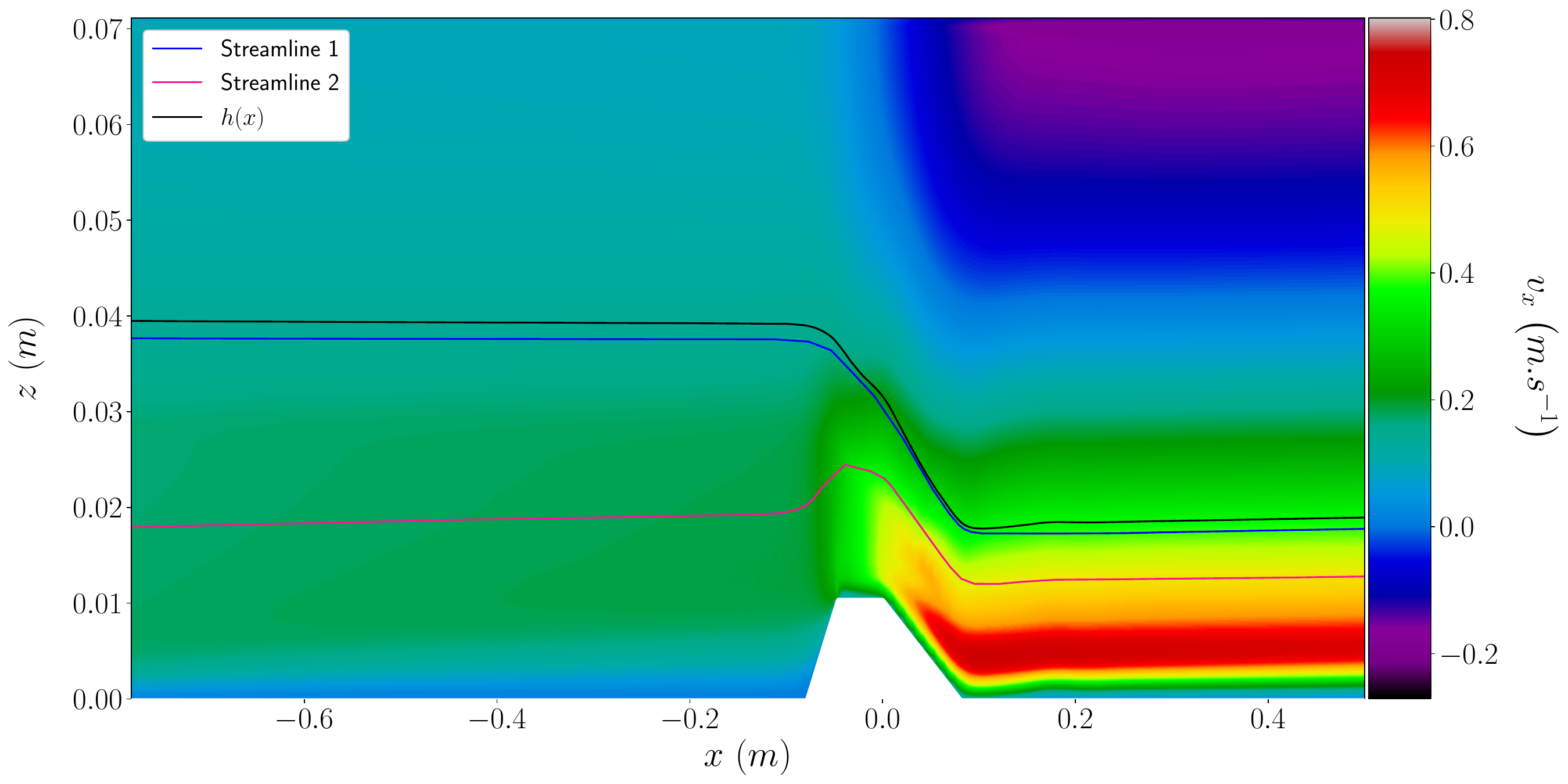}
    \caption{Values of the projection of the longitudinal speed ($v_x$) with its sign for a transcritical black hole flow in a numerical simulation. We can also see the water level (in black) at the free surface and two streamlines: the first close to the free surface (in blue) and the other in the center of the bulk flow (in pink).}
    \label{Simu Trans Ux}
\end{figure}

\begin{figure}[h!]
    \centering
    \includegraphics[scale=0.35]{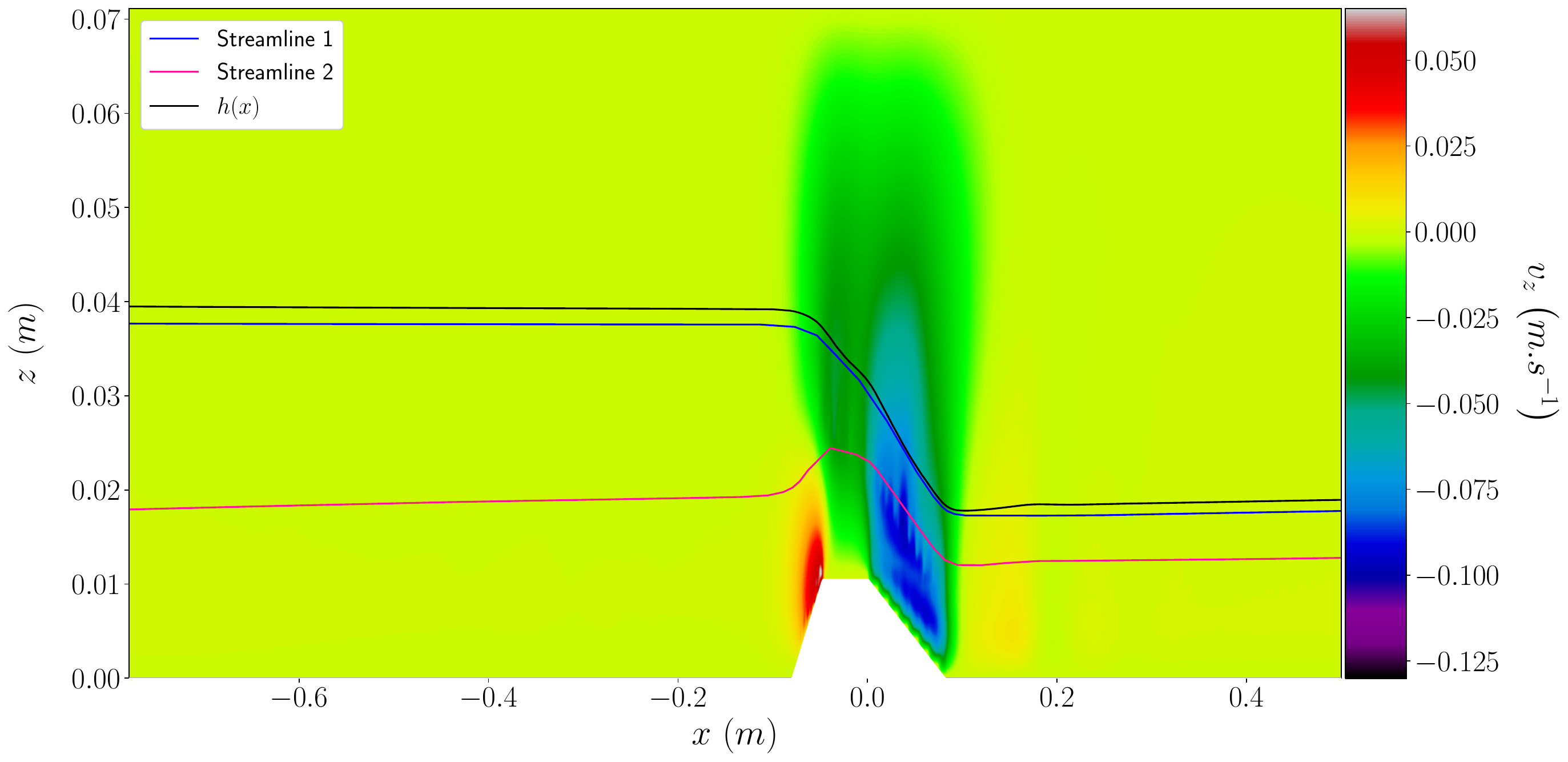}
    \caption{Values of the projection of the transverse speed ($v_z$) with its sign for a transcritical black hole flow in a numerical simulation. We can also see the water level (in black) at the free surface and two streamlines: the first close to the free surface (in blue) and the other in the center of the bulk flow (in pink).}
    \label{Simu Trans Uz}
\end{figure}

\subsubsection{Transcritical case}

For the transcritical regime ( which corresponds to the reproduction of the $\text{T}^\nearrow_{\text{\Romanbar{2}}}-\text{T}^{\searrow,\text{U}}_{\text{\Romanbar{4}}}$ regime in the figure \ref{Planches_trans_sans_porte}), we select the flow rate as $q=0.0047\: m^2.s^{-1}$ and the ACRI 2010 geometry (Figure \ref{ACRI2010}) with a maximum height of $b_\text{max}=0.0105\: m$. In order to carry out the numerical simulations, we make the following choices:
\begin{itemize}
    \item The computational domain $[-1;2.3]\times[0;0.0825]$ is discretized by a $2400\times60$ uniform grid, therefore the cell size equals $\delta\!x = \delta\!z = 0.001375\: m$; 
    \item The time step $\delta\!t = 0.00025\: s$ is constant;
    \item Initial conditions: at time $t=0$, upstream of the obstacle the water height equals $h = 0.0215\: m$ while $h = 0.011\: m$ elsewhere; in addition, we use the plug flow assumption $u = q / h$ for the velocity profile;
    \item Velocity boundary conditions (b.c.): no-slip b.c. ($\mathbf{v}=\mathbf{0}$) on the bottom of the channel and on the obstacle, advective b.c. on the outlet, Dirichlet b.c. on the inlet and slip b.c. ($\partial_{z v_x}=0$ and $v_z=0$) on the top of the channel;
    \item Levelset b.c.: the free surface is allowed to slide on the walls, i.e. the upstream and downstream water heights are not fixed. Only the inlet flow rate $q$ is enforced; 
    \item The level set redistancing algorithm is applied every $100$ time iterations and $1000$ fictitious iterations ensure that the air-water interface is represented as the zero level set of a signed distance function in the whole computational domain; 
\end{itemize}

\begin{figure}[h!]
    \centering
    \includegraphics[scale=0.35]{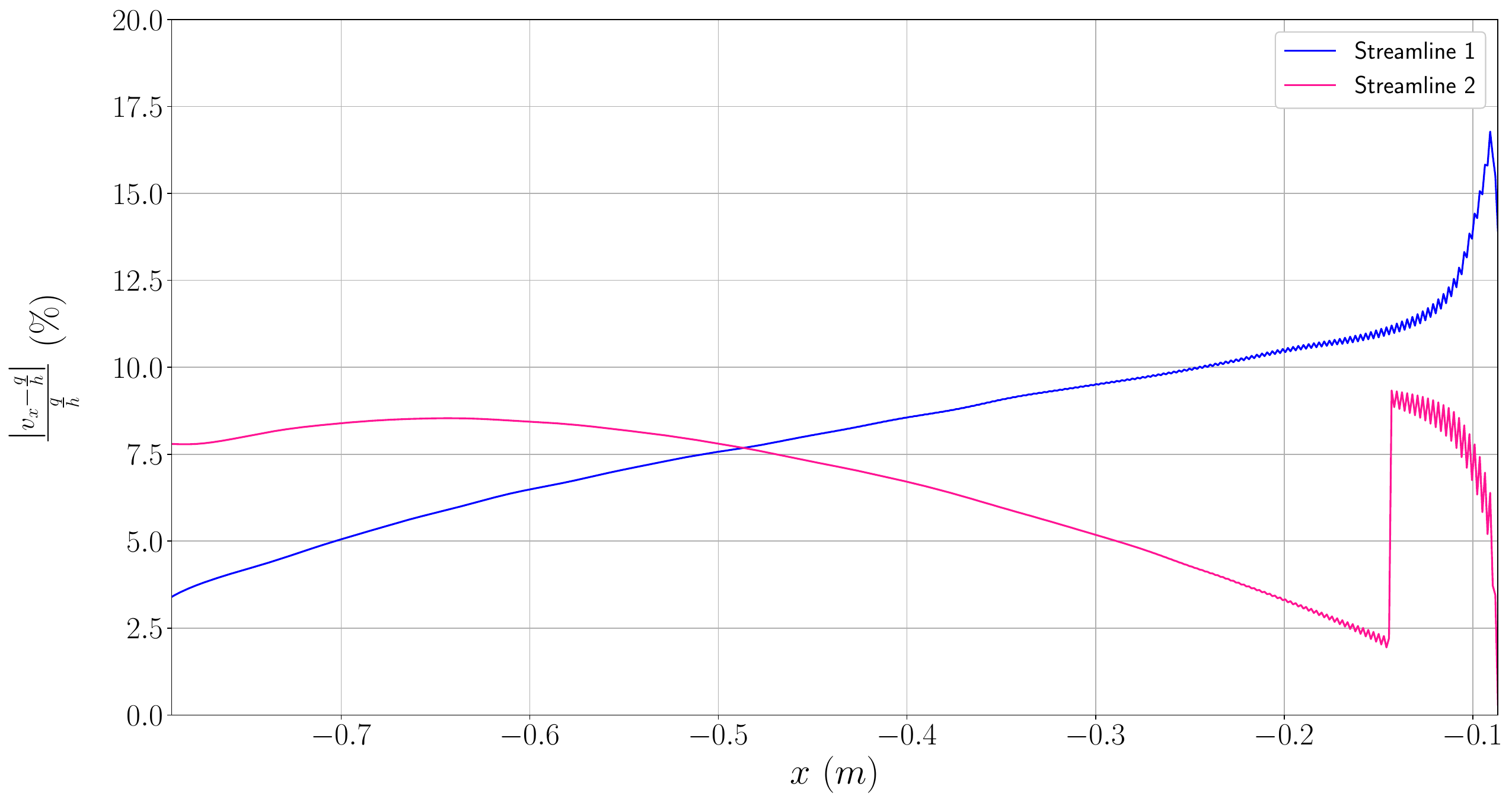}
    \caption{Relative difference between the longitudinal velocity on a streamline and the velocity model $v(x)=q/h(x)$ for a transcritical black hole flow in a numerical simulation in the upstream region. The average relative deviation for the closest streamline to the free surface (in blue) is about $7.2\%$. The mean relative deviation for the bulk streamline (in pink) is about $6.5\%$.}
    \label{Ecart}
\end{figure}

\begin{figure}[h!]
    \centering
    \includegraphics[scale=0.35]{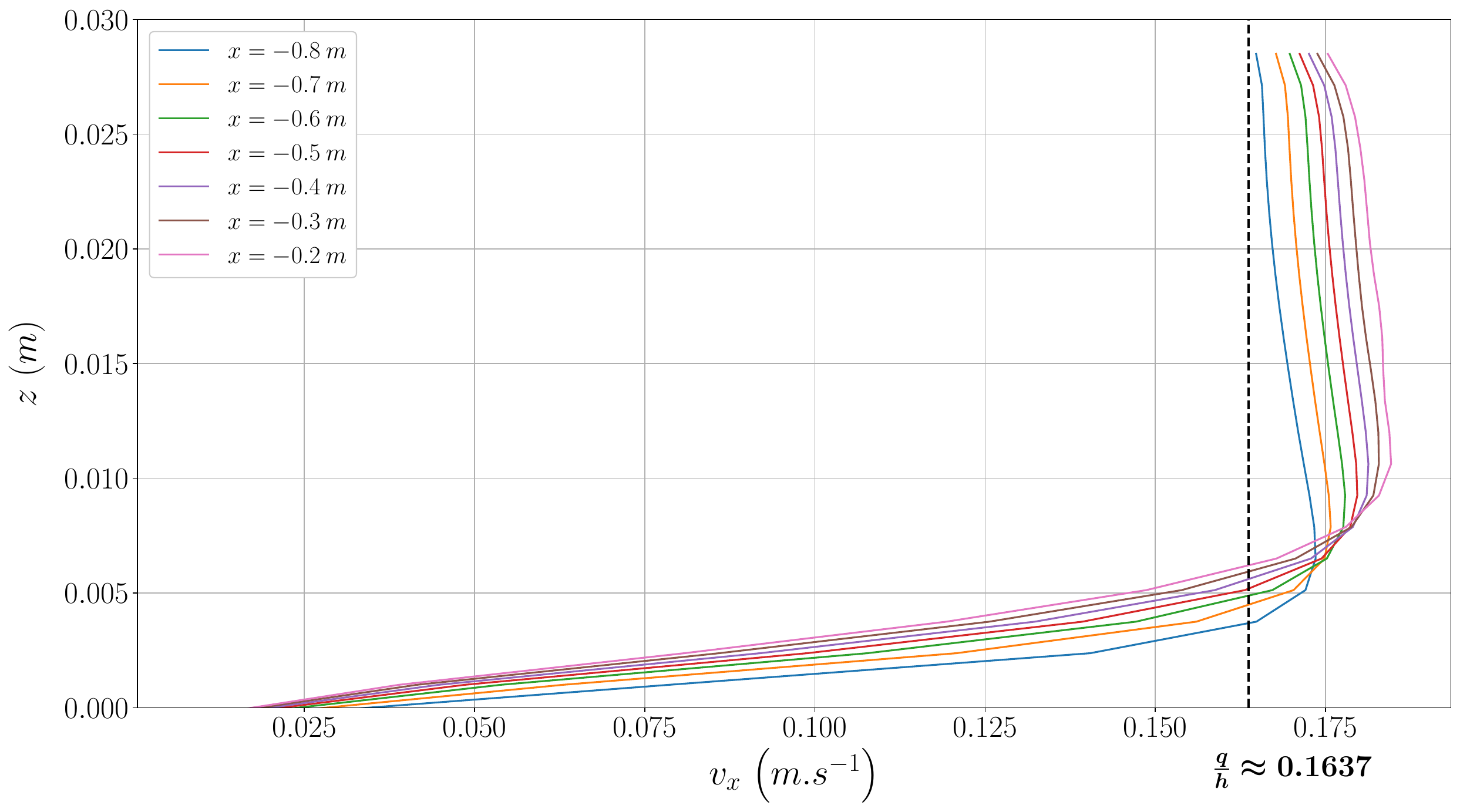}
    \caption{Longitudinal velocity ($v_x$) profile as a function of the depth upstream of the obstacle for a transcritical black hole flow in a numerical simulation.}
    \label{Profil}
\end{figure}

We observe that the CFL number $\frac{\delta\!t ||\mathbf{v}||_\infty}{\delta\!x}$ remains below $0.153$. 
Thanks to the BoomerAMG algebraic multigrid solver, the incompressibility constraint is well satisfied: the discrete divergence of the velocity equals approximately $10^{-12}$. The CPU time is divided like this: transport of the level set function and redistancing ($44\%$), prediction and projection ($56\%$). 
The Figures \ref{Simu Trans Ux}--\ref{Profil} show how the numerical simulations succeed in both reproducing the experiment and estimating the thickness of the boundary layer. They justify also the plug flow hypothesis in this case. In fact, the relative difference (figure \ref{Ecart}) between the speed resulting from the simulation and the speed model is around 7\% on average, upstream of the obstacle.

\subsubsection{Subcritical case}

\begin{figure}[h!]
    \centering
    \includegraphics[scale=0.35]{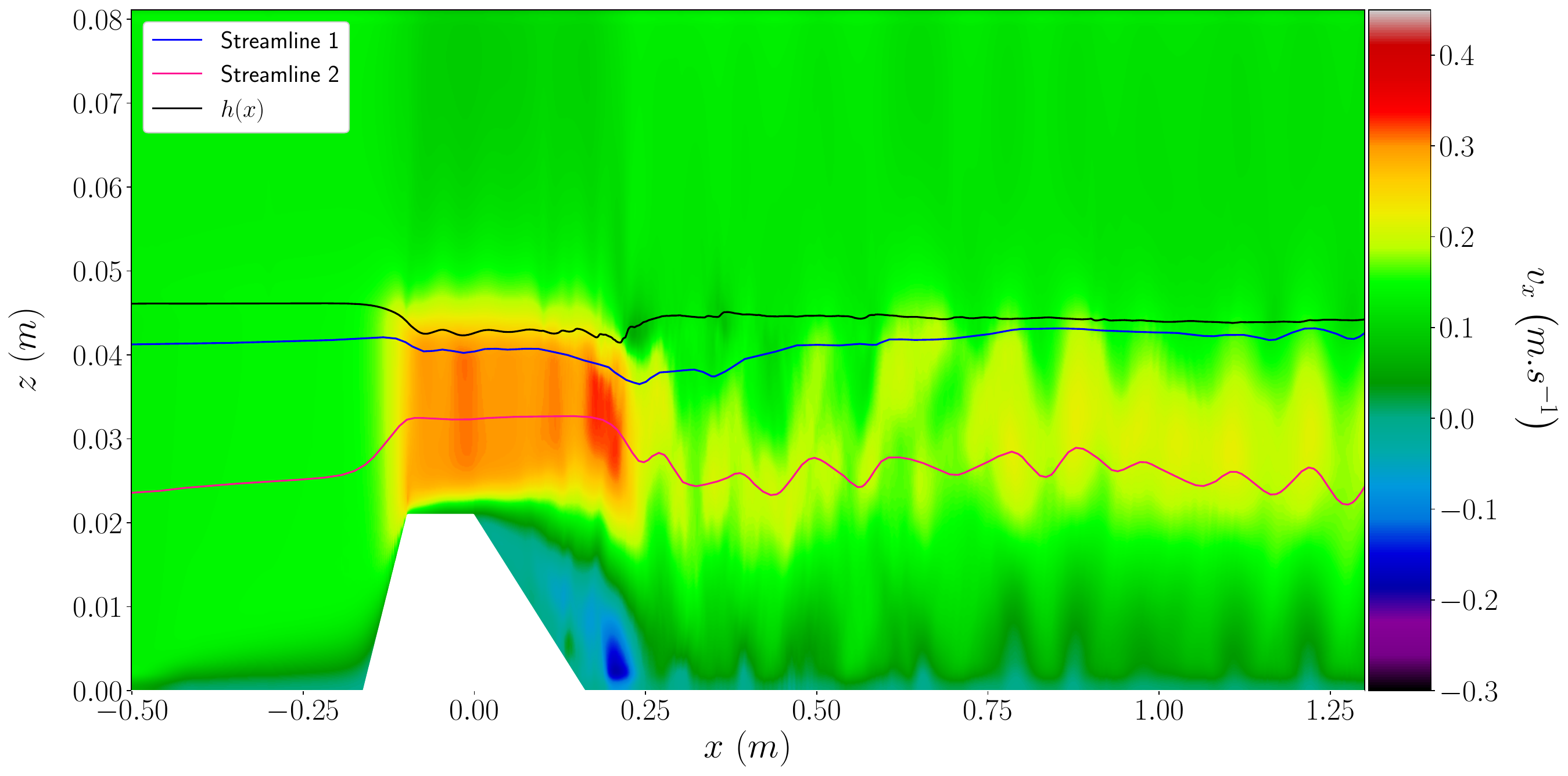}
    \caption{Values of the projection of the longitudinal speed ($v_x$) with its sign for a subcritical flow in a numerical simulation. We can also see the upstream water level (in black) and two current lines: a current line close to the free surface (in blue) and a current line in the middle of the flow (in pink).}
    \label{Simu sous Ux}
\end{figure}

\begin{figure}[h!]
    \centering
    \includegraphics[scale=0.35]{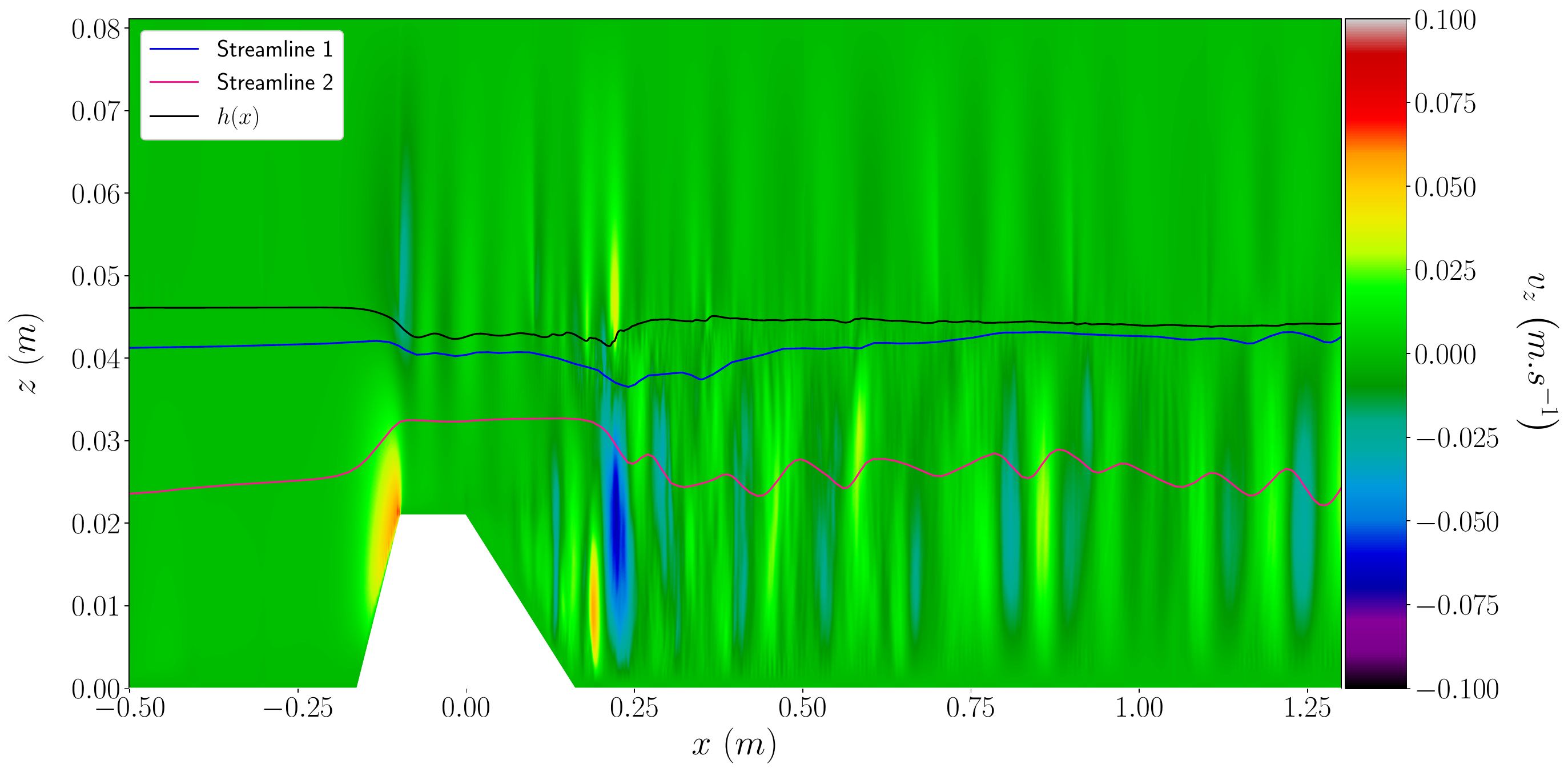}
    \caption{Values of the projection of the longitudinal speed ($v_z$) with its sign for a subcritical flow in a numerical simulation. We can also see the water level (in black) at the free surface and two streamlines: the first close to the free surface (in blue) and the other in the center of the bulk flow (in pink).}
    \label{Simu sous Uz}
\end{figure}

\begin{figure}[h!]
    \centering
    \includegraphics[scale=0.35]{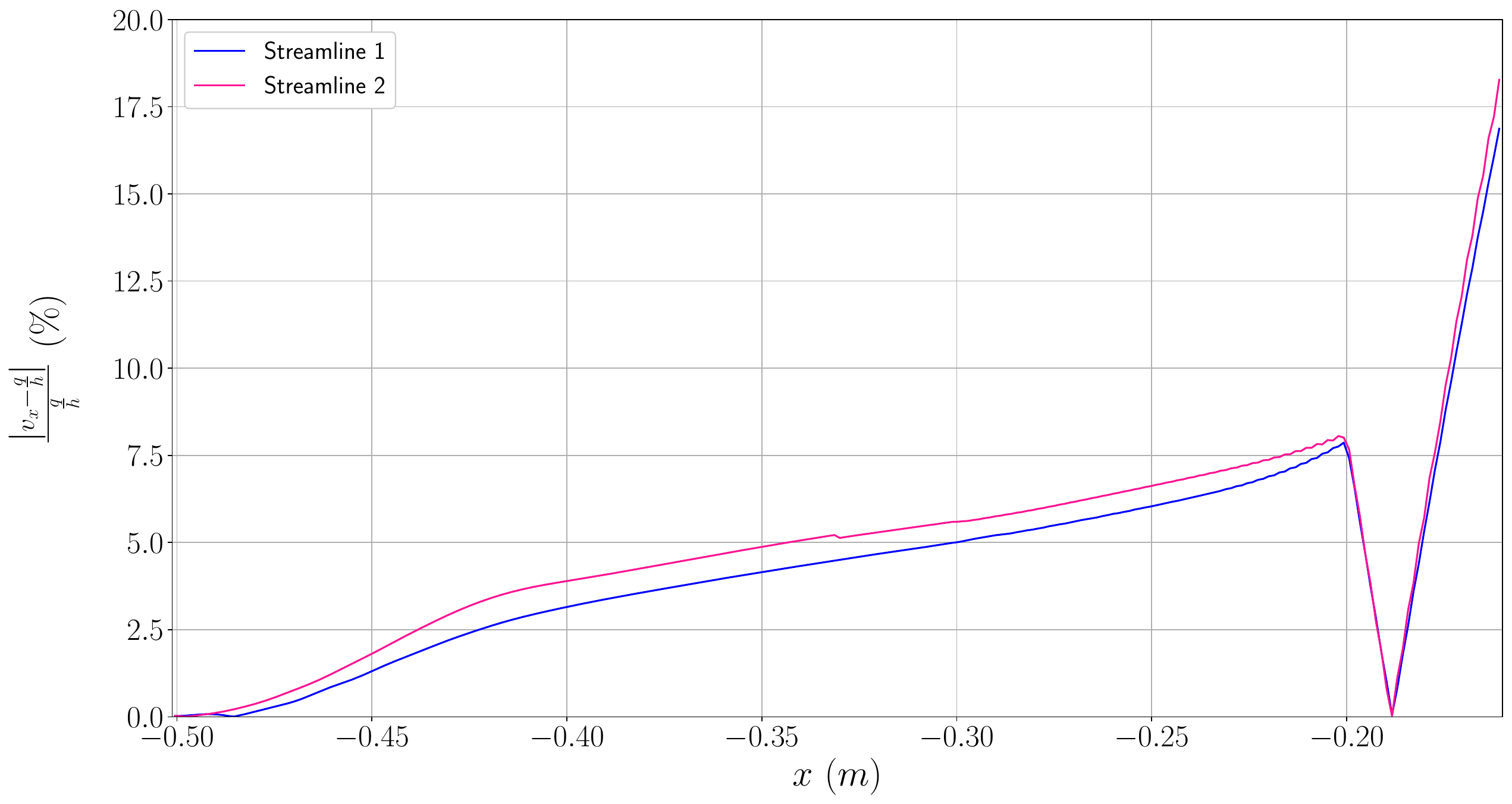}
    \caption{Relative difference between the longitudinal velocity on a streamline and the velocity model $v(x)=q/h(x)$ for a subcritical flow in a numerical simulation in the upstream region. The average relative deviation for the closest streamline to the free surface (in blue) is about $4.27\%$. The mean relative deviation for the bulk streamline (in pink) is about $4.83\%$.}
    \label{Ecart sous}
\end{figure}

\begin{figure}[h!]
    \centering
    \includegraphics[scale=0.35]{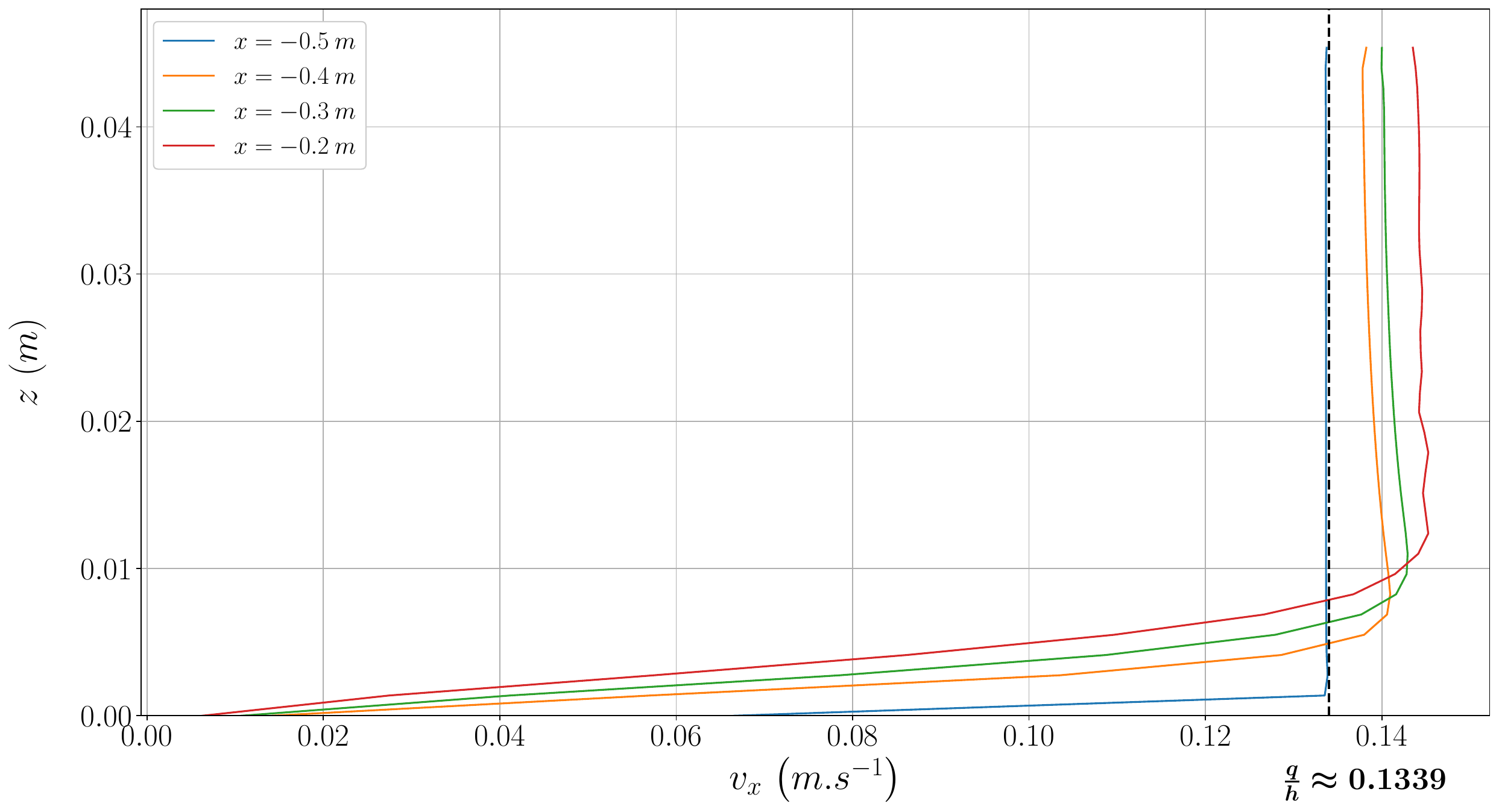}
    \caption{Longitudinal velocity ($v_x$) profile as a function of the depth upstream of the obstacle for a subcritical flow in a numerical simulation.}
    \label{Profil sous}
\end{figure}

For the subcritical regime ( which corresponds to the reproduction of the $D_{\Romanbar{2}^\prime}$ regime in the figure \ref{Planches_sous_sans_ondu}), the numerical simulation was produced with the following conditions: flow rate $q= 0.00616 \: m^2.s^{-1}$, ACRI 2010 geometry with a maximum height of $b_\text{max}=0.021\: m$ and the same parameters for the numerical method, except for time step $\delta\!t = 0.0001\: s$. Here again, the numerical simulations justify the plug flow hypothesis hence the acoustic metric formalism, a very important assumption in Analogue Gravity that validates the importance of this appendix (see Figures \ref{Simu sous Ux}--\ref{Profil sous}). In fact, the relative difference (figure \ref{Ecart sous}) between the speed resulting from the simulation and the speed model used in the theoretical part of the article is around 5\% on average, upstream of the obstacle.

\end{document}